\documentclass[manuscript]{aastex}
\usepackage{amsmath}
\usepackage{multirow}

\newcommand{\HI}{\ion{H}{1}}
\newcommand{\et}{et al.}
\newcommand{\kms}{km s$^{-1}$}

\begin{document}
\slugcomment{\today}

\title{The \HI\ Chronicles of LITTLE THINGS BCDs: Evidence for External Perturbations in the Morphology and Kinematics of Haro 29 and Haro 36}
\author{Trisha Ashley and Caroline E. Simpson}
 \affil{Department of Physics, Florida International University}
 \affil{11200 SW 8th Street, CP 204, Miami, FL 33199}
\email{trisha.ashley@gmail.com}
 \email{simpsonc@fiu.edu}

 \author{Bruce G. Elmegreen}
 \affil{IBM T. J. Watson Research Center, PO Box 218, }
 \affil{Yorktown Heights, New York 10598}
\email{bge@us.ibm.com}

 \begin{abstract}

We analyze high angular and velocity resolution \HI-line data of two LITTLE THINGS\footnote{``Local Irregulars That Trace Luminosity Extremes The \HI\ Nearby Galaxy Survey"  http://www2.lowell.edu/users/dah/littlethings/index.html} blue compact dwarfs (BCDs): Haro~29 and Haro~36.  Both of these BCDs are disturbed morphologically and kinematically.  Haro 29's \HI\ data reveal a kinematic major axis that is offset from the optical major axis, and a disturbed outer \HI\ component, indicating that Haro 29 may have had a past interaction.  Position-velocity diagrams of Haro~36 indicate that it has two kinematically separate components at its center and a likely tidal tail in front of the galaxy.   We find that Haro 36 most likely had an interaction in the past, is currently interacting with an unknown companion, or is a merger remnant.  

 \end{abstract}
 

 \section{Introduction}
Blue compact dwarf galaxies (BCDs) are some of the more puzzling galaxies in the universe.  They are undergoing a burst of star formation, often without an obvious trigger, resulting in bluer colors than other irregular galaxies \citep{kunth95}, hence the ``blue" in their name.   The dense, almost star-like appearance of BCDs on photographic plates gave them the ``compact" in their name \citep{zwicky66}.  It was originally speculated that BCDs might be young galaxies (forming stars for the first time) because of their low metallicities and because the young stellar population was the only component seen \citep{searle73}.  \citet{loose85} quelled that theory when they published data revealing that most BCDs have an extended, faint, older stellar disk, leading to the question: what do BCDs look like when they are not undergoing a starburst? 

	It is clear that BCDs have not been undergoing a starburst in isolation forever, as their gas consumption rates would have depleted the gas in $\thicksim$$10^9$ years without additional accretion \citep{gil05}.  BCDs could, however, have long quiescent stages in which little star formation occurs and then periodically have a burst of star formation without depleting gas reservoirs quickly \citep{searle73,  mas-hesse99}.  Alternatively, BCDs could have been forming stars all along at rates lower than their current star formation rate, and then only more recently have had an increase in activity \citep{schulte01, crone02}.  So if BCDs have not been undergoing a starburst forever, then what is the cause of the recent frenzy of star formation?   

	A popular theory is that BCDs are a result of interactions or mergers \citep{noeske01, pustilnik01, bekki08}, but a major weakness of the theory is that many BCDs are found to be isolated \citep{campos93, telles95, taylor97}, therefore it is difficult to find a source for the external disturbance. If BCDs are not a phase in an isolated evolutionary process, but are instead the result of an external perturbation, then their place in the spectrum of dwarf galaxies would be better understood.  Deep optical observations have been able to find faint perturbers, as in the case of NGC 4449, where gas and stars appear disturbed, as if an interaction occurred.  This eventually led to a deep optical search around NGC 4449, resulting in the discovery of a faint dwarf spheroidal companion \citep{delgado12}. The first clues to an external perturbation could then be in the gas and stars of a galaxy.  
	
	External perturbations can also leave morphological and kinematic signatures in the neutral atomic hydrogen (\HI) gas.  The morphology of the gas should contain signatures of possible external perturbations such as tidal tails and bridges.  A bridge can indicate the early stages of a merger or an ongoing interaction. A narrow, curved tidal tail will denote a past close encounter with another galaxy \citep{toomre72}.  The kinematics of the gas can distinguish violent (e.g., mergers) from more passive processes (e.g., slow accretion of a gaseous halo): for example, more violent processes can cause warps in the disk \citep{hunter69, shang98, sanchez05}.  			

	In this paper we analyze the Karl G. Jansky Very Large Array (VLA\footnote{The VLA is a facility of the National Radio Astronomy Observatory (NRAO). The NRAO is a facility of the National Science Foundation operated under cooperative agreement by Associated Universities, Inc. These data presented in this paper were collected during the upgrade of the VLA to the Extended VLA or EVLA.}) \HI-line integrated maps and intensity-weighted velocity field maps of two BCDs from the LITTLE THINGS sample: \objectname{Haro~29} (=\objectname{Mrk 209}=\objectname{I Zw 36}=\objectname{UGCA 281}) and \objectname{Haro~36} (=\objectname{UGC 7950}) to build not only a more complete picture of these two galaxies, but also to look for signs of external perturbations in the neutral gas. The LITTLE THINGS survey is studying star formation in forty-one dwarf irregular galaxies, including six BCDs.  The survey was granted 367 hours at the VLA in B, C, and D arrays to observe the \HI-line of the sample.  These data were added to archival data, when available, to create combined B+C+D array maps of high angular resolution (reaching as high as 6\arcsec) and high velocity resolution (2.6 \kms\ or 1.3 \kms).  Some basic information for Haro 29 and Haro 36 (distance, systemic velocity, radius, etc.) is listed in Table~\ref{tab:galinfo} and some basic observing information (time on source, project ID, etc.), is listed in Table~\ref{tab:obsinfo}. The \HI-line data presented here have a minimum velocity resolution $\thicksim$1.3 \kms, and a minimum angular resolution of $\thicksim$7\arcsec, resulting in a linear resolution of 200 pc for Haro 29 (at a distance of 5.8 Mpc) and 310 pc for Haro 36 (at a distance of 9.3 Mpc).  High resolution data like these should be able to reveal velocity and morphological signatures of external perturbations in the dense regions of these galaxies.
	
	Despite the LITTLE THINGS galaxies being chosen so that they are a sample of relatively isolated galaxies, many show disturbed morphologies and kinematics \citep{hunter12} that are a possible sign of external perturbations.  Both Haro 29 and Haro 36 were chosen for this paper because they contain dramatically disturbed morphology and kinematics in their \HI, making them likely candidates for external perturbations despite their apparent isolation.  Both galaxies also currently lack in-depth, high resolution, individual \HI\ studies, and have no obvious reason for their current starburst. Haro 29's closest known companion is NGC 4736 \citep{karach04} at a deprojected distance of $\thicksim$1.4 Mpc from Haro 29 (using the tip of the red giant branch distance for NGC 4736 given by \citealt{jacobs09}) and the closest known companion to Haro 36 is NGC 4707 at a deprojected distance of $\thicksim$0.8 Mpc from Haro 36 \citep{hunter04}.
	
	Haro~29's \HI\ has been studied previously with the Westerbork Synthesis Radio Telescope at a velocity resolution of 2.06 \kms\ and an angular resolution of $\thicksim$13\arcsec\ or linear resolution of 370 pc using a distance of 5.8 Mpc \citep{stil02a}.   The \HI\ was also studied by \citet{viall83} using the Westerbork Synthesis Radio Telescope, but at a lower resolution of 37\arcsec.2.  The stellar population of Haro~29 has been more extensively studied than the \HI.   The older stellar population has an age of at least $\thicksim$2 Gyrs \citep{schulte01} and the younger stellar population contains stars at least as young as a few Myrs, as evident by its Wolf-Rayet stars \citep{izotov97}. The dominant younger stellar population has four super star clusters (SSCs; \citealt{thuan05}).  To the east of the SSCs, there is weak H-$\alpha$ emission that lacks near-infrared (NIR) emission.  This lack of NIR emission could be due to the area not having had time to form red supergiants yet, making the recent burst very young.  \citet{schulte01} studied the NIR in detail using Hubble Space Telescope data and creating color-magnitude diagrams (CMDs) of the SSC regions with the goal of separating out the truly ancient stellar population of the galaxy.  They discovered that in Haro 29, the red, ancient stars (or ``Baade's sheet") have a nearly uniform distribution in Haro 29.  They also modeled the star formation history of Haro 29, revealing that the most likely scenario included stars older than 1 Gyr, up to 15 Gyr old, forming at a low star formation rate of $4.3\times10^{-4}\ \rm{M}_{\sun}\ yr^{-1}$ and stars younger than 1 Gyr old forming at an average, moderate rate of $5.7\times10^{-3}\ \rm{M}_{\sun}\ yr^{-1}$, allowing the current gas supplies to last another 10 Gyr.   The current SFR in Haro 29, measured from the FUV and normalized to the area swept out in a circle with a radius defined by the V-band disk scale length, is $8.51\times10^{-2}\ \rm{M}_{\sun}\ yr^{-1}\ kpc^{-2}$ \citep{hunter12}, an order of magnitude larger than the average rate over the last gigayear estimated by \citet{schulte01}.  Higher star formation rates, like the current one, are considered to be short fluctuations by \citet{schulte01}.  

	Haro~36 has been included in several larger surveys \citep[e.g.,][]{kennicutt08, dale09, lee11} and has been studied in \HI\ using the VLA C-array \citep{simp00} at a velocity resolution of 12.5 \kms\ and an angular resolution of $\thicksim$26\arcsec.7 or a linear resolution of 1.2 kpc using a distance of 9.3 Mpc.   The stellar population is briefly discussed in \citet{hunter06}, where they noted that the blue stellar component is elongated along the major axis.  The current SFR, measured from the FUV and normalized to the area swept out in a circle with a radius defined by the V-band disk scale length, is $2.82\times10^{-2}\ \rm{M}_{\sun}\ yr^{-1}\ kpc^{-2}$.
	
In section~\ref{sec:obser} we describe the observations and data reduction.  In sections~\ref{sec:results29}~and~\ref{sec:results36} we present the \HI\ moment maps and position-velocity diagrams.  We discuss the results in sections~\ref{sec:disc29}~and~\ref{sec:disc36}, with the conclusions stated in section~\ref{sec:concl}.
	
  \section{Observations and Data Reduction}\label{sec:obser}

	The data were obtained during the VLA upgrade to the Extended VLA (EVLA).  The combination of VLA and EVLA antennas resulted in several unique problems with the data that are discussed in detail in \citet{hunter12} including the ``aliasing problem"\footnote{https://science.nrao.edu/facilities/vla/obsolete/aliasing} and an inability to Doppler track.  The EVLA-EVLA baselines were significantly affected by the aliasing problem resulting in all of them being flagged during calibration. Since Doppler tracking was not possible, observations occurred at a fixed frequency and checks were made to ensure that the galaxy did not move out of the channel.  
	
	The \HI\ data were calibrated and mapped using the LITTLE THINGS reduction recipe in AIPS \citep[see][]{hunter12}. Mapping was \textit{not} done using the classical \textsc{clean} algorithm, which treats each source as a sum of point sources, removing the brightest first, and then working its way down to a set limit.  Instead, mapping was done using the AIPS Multi-Scale (\textsc{m-s}) \textsc{clean} algorithm option in \textsc{imagr}.  This algorithm first convolves the data cube by multiple beams; for LITTLE THINGS data these beam sizes were chosen to be 0 (or no convolution), 15, 45, and 135 arcsec.  The larger synthesized beams result in a lower angular resolution and higher surface brightness sensitivity, mapping the more tenuous, large-scale features with a higher signal-to-noise than the smaller beams, which result in a higher angular resolution but lower surface brightness sensitivity.  \textsc{ms-clean} then chooses the convolved cube with the highest peak flux, images it, cleans the image (in a classic \textsc{clean} manner) down to a user-specified level, and then begins the process over again.  \textsc{ms-clean} therefore still has the ability to image high resolution details of the galaxy, but at the same time, it recovers more faint emission from the large-scale structures than classic \textsc{clean}.  Recovering this emission gives \textsc{ms-clean} the distinct advantage of more accurate HI flux measurements as shown in \citet{hunter12}.  The benefits of \textsc{ms-clean} are given in more detail by \citet{cornwell08}. 

	\textsc{imagr} was used to produce two sets of maps, the natural-weighted maps and the robust-weighted maps (robust value of 0.5).  The natural-weighted maps have a larger synthesized beam (higher sensitivity and lower resolution) than the robust-weighted maps, which have a resolution and sensitivity that falls between the natural and uniform weighted maps.  The time on source for Haro 29 and Haro 36 are 18.467 hours and 18.275 hours respectively and their expected r.m.s. values are 0.60 mJy and 0.61 mJy, which are consistent with the measured RMS values.  Information for each set of maps (resolution, beam sizes, and noise levels) is listed in Table~\ref{tab:mapinfo}.  For more details on the mapping process, see \citet{hunter12}. 
	
	The LITTLE THINGS team has collected ancillary data for all of the galaxies, including V-band and far ultraviolet (FUV) data.  The V-band data were obtained at Lowell Observatory \citep{hunter06} and provide information about the integrated star formation over the past 1~Gyr.  The FUV data were obtained from the Galaxy Evolution Explorer satellite (\textit{GALEX}) and provide information about the star formation occurring over the past 100-200 Myr \citep{hunter10}.

\section{Results: Haro 29}\label{sec:results29}

\subsection{Stellar Component}

\begin{figure}[!ht]
\centering
\epsscale{1.11}
\plottwo{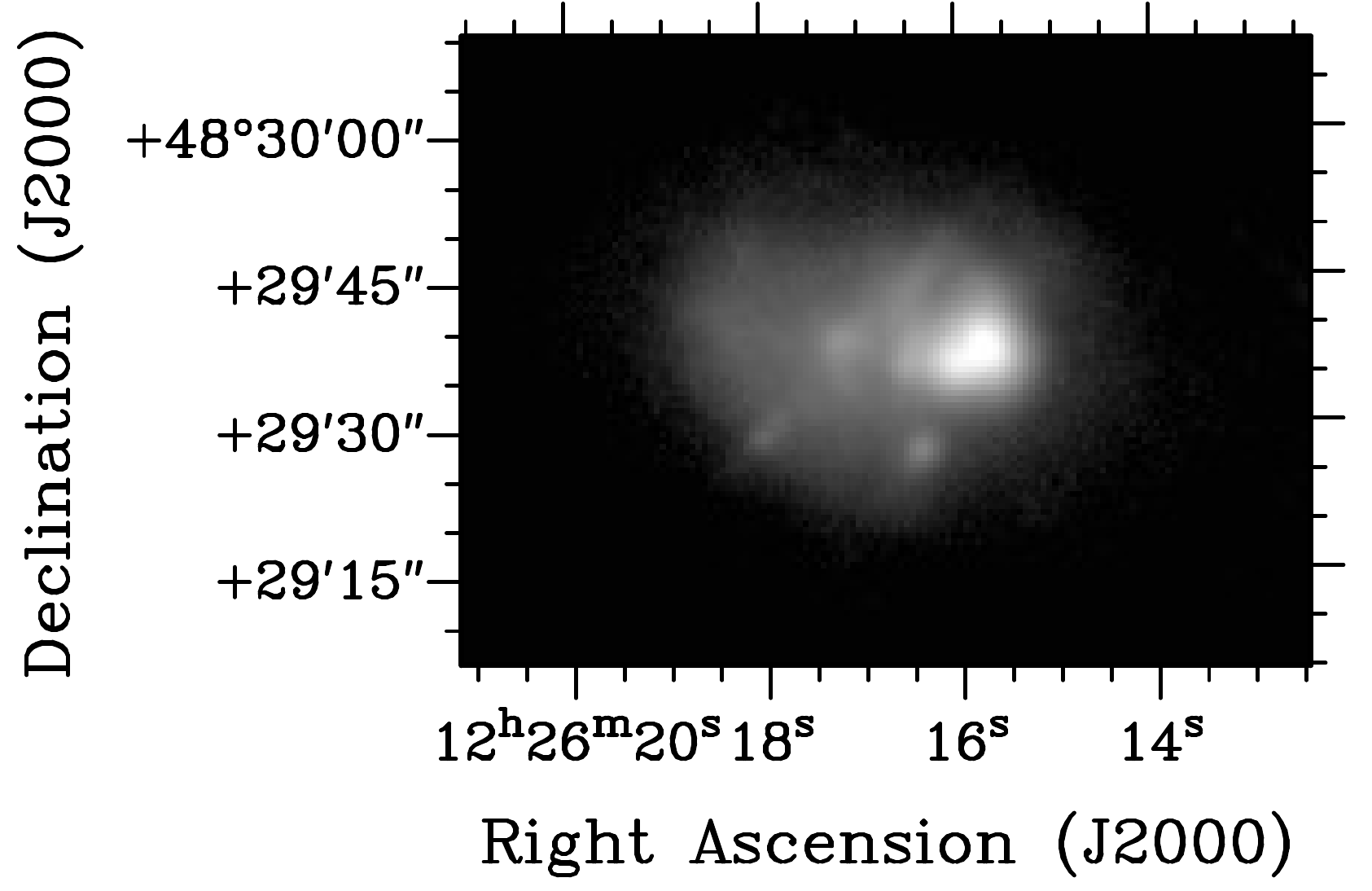}{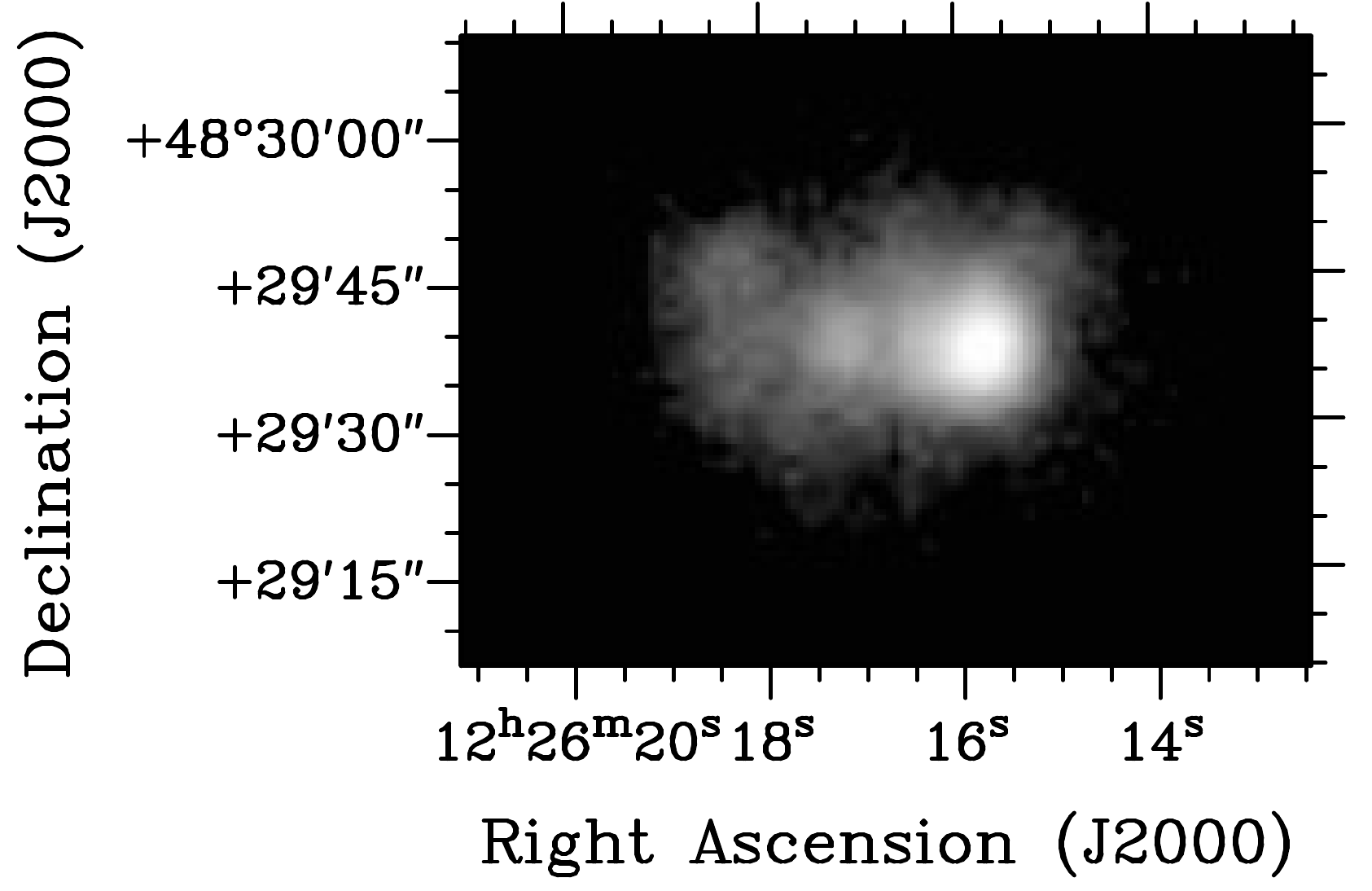}
\figcaption{Haro~29\ \ \textit{Left:} V-band;\ \ \textit{Right:} FUV \label{fig:h29v_fuv}}
\end{figure}

The FUV and V-band data for Haro~29 are shown in Figure~\ref{fig:h29v_fuv}.  The V-band and FUV follow nearly the same morphology. Since the morphologies of the two sets of data are so similar, it is likely that the younger stellar population is dominating the V-band emission in the central regions.  

\subsection{\HI\ Morphology}  \label{sec:results29morph}

\begin{figure}
\epsscale{.492}
\plotone{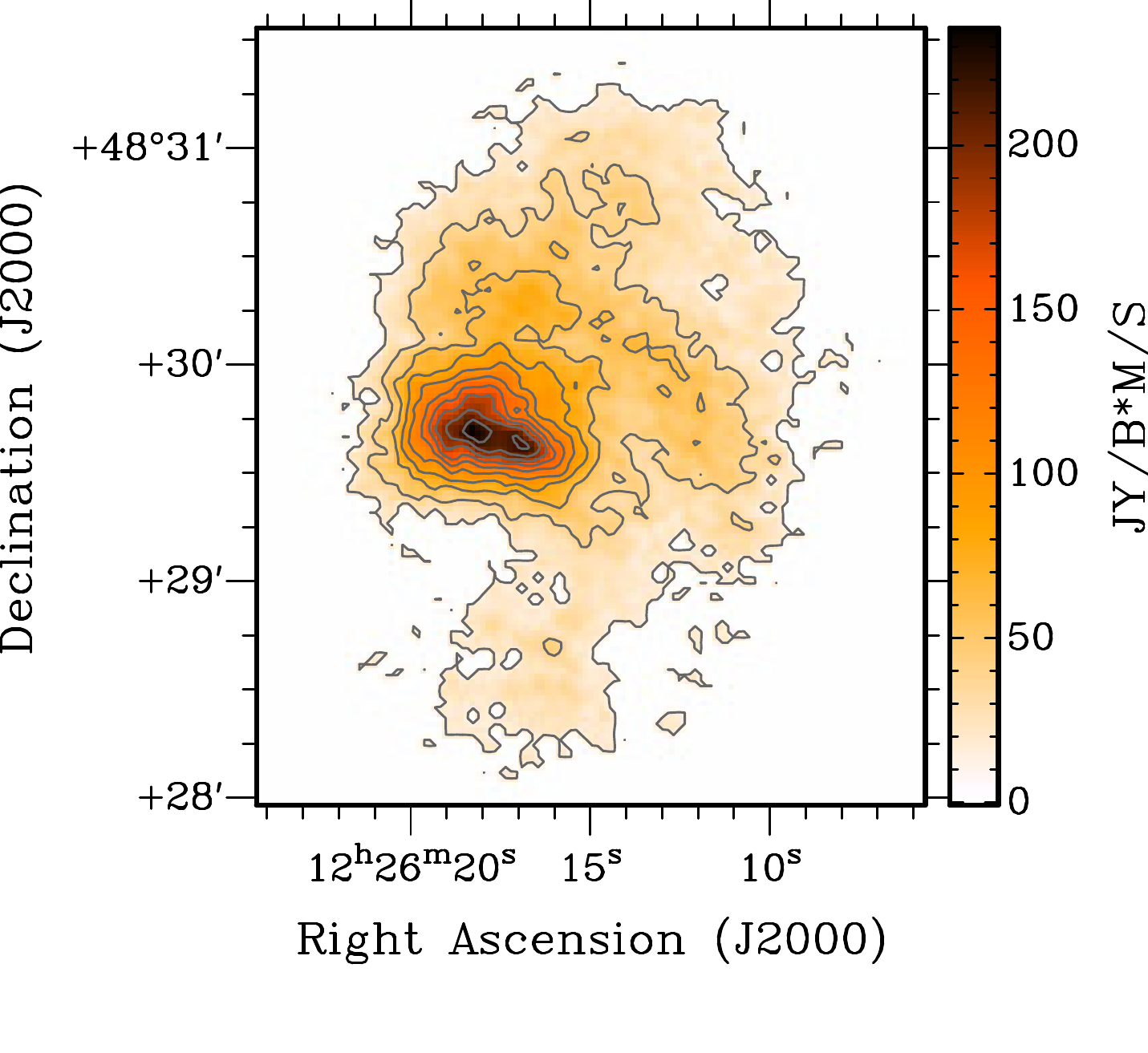}
\plotone{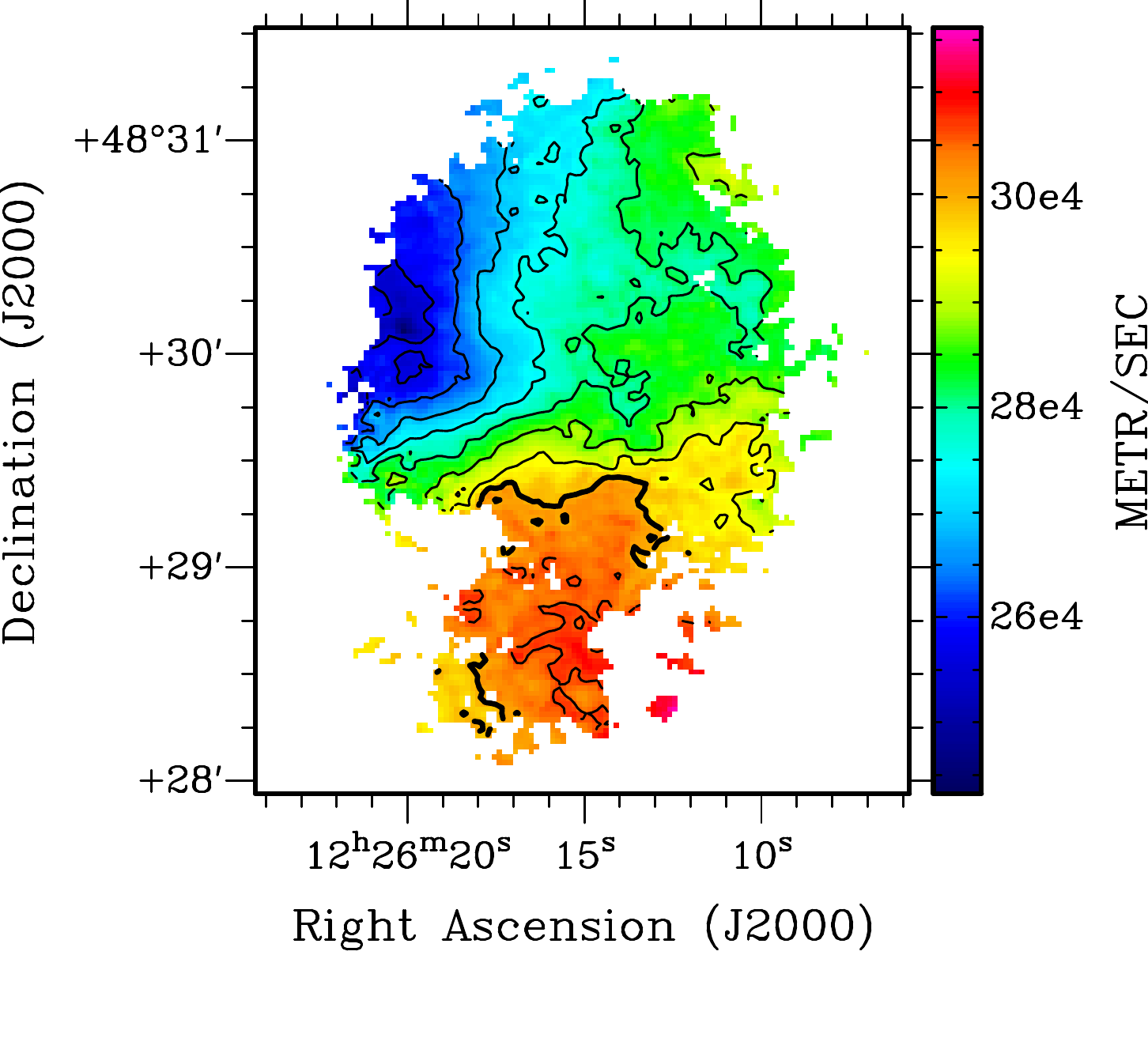} \\
\plotone{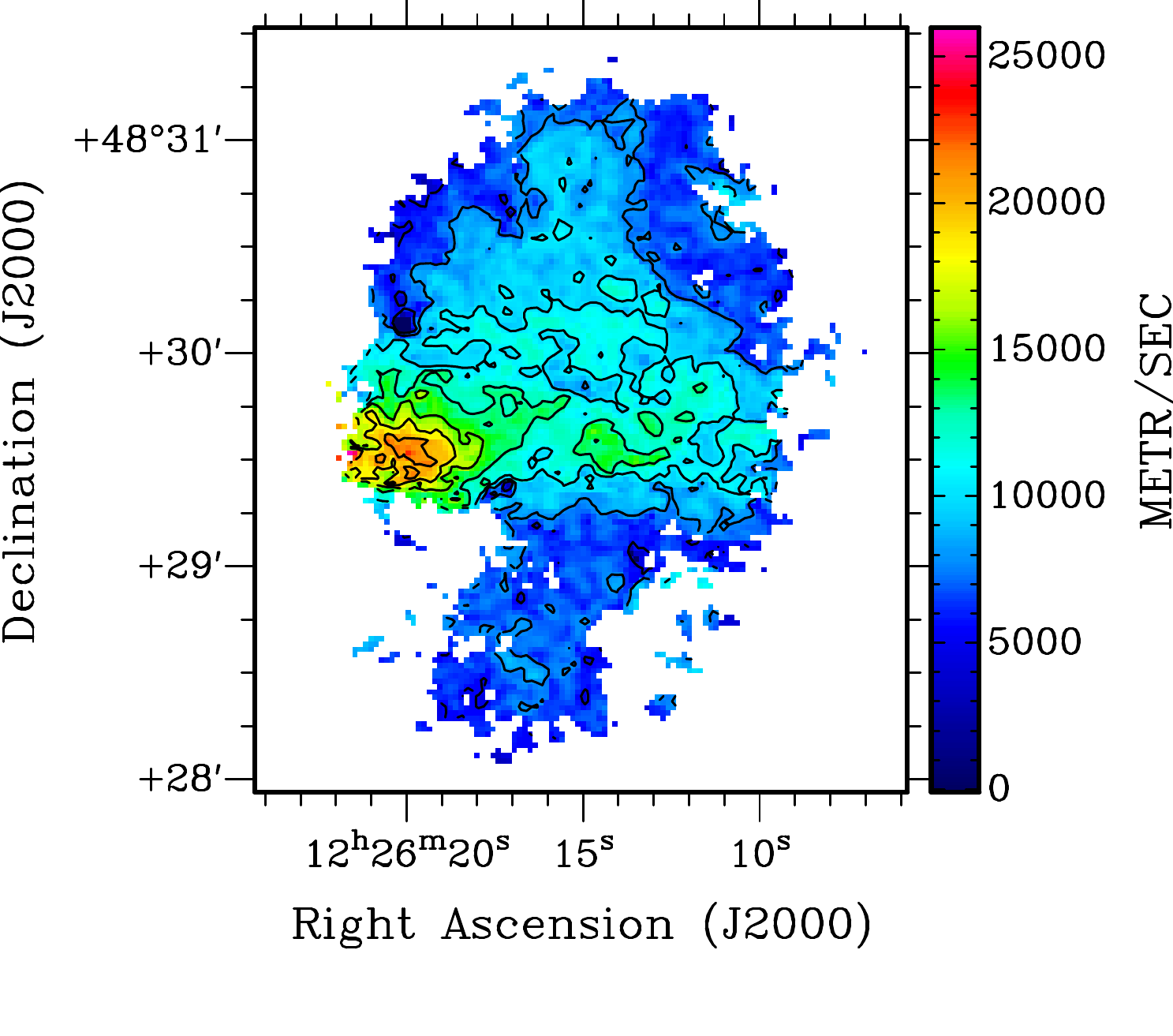} 
\figcaption{Haro~29's integrated \HI\ natural-weighted moment maps. \textit{Upper left:} Integrated \HI\ intensity map; contour levels are 1$\sigma\times$(2, 5, 8, 11, 14, 17, 20, 23, 26, and 29) where 1$\sigma=8.47\times10^{19}\ \rm{atoms}\ \rm{cm}^{-2}$. \textit{Upper right:} Velocity field; contour levels are 256.25, 262.5, 268.75, 275,  281.25, 287.5, 293.75, 300, 306.25 \kms. \textit{Bottom:} Velocity dispersion field; contour levels are 5.18, 7.77, 10.36, 12.95, 15.54, 18.13, 20.72 \kms. \label{fig:h29x012}}
\end{figure}

\begin{figure}
\epsscale{.492}
\plotone{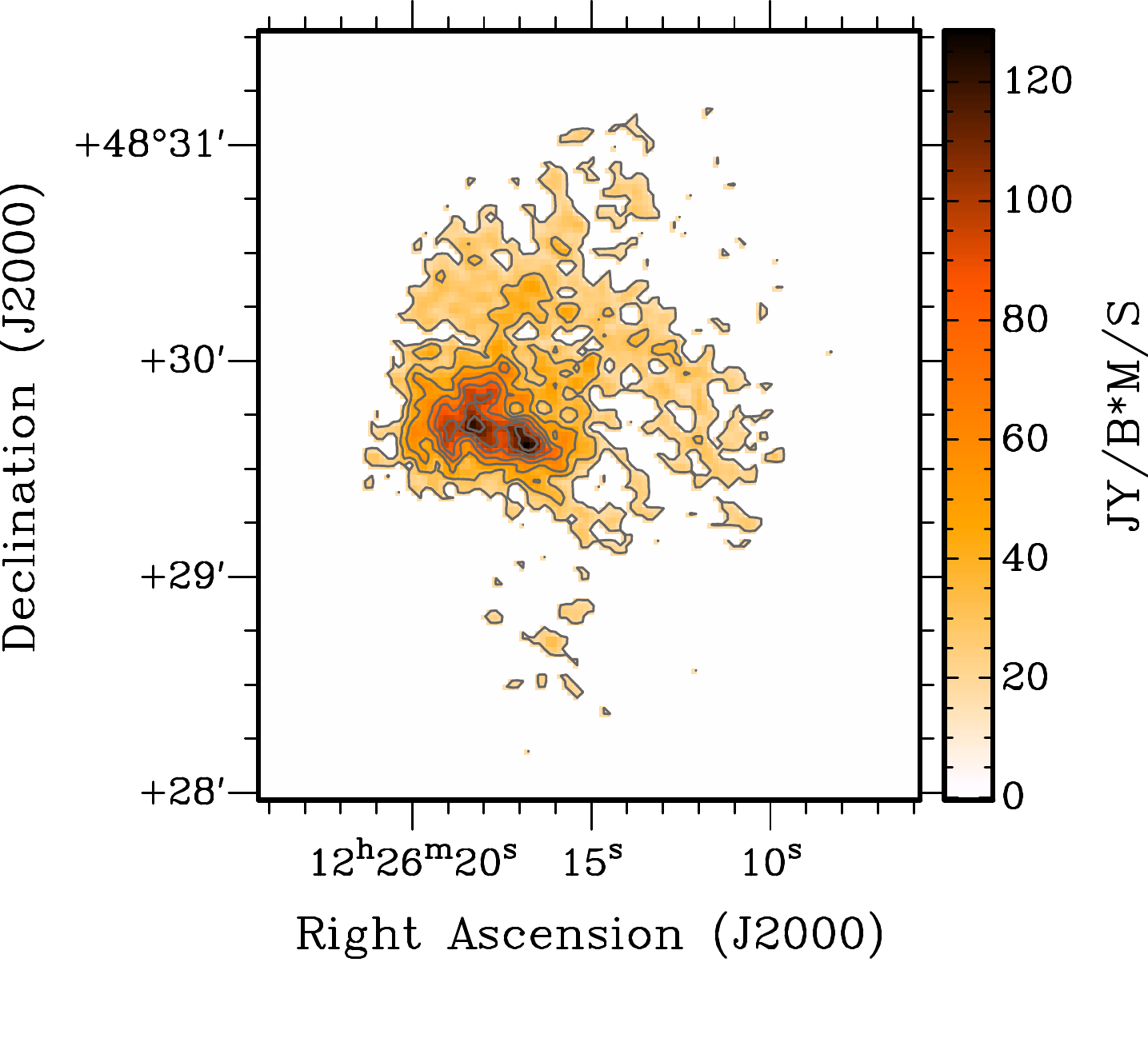}
\plotone{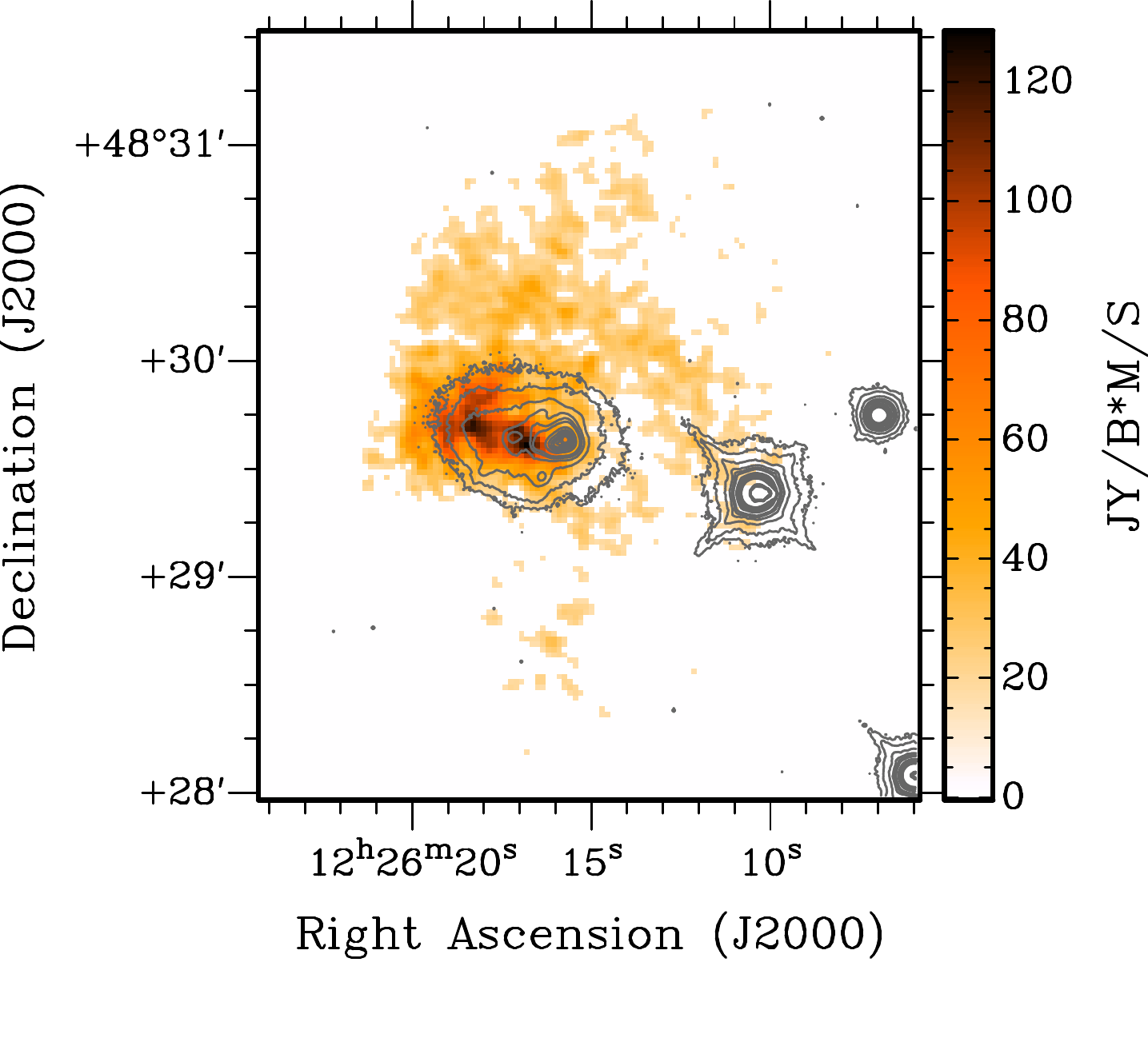}\\  
\plotone{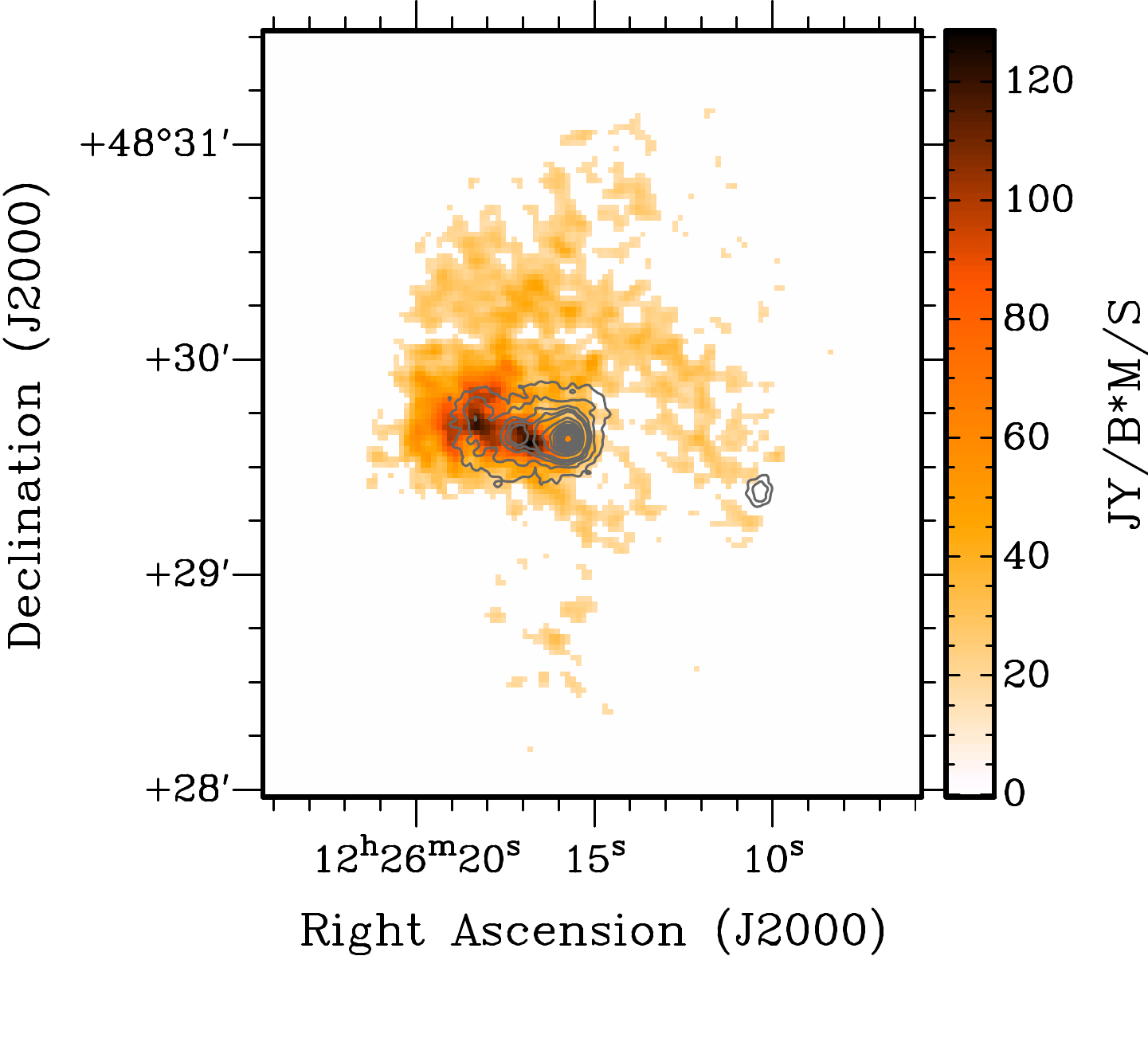}
\figcaption{Haro~29's integrated \HI\ robust-weighted moment maps. \textit{Upper left:} Integrated \HI\ intensity map; contour levels are 1$\sigma\times$(2, 4, 6, 8, 10, 12, 14) where 1$\sigma=2.37\times10^{20}\ \rm{atoms}\ \rm{cm}^{-2}$. \textit{Upper right:} Integrated \HI\ intensity map greyscale and V-band contours. The large sources around the main V-band emission are foreground stars.  \textit{Bottom:} Integrated \HI\ intensity map greyscale and FUV contours. \label{fig:h29x0r}}
\end{figure}

The integrated \HI\ natural-weighted and robust-weighted maps for Haro~29 are shown in Figures~\ref{fig:h29x012} and \ref{fig:h29x0r} respectively.   In the natural-weighted image two \HI\ peaks are visible and the morphology becomes increasingly disordered with increasing distance from the peaks. In the outer parts of Haro 29, there is a large-scale curved feature that begins north of the densest region, curves around to the west (right), and eventually to the south.  There is also an extension of diffuse gas to the south of the densest region. The two central peaks in Haro 29 appeared in the data of \citet{stil02a}, but the outer, curved feature was not visible in their maps (see their Figure 4).  \citet{viall83} present a low resolution \HI\ image of Haro 29 (see their Figure 3); \citet{viall83} split the \HI\ emission in Haro 29 into 7 components, their components 2, 4, and 5 comprise our natural weighted image in Figure~\ref{fig:h29x012}.  The rest of the \HI\ components seen in \citet{viall83} may have been resolved out by the VLA in our study.  

In the robust image of Haro 29 shown in Figure~\ref{fig:h29x0r}, the two central peaks are again evident in the \HI. Also in Figure~\ref{fig:h29x0r} are Haro~29's V-band and FUV data contours plotted over the greyscale of the integrated \HI\ robust-weighted map. Both the young and integrated stellar populations are offset to the west of the two \HI\ peaks; off-peak star formation is a trend seen in many other dwarf galaxies and is thought to be due to the current stars consuming, ionizing, and dispersing gas \citep{vanzee97, hunter98}.  The offset from the optical center ($12^{\rm{h}}26^{\rm{m}}16^{\rm{s}}.7, 48\degr29\arcmin38\arcsec$; as determined by  \citealt{hunter06}) is $\thicksim$400 pc away from the \HI\ peak in the east and $\thicksim$80 pc away from the \HI\ peak in the west.

\subsection{\HI\ Velocity Field} 

In Haro~29's \HI\ intensity-weighted velocity field (moment 1 map; Figure~\ref{fig:h29x012}), the velocity field appears most organized in the region of highest \HI\ density; this behavior was also seen in both \citet{viall83} and \citet{stil02a}.  The kinematic major axis is not well aligned with the optical major axis in this high density region.  The isovelocity contours begin to separate from each other with increasing distance from the high column density region, demonstrating how disturbed the outer gas is in Haro 29.  

The velocity dispersion field (moment 2 map) of Haro~29 is also shown in Figure~\ref{fig:h29x012}.  Haro 29 has a maximum velocity dispersion of $\thicksim$30 \kms\ occurring to the southeast of the highest density region.  From there the velocity dispersions get progressively smaller as the outer regions of the galaxy are reached.  The higher velocity dispersions occur where the isovelocity field lines are tightly packed with respect to the beam size, as opposed to the outer region, where the isovelocity field lines become more spread out.  The velocity dispersion map in Figure~\ref{fig:h29x012} has not been corrected for beam smearing effects.  

To determine the turbulent speed, we correct for beam smearing effects using the following equation:
\begin{equation}\label{equation1}
\sigma_{\rm{turb}}=\sqrt{\sigma_{\rm{meas}}^{2}-\sigma_{\rm{beam}}^{2}-\sigma_{\rm{instr}}^{2}}
\end{equation} 
where the $\sigma_{\rm{turb}}$ is the intrinsic velocity dispersion due to turbulence, the $\sigma_{\rm{meas}}$ is the total velocity dispersion measured from the velocity dispersion field map, $\sigma_{\rm{beam}}$ the velocity shear across the synthesized beam's FWHM, and $\sigma_{\rm{instr}}$ is the instrumental contribution to the dispersion, which is the velocity channel width (given in Table~\ref{tab:obsinfo}).  $\sigma_{\rm{beam}}$ was estimated by measuring the approximate variation in the velocity field by eye. Beam smearing affects Haro 29's velocity dispersion map in the following manner: velocity dispersions less than 15.54 \kms\ were reduced to dispersions of 9.7 \kms\ or less and velocity dispersions greater than 15.54 \kms\ were decreased by $\thicksim$5 \kms.  A velocity dispersion of 10 \kms\ is typical for dwarf irregular galaxies \citep{tamburro09, warren12}, therefore Haro 29 does appear to have some gas that is disturbed to the southeast of the highest \HI\ density region, in the orange region of the bottom map of Figure~\ref{fig:h29x012}.

\subsection{\HI\ Mass} \label{sec:h29mass}

The total galaxy \HI\ masses are measured using the task \textsc{ispec} in AIPS.  \textsc{ispec} adds the flux, from every channel in the data cube, in a user-specified box and gives the total flux inside the box in units of Jy.  The masses of the galaxies' individual morphological features are measured using \textsc{blsum}.  Just like \textsc{ispec}, \textsc{blsum}  measures the flux in every channel over a specified area, but rather than a simple box the area is traced out by the user on an image of the galaxy.  This is useful for isolating the mass in the individual morphological features.

The summed flux can then be used to calculate the \HI\ mass (in units of solar masses) inside the specified area with the following equation:
\begin{equation}
\rm{M}(\rm{M}_{\sun})=235.6\rm{D}^{2}\sum_{i}\rm{S}_{\rm{i}}\Delta \rm{V}
\end{equation}
where D is the distance of the galaxy in units of Mpc, $\rm{S}_{\rm{i}}$ is the flux in mJy in channel i, and $\Delta$V is the channel width in \kms.  The natural-weighted cubes are used to determine \HI\ masses so that masses of tenuous features can be separately determined; the robust-weighted cubes are unable to map the tenuous features completely because they have a lower signal-to-noise.  
	
	Haro~29's total \HI\ mass measured from the natural-weighted cube is $5.8\times10^{7}\ \rm{M}_{\sun}$.  We have taken the total \HI\ masses measured by \citet{viall83} and \citet{stil02a} and recalculated them using our distance of 5.8 Mpc to get a total \HI\ mass of $6.5\times10^{7}\ \rm{M}_{\sun}$ and $5.6\times10^{7}\ \rm{M}_{\sun}$ respectively.  The mass measured by \citet{stil02a} is reasonably close to our own, while the mass measured by \citet{viall83} appears a bit high. However, \citet{viall83} used a larger synthesized beam of $27\arcsec.9\times37\arcsec.2$; a beam this size will have a higher sensitivity and therefore will capture a larger extent of the flux.  Also, the lower end of the \HI\ mass error stated in \citet{viall83} puts their estimate near our own \HI\ mass estimate. 
	
	The \HI\ mass of the inner region is also measured; the outline of the fourth contour (counting from the lowest contour) of the natural-weighted \HI\ intensity image is used to define the inner region.  This contour corresponds to $9.32\times10^{20}\ \rm{atoms\ cm}^{-2}$ and is chosen as a limit because it does not contain any prominent curved morphological features that define the outer region.  It also corresponds to the region where the velocity field is the most organized at the resolution of our observations.  This inner region has an \HI\ mass of $1.4\times10^{7}\ \rm{M}_{\sun}$ which is 24\% of the total \HI\ mass. The outer region created by the curved features therefore accounts for 76\% of the \HI\ mass of the galaxy.

\subsection{Position-Velocity Diagrams} \label{sec:pv}

\begin{figure}
\epsscale{.94}
\begin{center}
\plottwo{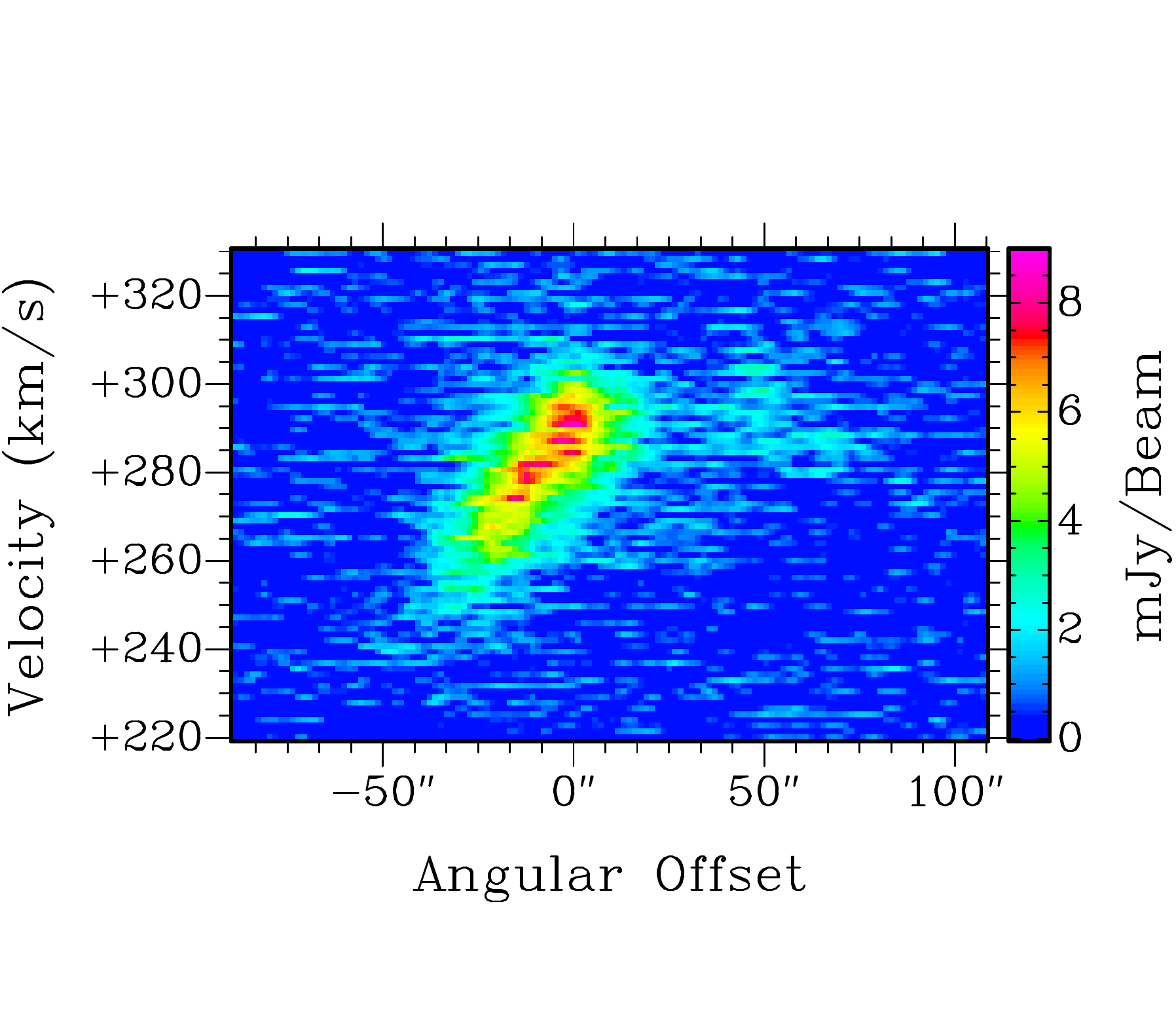}{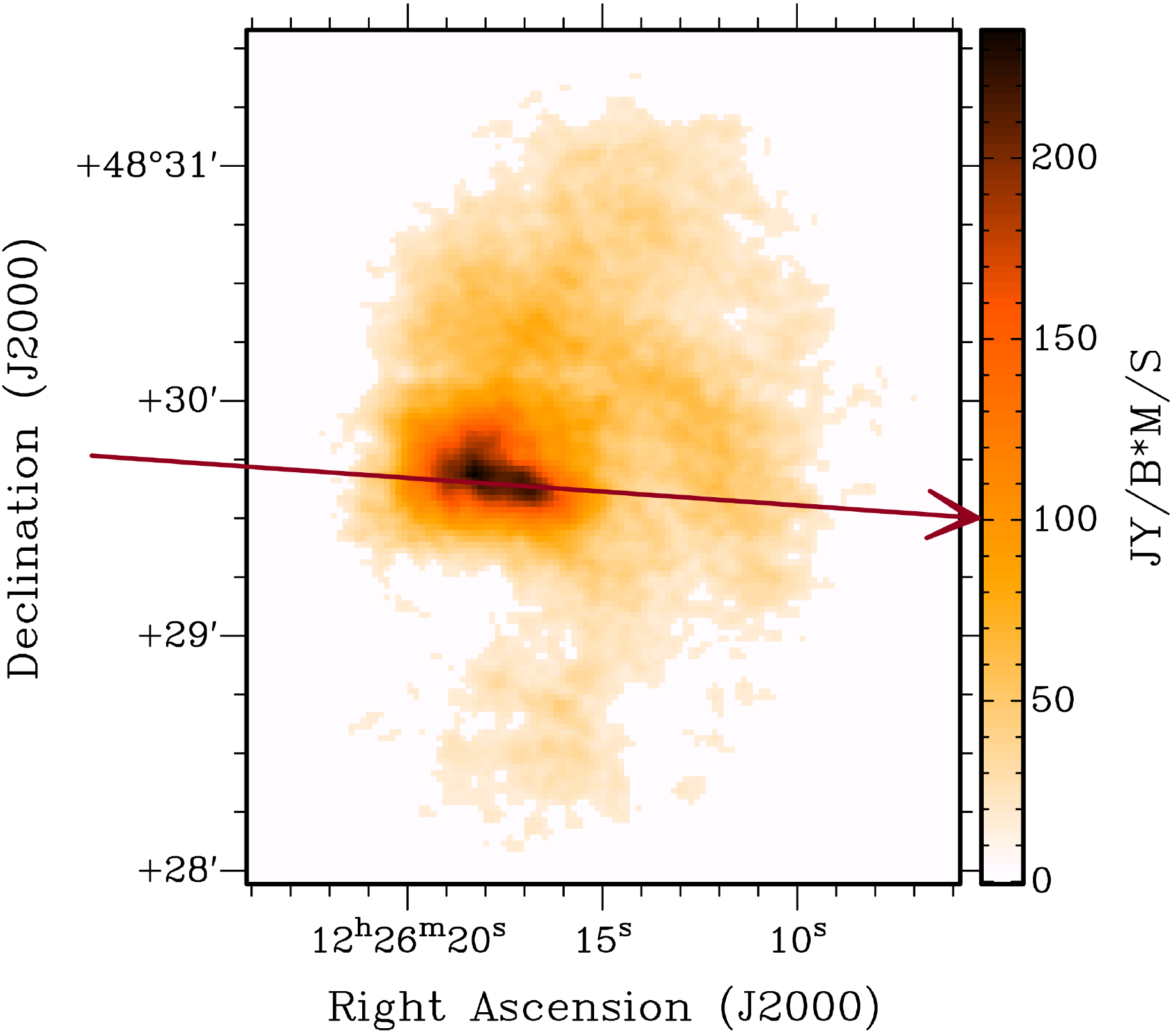}
\plottwo{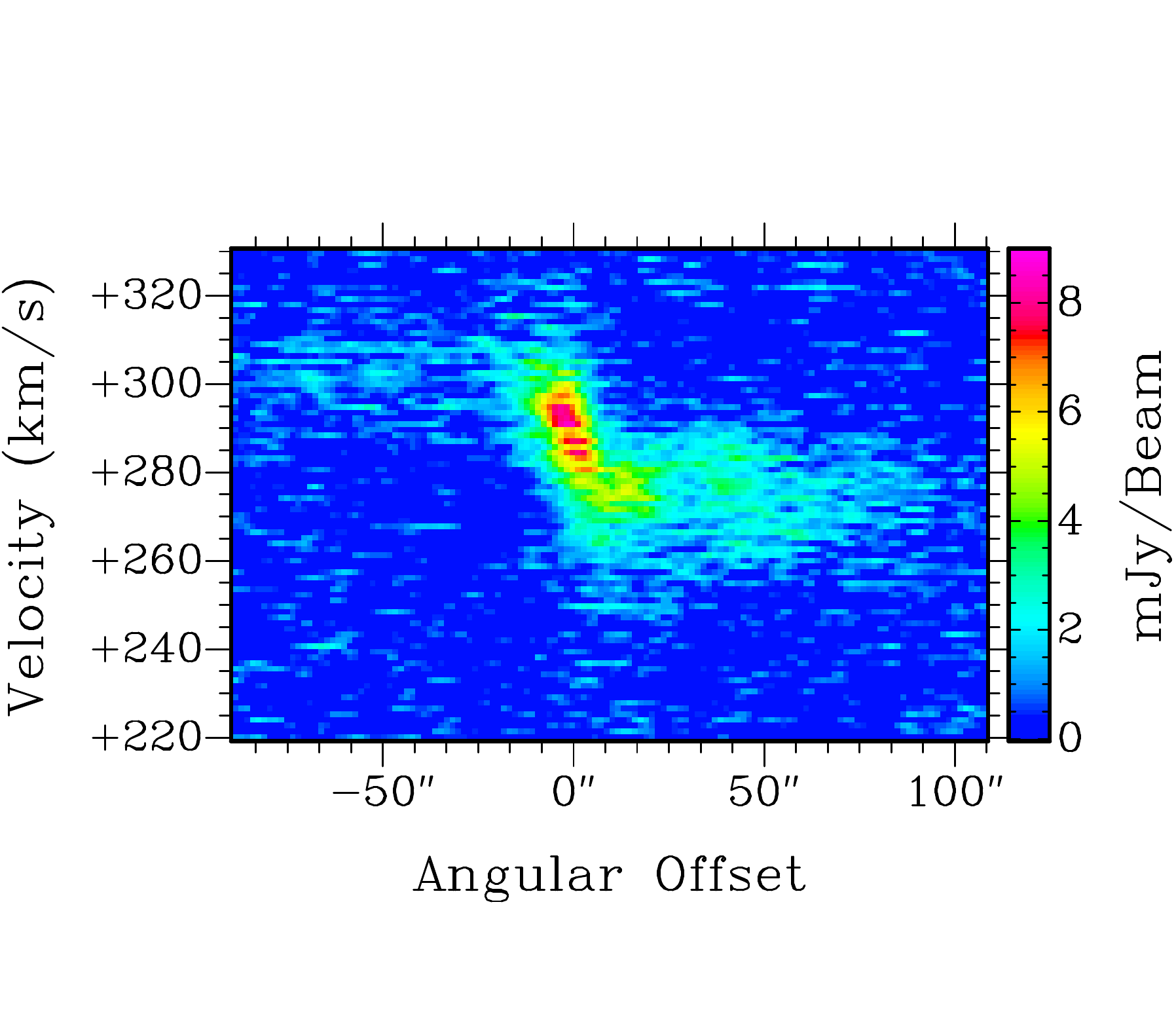}{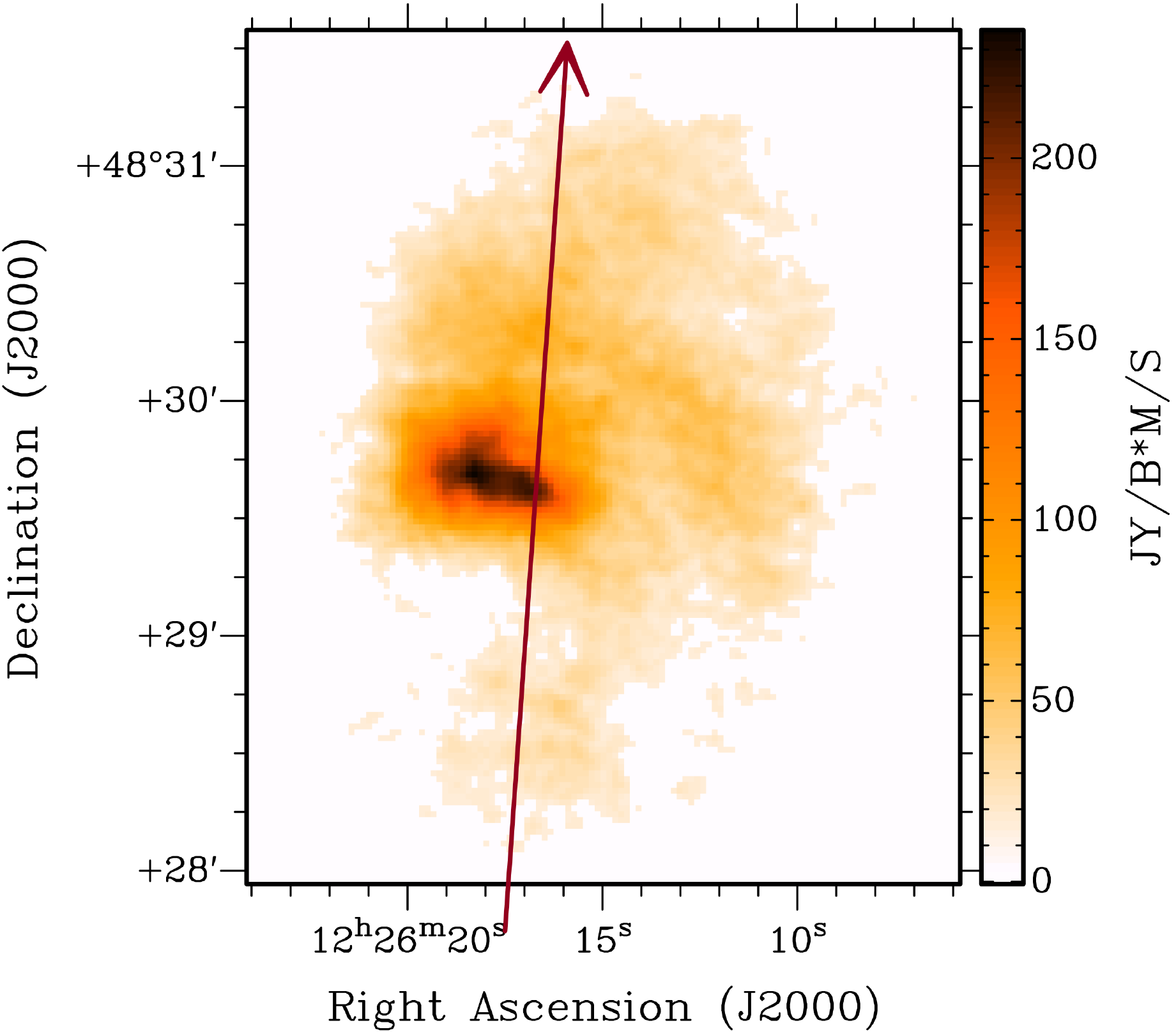}
\plottwo{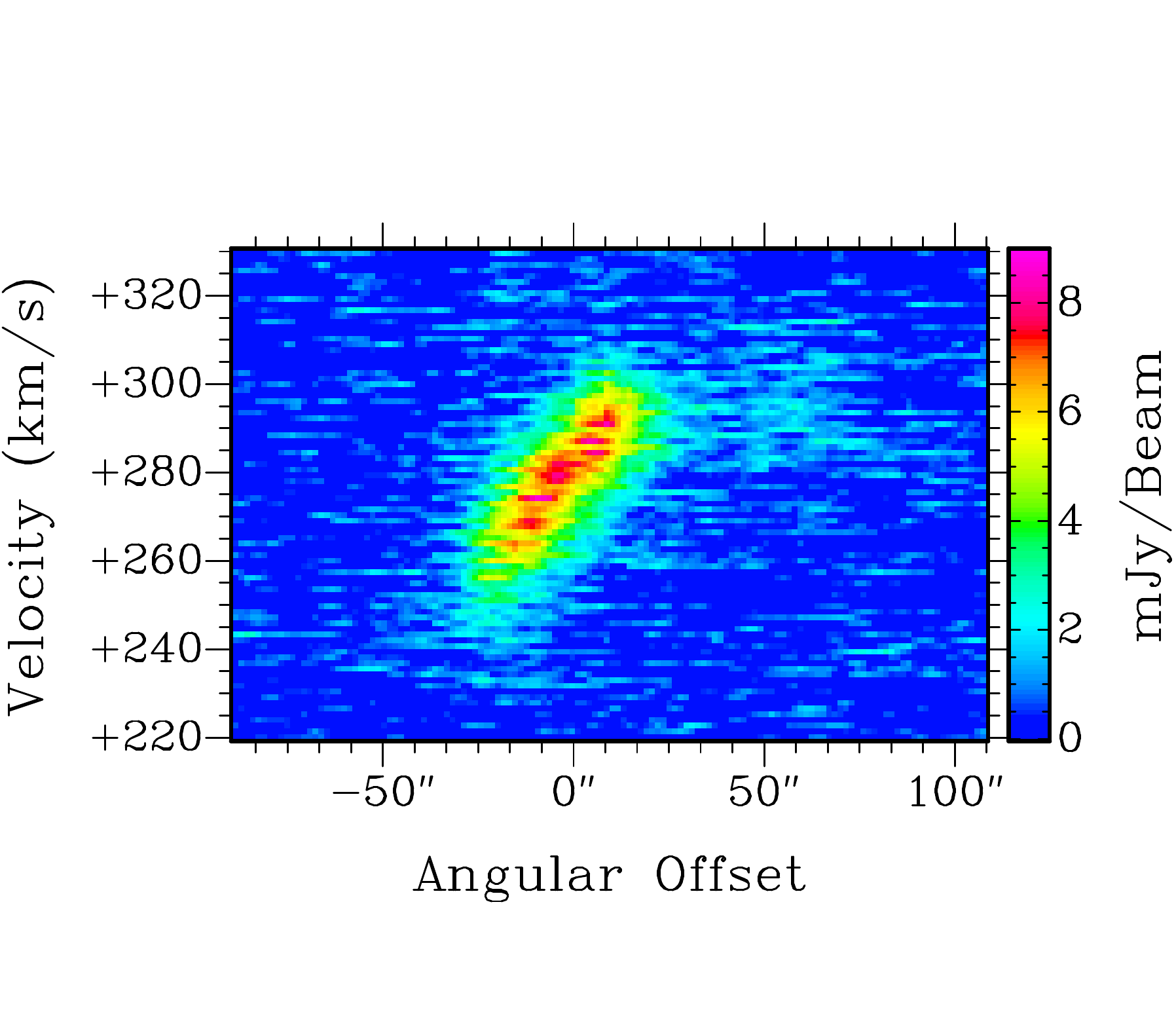}{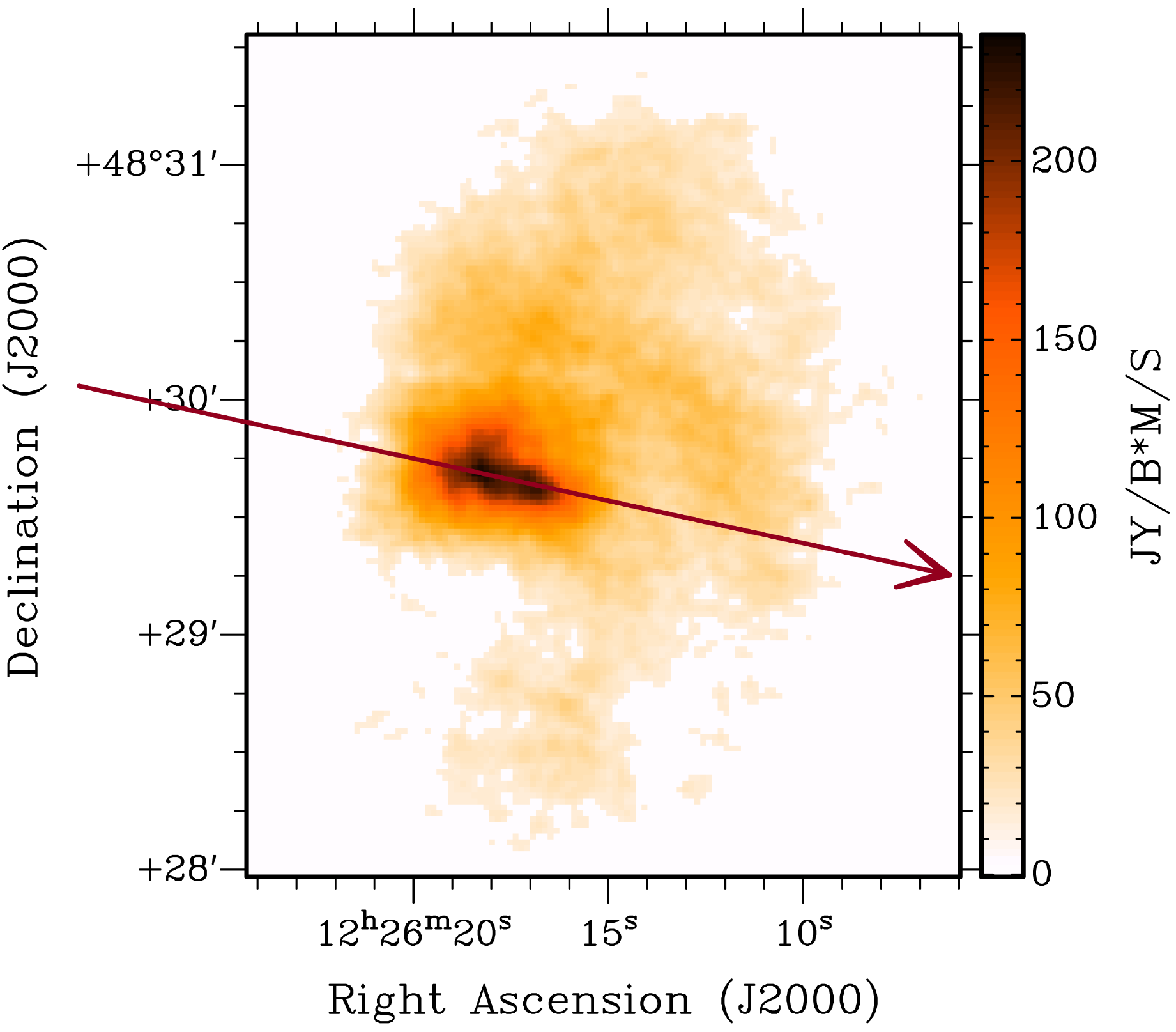}
\end{center}
\figcaption{Haro 29: The left column contains the P-V diagrams and the right column contains the natural-weighted integrated \HI\ map with a red arrow indicating the location of the corresponding slice through the galaxy and pointing in the direction of positive offset. The first row corresponds to the optical major axis, the second row to the optical minor axis, and the third row is a slice through both central peaks. \label{fig:h29p-v}}
\end{figure}

Position-velocity (P-V) diagrams are presented in an attempt to better understand the kinematic details of the galaxies and look for possible kinematic disturbances.  All of the P-V diagrams are made in KARMA using \tt kpvslice \rm on the non-Hanning-smoothed, natural-weighted data cubes.  The non-Hanning-smoothed cubes are chosen for the P-V diagrams because they have a velocity resolution of 1.81 \kms, as opposed to the Hanning-smoothed resolution of 2.58 \kms.  These data are not used for the rest of the analysis because higher velocity resolution comes at the cost of lower signal-to-noise, but since the velocity is what is of interest to us in a P-V diagram and the drop in signal-to-noise is small, better velocity resolution is chosen over signal-to-noise for this task.  The signal-to-noise, velocity resolution, and noise of these maps are listed in Table~\ref{tab:mapinfo}.  In the P-V diagram figures (Figures~\ref{fig:h29p-v}, \ref{fig:h36p-v}, and \ref{fig:h36p-v2}), the column to the left is composed of P-V diagrams, and the column to the right is composed of the corresponding slices (in red) across the galaxy.  The slices are shown on the \HI\ Hanning-smoothed, natural-weighted, intensity maps so that P-V diagram features can be more easily associated with the figures presented earlier.   The direction of the slice (indicated by the arrow) was also chosen to read from left to right on the \HI\ intensity map, so that the P-V diagram and intensity map could be more easily associated.

The first row in Figure~\ref{fig:h29p-v} is a P-V diagram along the optical major axis of Haro~29 (position angle (PA) of 86\degr) and the second row is a slice through Haro~29's optical minor axis (both are centered at the galaxy's optical central coordinates of $12^{\rm{h}}26^{\rm{m}}16^{\rm{s}}.7, 48\degr29\arcmin38\arcsec$).  Along the optical major axis, the P-V diagram shows near solid body rotation from $-$42\arcsec\ to 16\arcsec\ and spread in velocity between $\thicksim$245 \kms\ and 310 \kms. Toward higher offsets there is diffuse emission from 16\arcsec\ to 75\arcsec\ and spread in velocity between $\thicksim$280 \kms\ and 310 \kms.  The optical minor axis shows an interesting feature just outside the densest region of \HI; the slice first moves through the southern extension with a slope of about zero at $\thicksim$300 \kms, then as the slice moves through the dense region from the south toward the north, the velocity decreases in a solid body fashion from 305 \kms\ to 260 \kms.  At an angular offset of 8\arcsec, as it begins to move through the curved tenuous emission in the north, the velocity becomes very spread out between 260 \kms\ and 290 \kms\ around a nearly constant median.  This P-V diagram indicates (as seen in the intensity weighted velocity field) that the inner region and outer region are kinematically different.

The third row of Figure~\ref{fig:h29p-v} is a slice through both central peaks (PA= 77.8\degr\ and centered between the peaks at $12^{\rm{h}}26^{\rm{m}}17^{\rm{s}}$.40, 48\degr29\arcmin39\arcsec.35).  The position angle of this diagram is very close to that of the optical major axis, but since the optical major axis did not go through the center of the \HI\ peak to the northeast, this new slice is made to examine the kinematics between the two peaks.  This diagram shows that the two central peaks appear to be participating in solid body rotation together.

\section{Results: Haro 36}\label{sec:results36}

\subsection{Stellar Component}

\begin{figure}[!ht]
\centering
\epsscale{1.11}
\plottwo{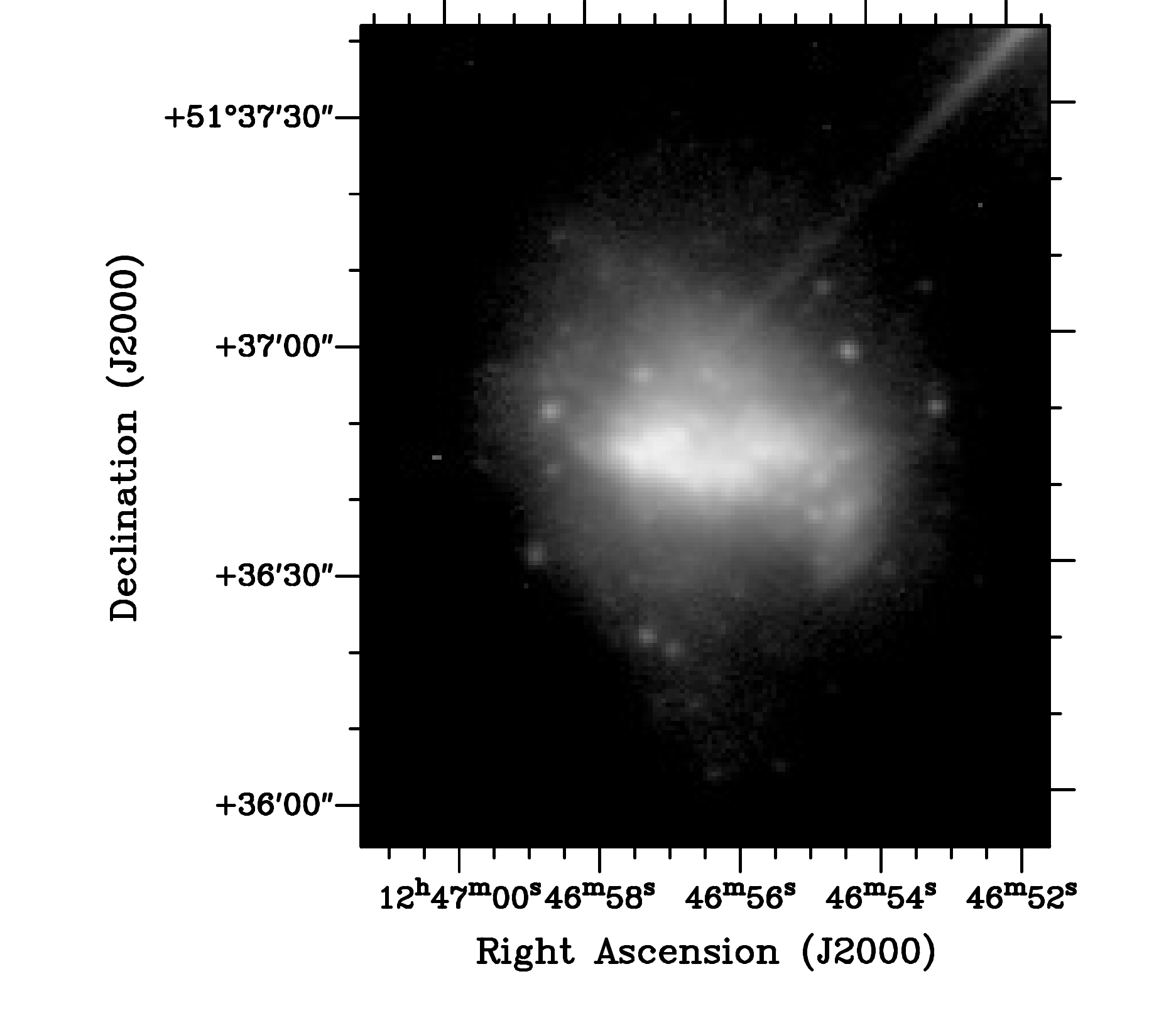}{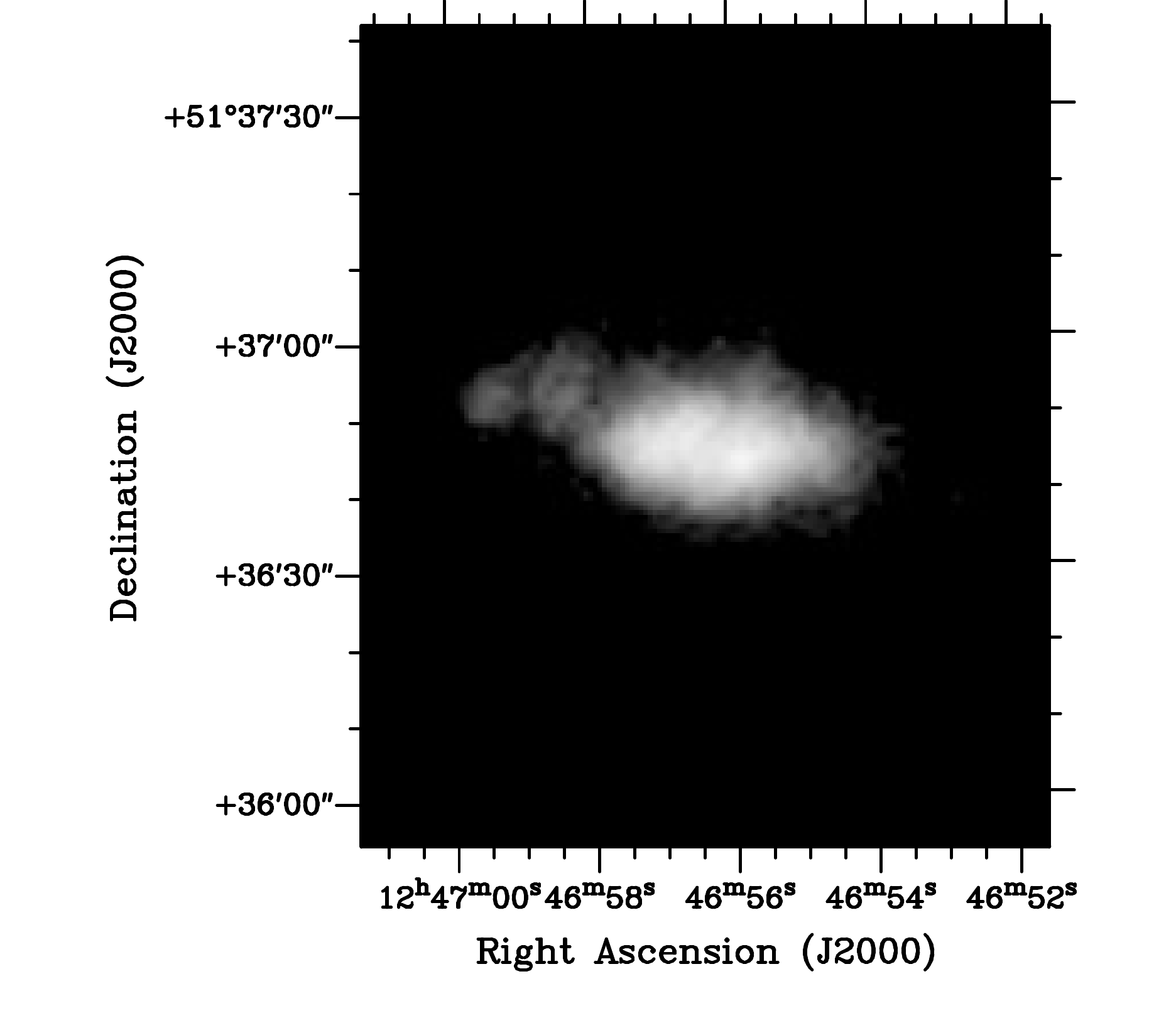}
\figcaption{Haro~36\ \ \textit{Left:} V-band, the streak coming from the top-right of the image is due to a foreground star in the image;\ \ \textit{Right:} FUV  \label{fig:h36v_fuv}}
\end{figure}

The FUV and V-band data for Haro~36 are shown in Figure~\ref{fig:h36v_fuv}.  Just like Haro~29, the FUV and V-band follow nearly the same morphology.  However, the faint outer emission of the V-band elongates the V-band structure in the north-south direction, while the FUV has no such feature at the sensitivity limit of the image.

\subsection{\HI\ Morphology}  

\begin{figure}
\epsscale{0.492}
\plotone{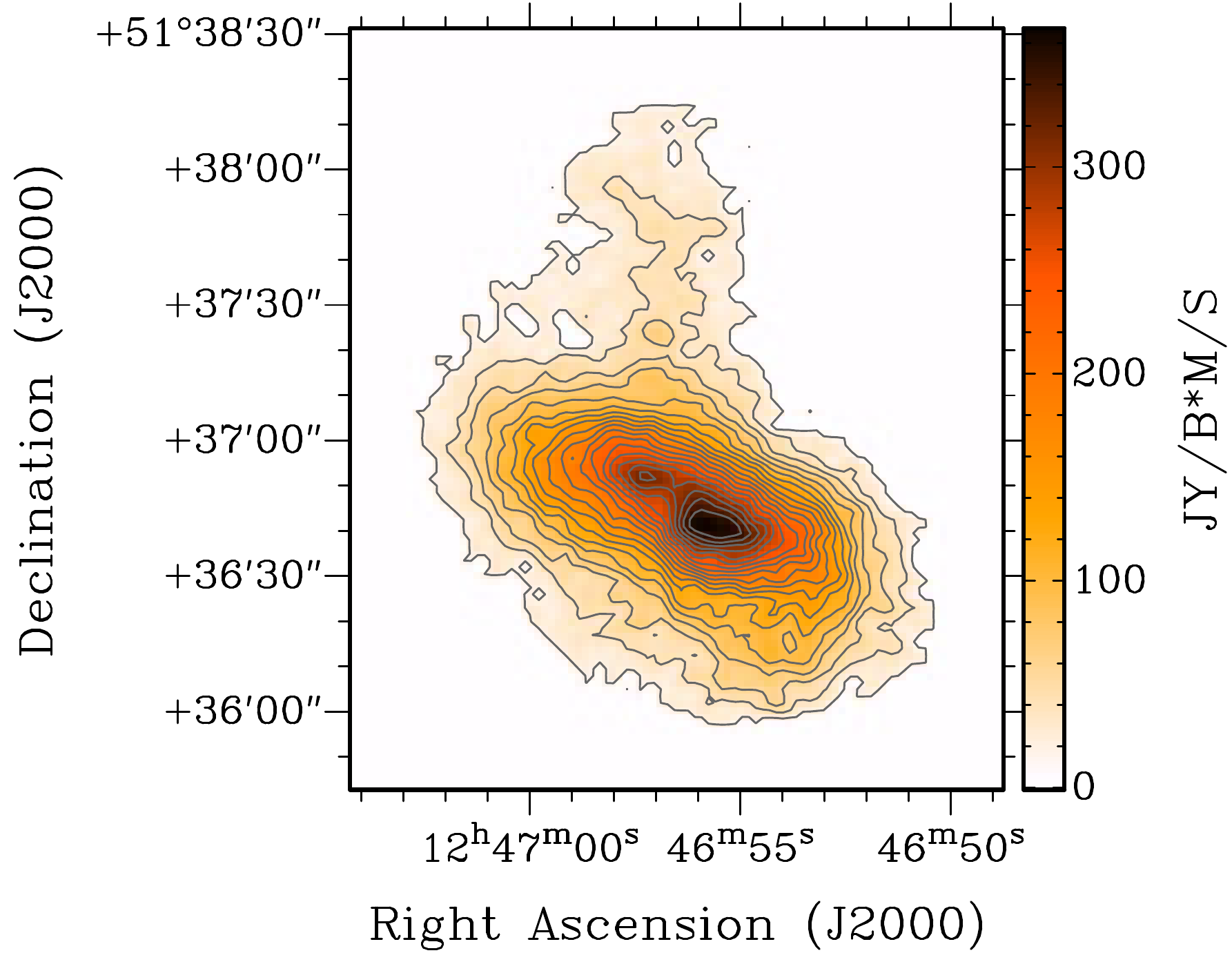}
\plotone{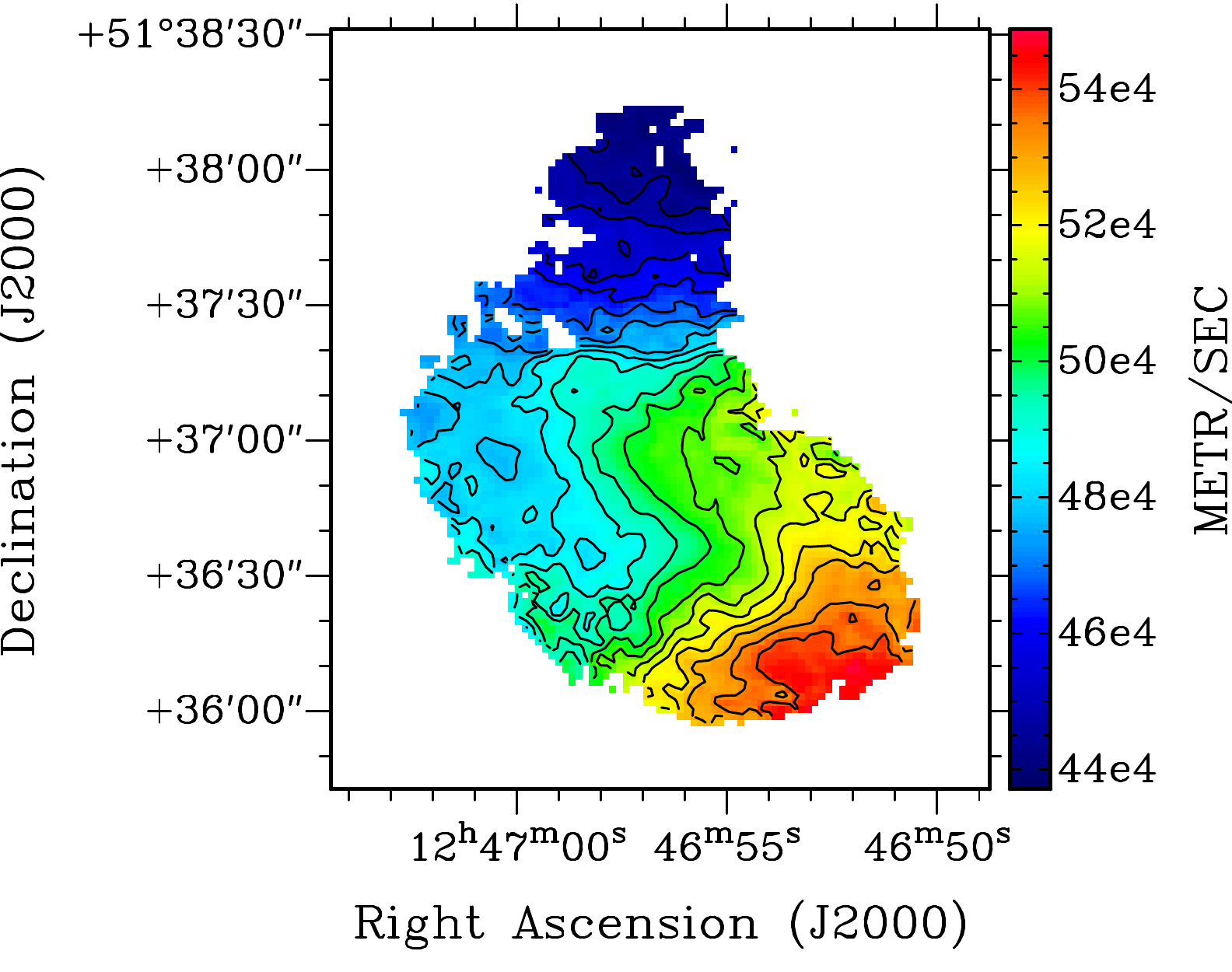}
\plotone{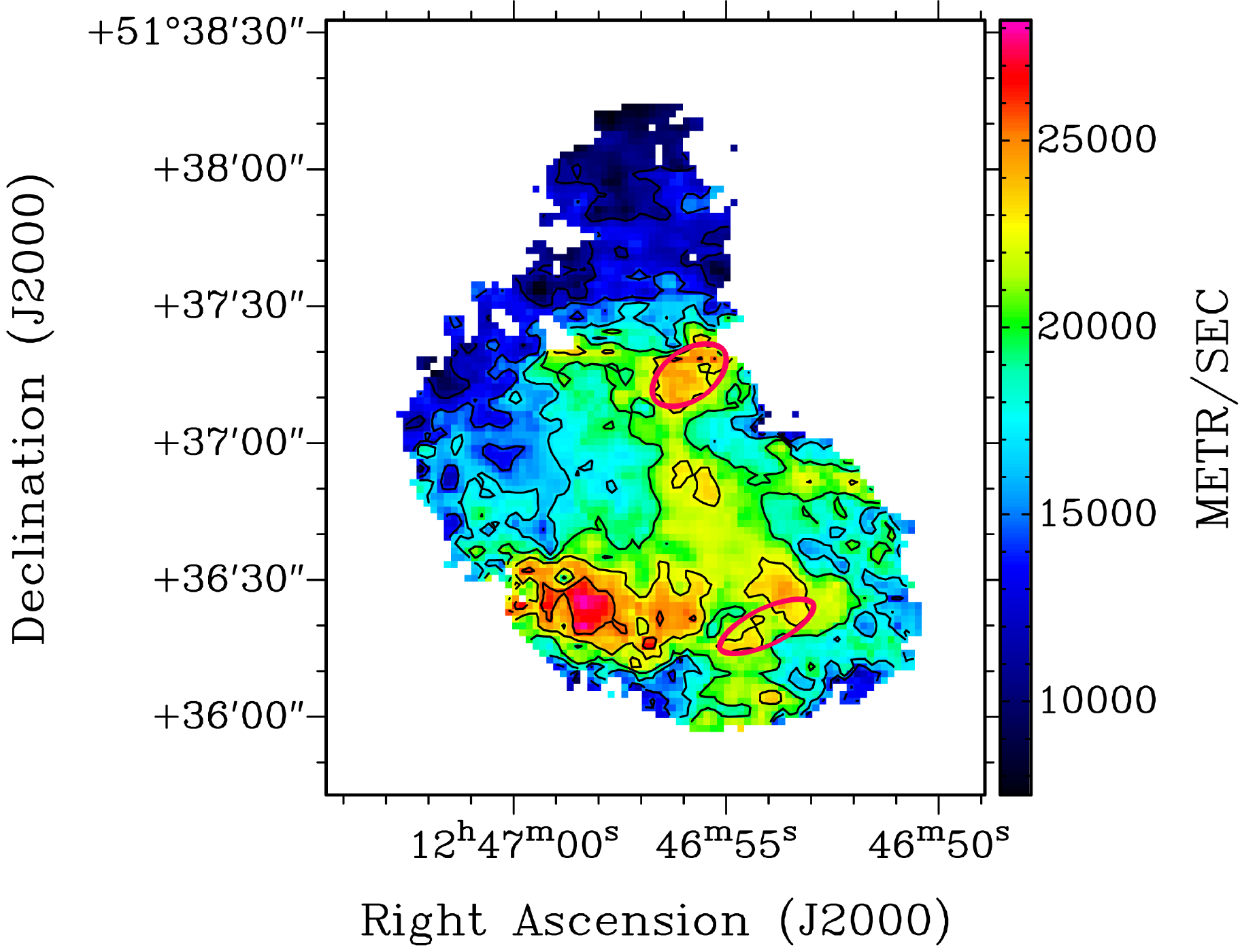}
\figcaption{Haro~36's natural-weighted moment maps. \textit{Upper left:} Integrated \HI\ intensity map; contour levels are 1$\sigma\times$(2, 4, 6, 8, 10, 12, 14, 16, 18, 20, 22, 24, 26, 28, 30, 32, 34, 36, 38) where 1$\sigma=9.15\times10^{19}\ \rm{atoms}\ \rm{cm}^{-2}$.  \textit{Upper right:} Intensity weighted velocity field; contour levels are 445.5, 451, 456.5, 462, 467.5, 473, 478.5, 484, 489.5, 495, 500.5, 506, 511.5, 517, 522.5, 528, 533.5, and 539 \kms. \textit{Bottom:} Velocity dispersion field; contour levels are 11.28, 14.1, 16.92, 19.74, 22.56, and 25.38 \kms. \label{fig:h36x012}}
\end{figure}

\begin{figure}
\plotone{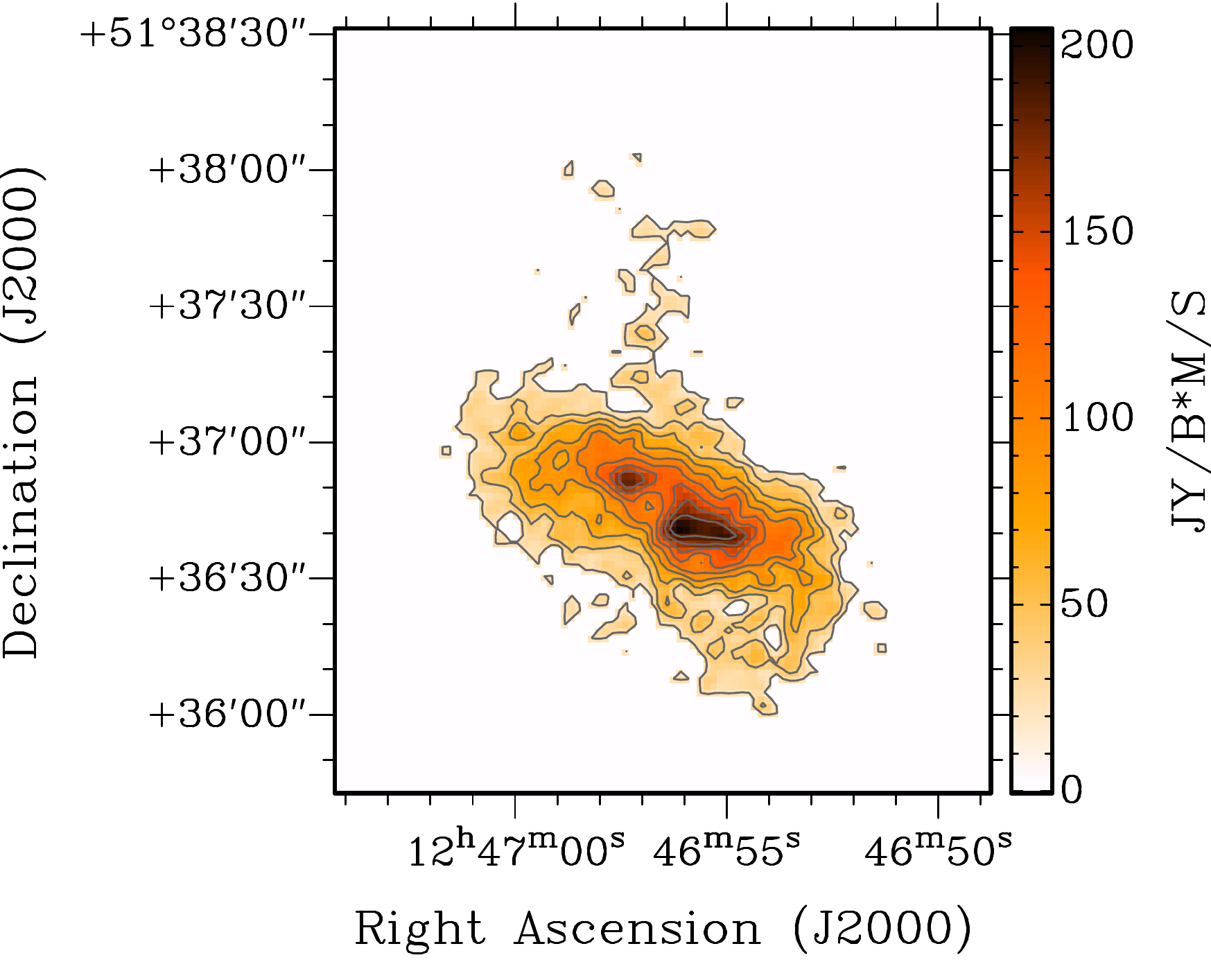}
\plotone{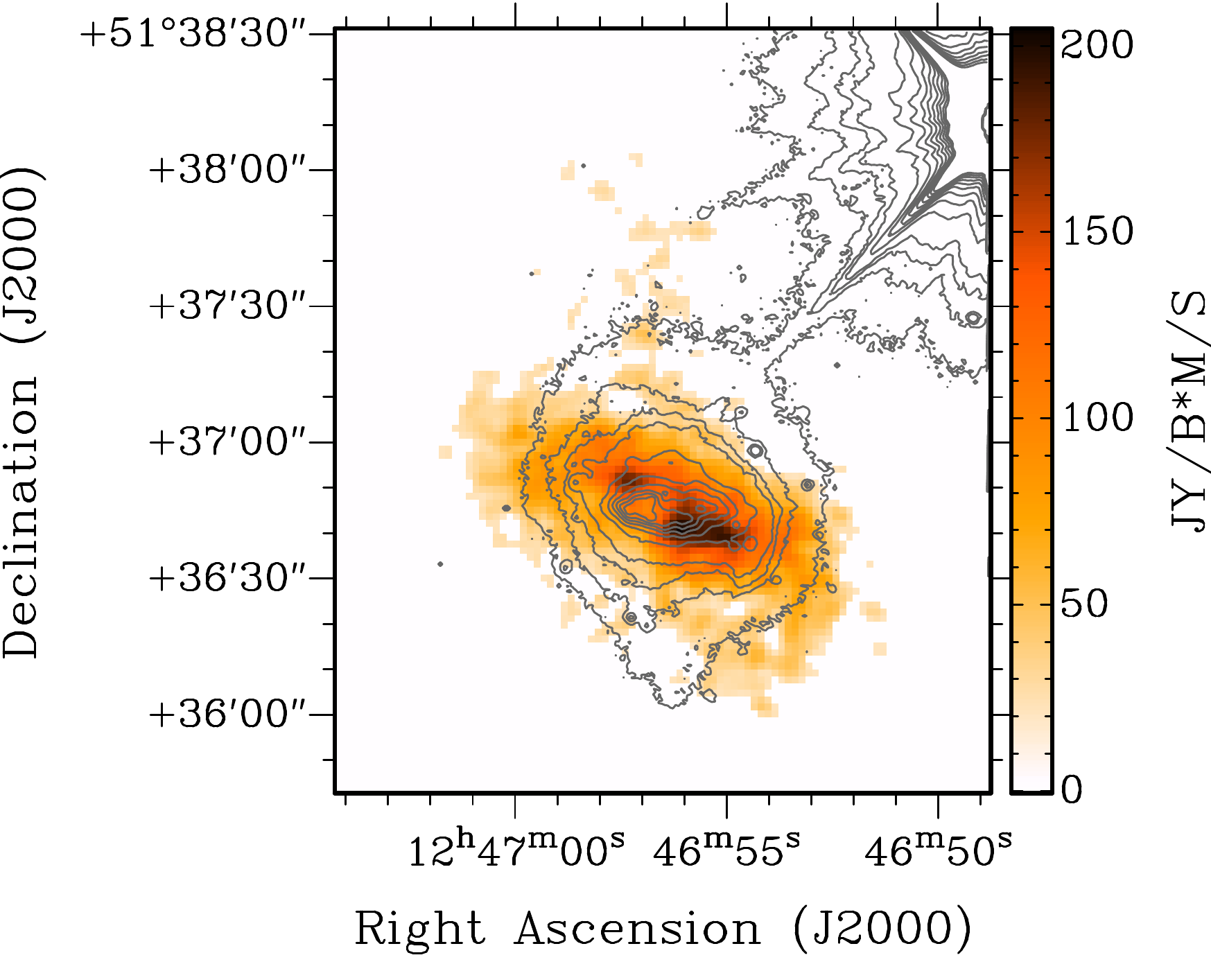}
\plotone{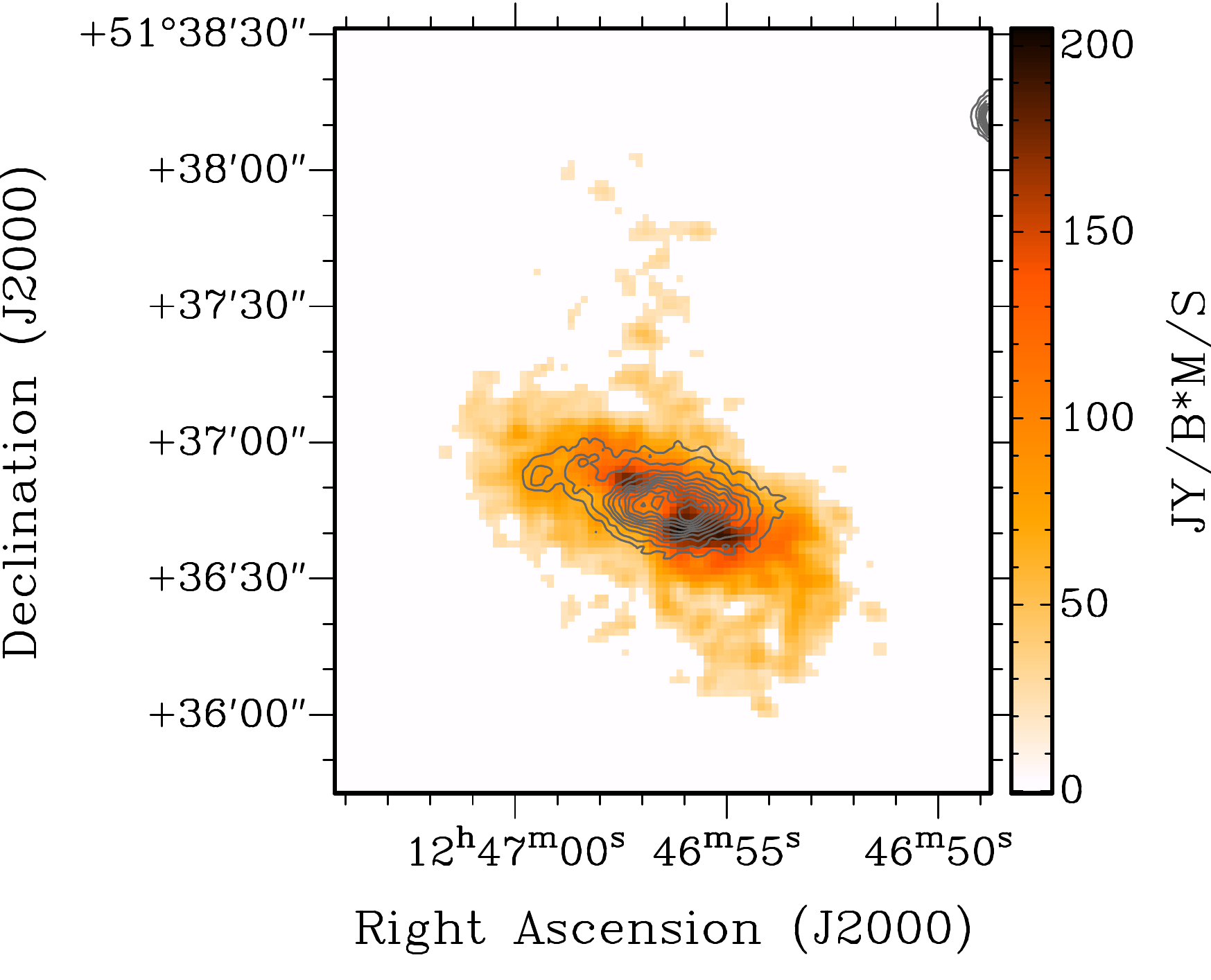}
\figcaption{Haro~36's robust-weighted moment maps. \textit{Upper left:} Integrated \HI\ intensity map; contour levels are 1$\sigma\times$(2, 4, 6, 8, 10, 12, 14, 16, 18) where 1$\sigma=2.79\times10^{20}\ \rm{atoms}\ \rm{cm}^{-2}$. \textit{Upper right:} Integrated \HI\ intensity map greyscale and V-band contours. \textit{Bottom:} Integrated \HI\ intensity map greyscale and FUV contours. \label{fig:h36x0r}}
\end{figure}

	The integrated \HI\ natural-weighted and robust-weighted maps for Haro~36 are shown in Figures~\ref{fig:h36x012}~and~\ref{fig:h36x0r}.  Haro~36 has two discernible components in the natural-weighted map; the main body with dense \HI\ emission and an extension of tenuous emission to the north of the main body.  The main body has two central \HI\ peaks visible in both the robust-weighted and the natural-weighted maps that are separated by a projected distance of $\thicksim$860 pc (measured from their approximate centers in the robust map).  \citet{simp00} saw hints of the two \HI\ peaks in Haro 36, but were unable to resolve them and were also unable to pick up the tenuous northern extension with the VLA C-array data.  The V-band emission and FUV emission are concentrated between the two central \HI\ peaks, as can be seen in Figure~\ref{fig:h36x0r}. 

\subsection{Velocity Fields} 

The velocity field for Haro~36 (Figure~\ref{fig:h36x012}) looks unsettled with two kinematic components; one associated with the main body and the other associated with the tenuous emission to the north of the main body.  The iso-velocity contours within each component also look unsettled.

 The velocity dispersion field of Haro~36 is also shown in Figure~\ref{fig:h36x012} and has not been corrected for beam smearing.  The northern extension uniformly has some of the lowest velocity dispersions in the galaxy with an average of approximately 10 \kms, while the main body has an approximate average dispersion of 20 \kms. The effect of beam smearing on the trends seen in Haro 36's velocity dispersion map is small: velocity dispersions less than 16.92 \kms\ are decreased to values of 11 \kms\ or less.  Velocity dispersions greater than 16.92 \kms\ are decreased to velocities of $\thicksim$15-19 \kms.  In two small regions of the map we are unable to determine the turbulent velocities due to significant velocity gradients across the beam; both region's approximate locations are outlined by pink ovals on the dispersion map in Figure~\ref{fig:h36x012}.  These two regions however, are not very large and do not significantly change the overall trends seen in the dispersion map.  Therefore, the gas in the main body of Haro 36 does indeed have high velocity dispersions relative to undisturbed dwarf irregular galaxies, such as Holmberg II, NGC 4214, and IC 2574.  These three dwarf irregular galaxies all appear not to be interacting and are undisturbed with velocity dispersions of 8.9 to 9.7 \kms\ \citep{tamburro09}.

\subsection{\HI\ Mass}

The total \HI\ mass of Haro 36 is measured to be $1.3\times10^{8}\ \rm{M}_{\sun}$.  \citet{simp00} measured $6.2\times10^{7}\ \rm{M}_{\sun}$ for the total \HI\ mass of Haro 36 (when adjusted for the distance discrepancy between our paper and theirs), which is lower than our measured value.  Their \HI\ mass was measured from the VLA C-array data alone, while our data was composed of B, C, and D array data, making their data not as sensitive as ours.  This is apparent in Figure 17 of \citet{simp00}, which shows the galaxy at a 2$\sigma$ level of $4.5\times10^{20}\ \rm{atoms}\ \rm{cm}^{-2}$ vs. our natural weighted image at a 2$\sigma$ level of $1.84\times10^{20}\ \rm{atoms}\ \rm{cm}^{-2}$.  

	\textsc{blsum} gives a mass for the northern extension of $1.3\times10^{7}\ \rm{M}_{\sun}$ or 10\% of the total mass.  The masses of the two central \HI\ peaks are also measured using areas confined by the second highest contour of the peak to the northeast (top-left) and by the fourth highest contour of the peak to the southwest.  These contours are where the two peaks are still distinguishable and both contours correspond to a column density of $2.9\times10^{21}\ \rm{atoms\ cm}^{-2}$.  The mass of the peak to the northeast is $2.0\times10^{6}\ \rm{M}_{\sun}$, and the mass of the peak to the southwest is  $1.1\times10^{7}\ \rm{M}_{\sun}$, making the northeast peak only 18\%\ of the mass of the southwest peak.

\subsection{Position-Velocity Diagrams} 

\begin{figure}
\epsscale{0.9}
\begin{center}
\plottwo{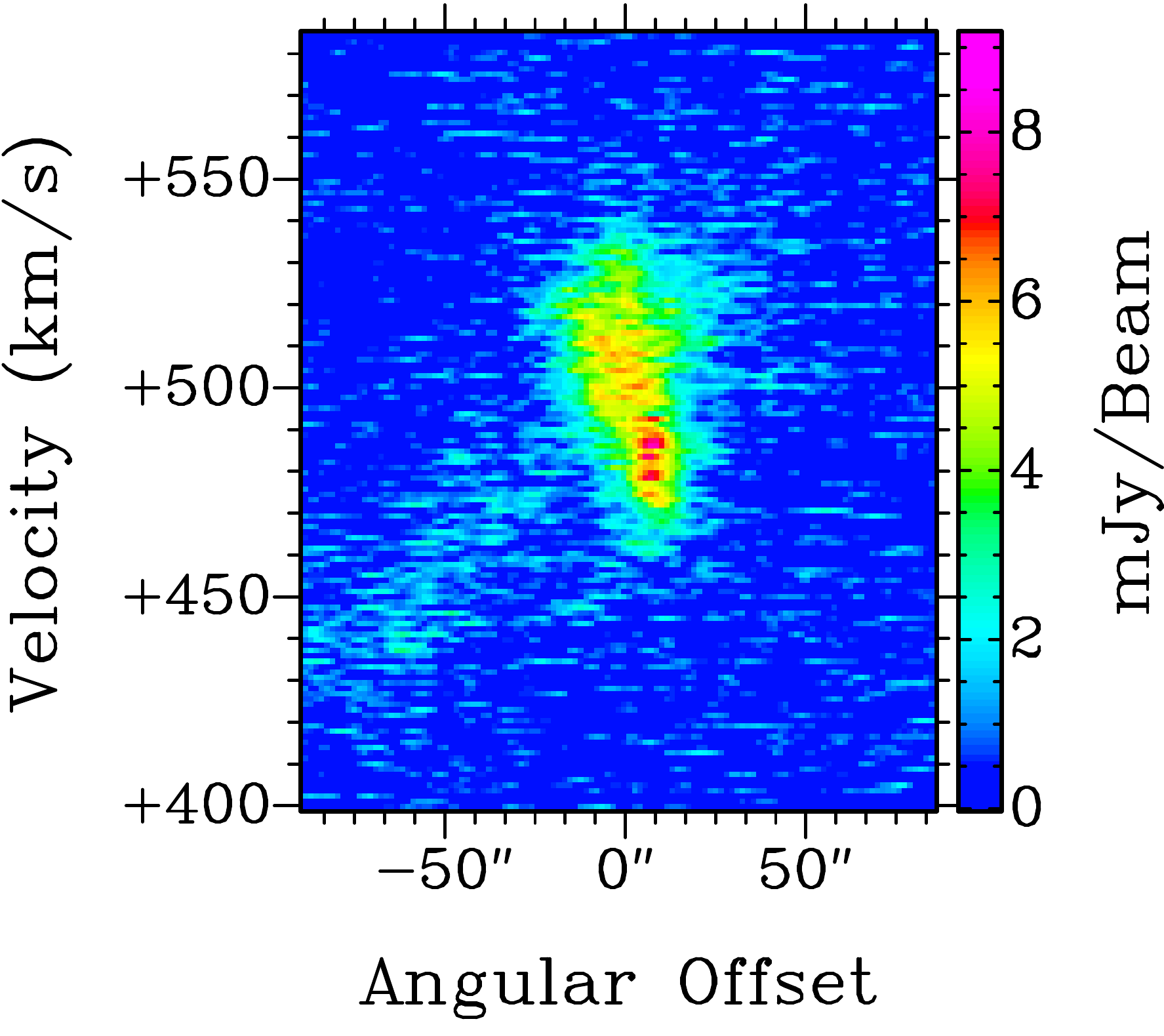}{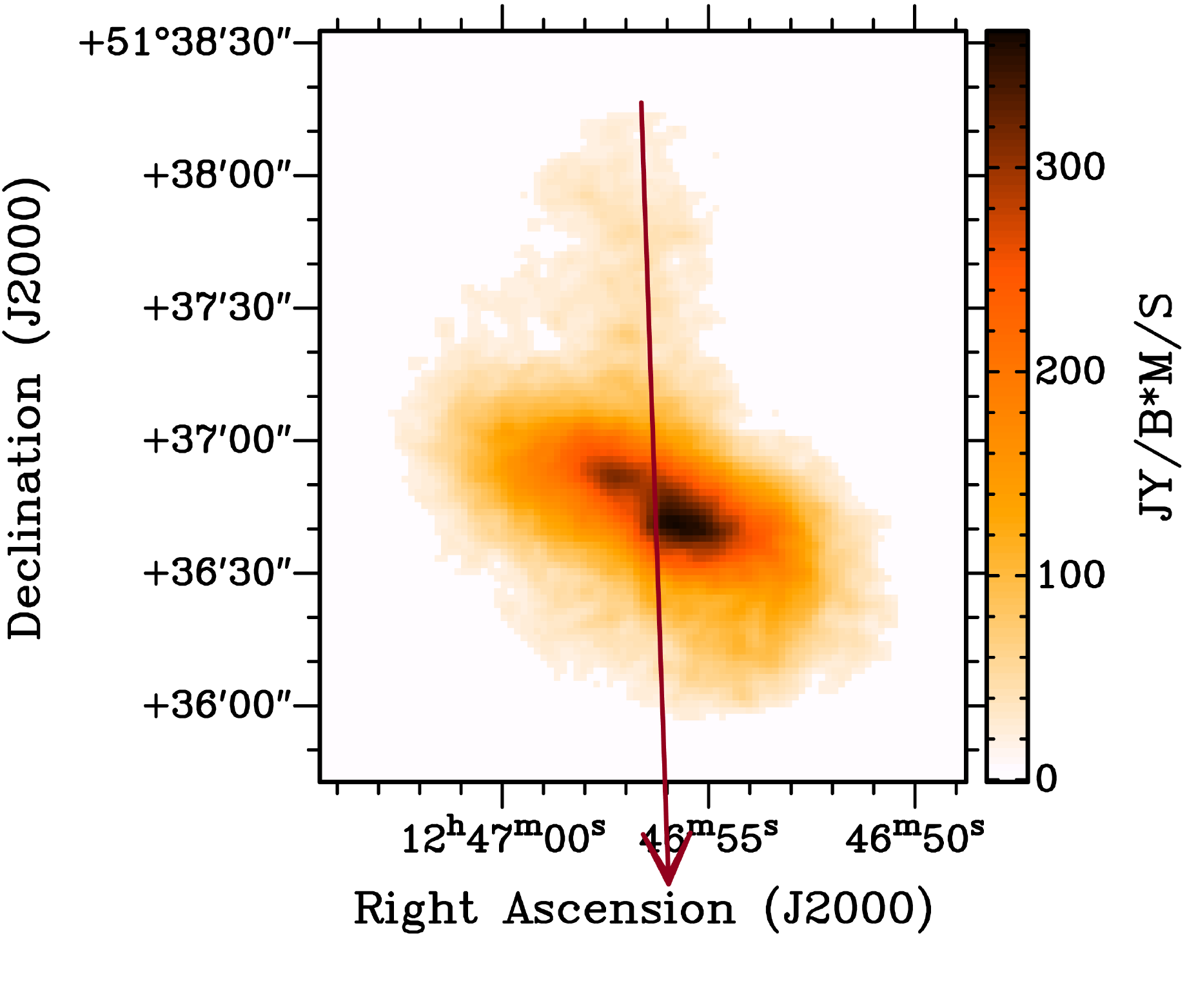}
\plottwo{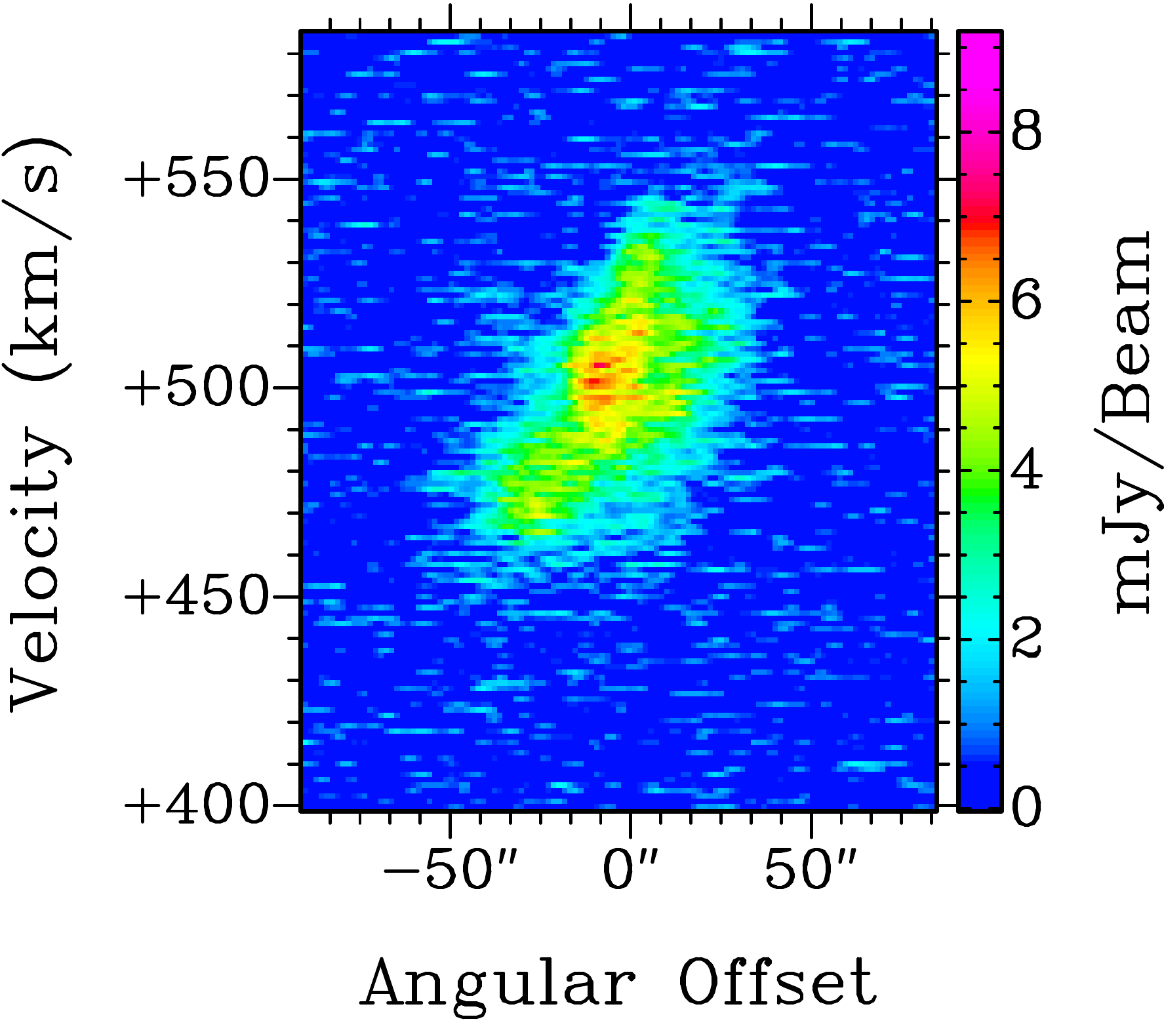}{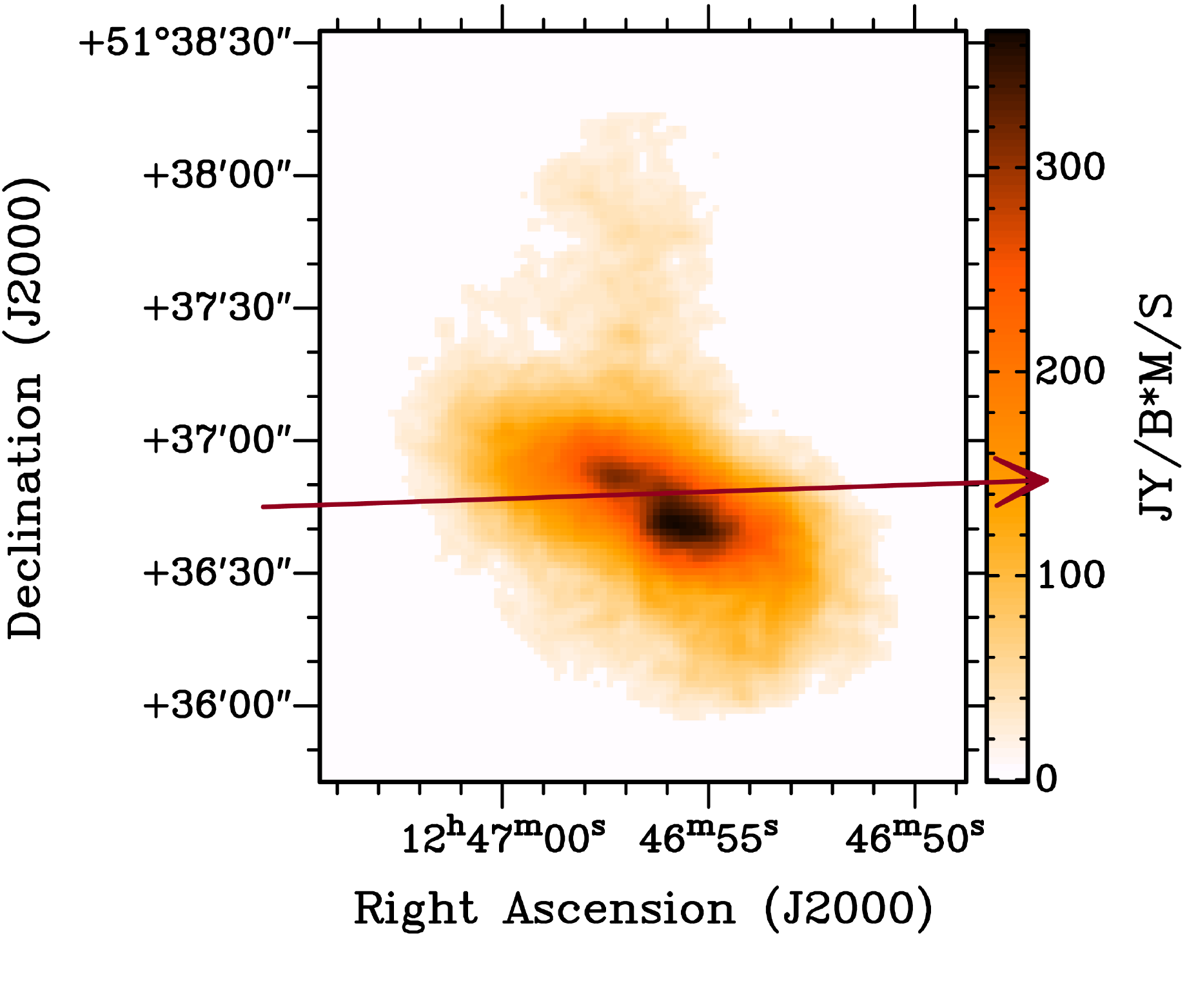}
\plottwo{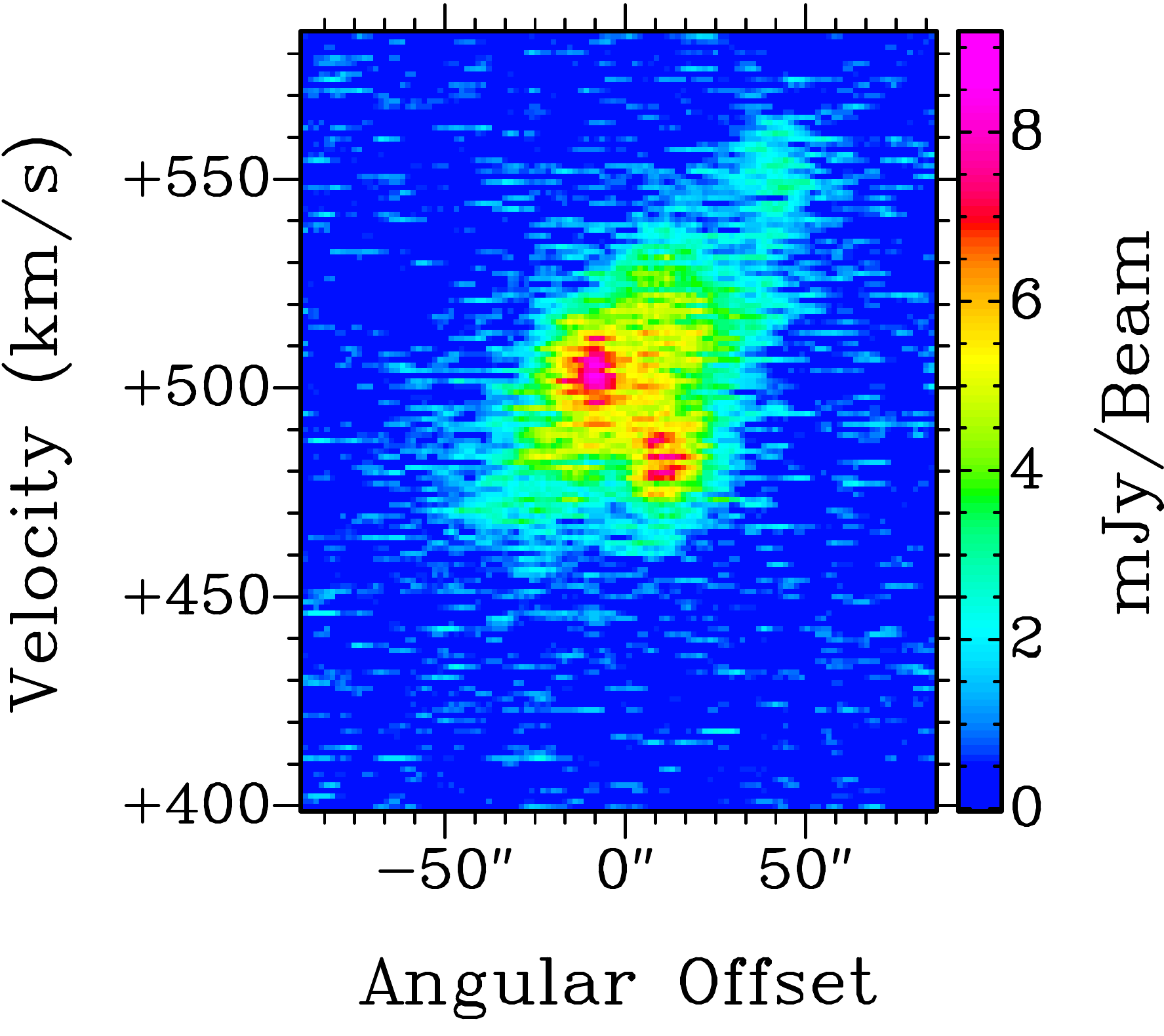}{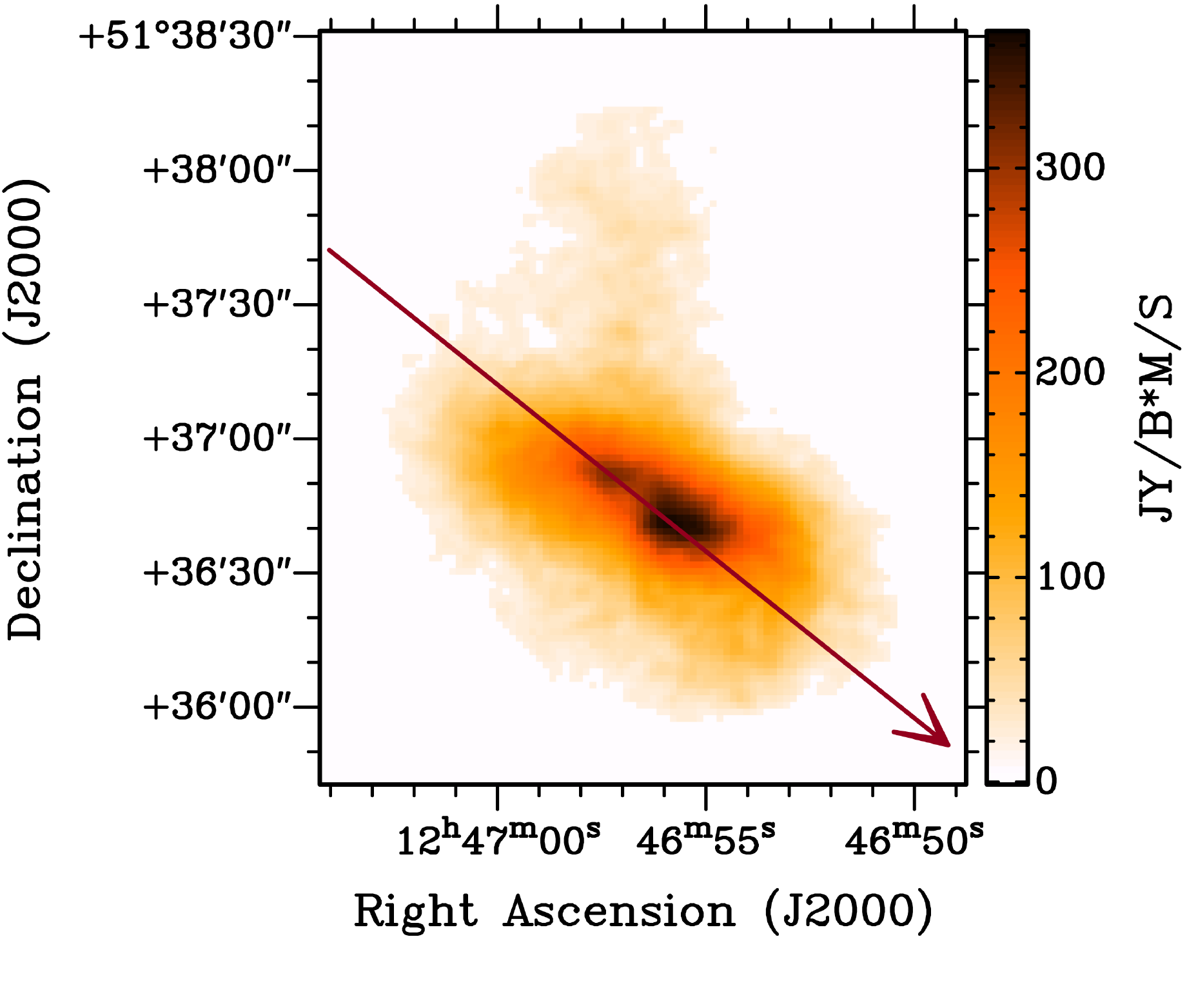}
\end{center}
\figcaption{Haro 36: The left column contains the P-V diagrams and the right column contains the natural-weighted integrated \HI\ map with a red arrow indicating the location of the corresponding slice through the galaxy and pointing in the direction of positive offset. The first row corresponds to the optical major axis, the second row to the optical minor axis, and the third row is a slice through both central peaks (i.e., the ``mouse diagram"). \label{fig:h36p-v}}
\end{figure}

Haro~36's P-V diagram of the optical major axis (PA=2\degr\ and centered on the galaxy's optical central coordinates of $12^{\rm{h}}46^{\rm{m}}56^{\rm{s}}.3, 51\degr36\arcmin48\arcsec$; as determined by \citealt{hunter06}) is shown in the first row of Figure~\ref{fig:h36p-v}.  The optical major axis was chosen to match the faint outer isophotes of the galaxy that are evident from the greyscale of Figure~\ref{fig:h36v_fuv} and from the optical-light contours in the top right of Figure~\ref{fig:h36x0r} \citep{hunter12}.  The optical major axis is not aligned with the \HI\ kinematic or morphological major axis.  The high slope in the optical major axis P-V diagram appears to be due to the choice in PA of this particular slice and the distorted velocity field along the slice, as can be seen in Figure~\ref{fig:h36x012}.

One notable feature of this P-V diagram is that the northern extension is significantly kinematically different from the main body. The northern extension is seen in the P-V diagram between $-$100\arcsec\ and $-$16\arcsec, varying from 420 \kms\ to 490 \kms\ with increasing velocity towards increasingly positive offsets.  The main body kinematics are distinct from the northern extension: the velocities decrease with increasingly positive offsets.  The second row is associated with the optical minor axis (also centered on the galaxy's optical coordinates); the main body shows some spread in the velocities closer to the center, but appears overall to be participating in solid body rotation.  
 
The most surprising P-V diagram comes from a slice directly through both peaks of Haro~36 resulting in a mouse-like shape (bird's eye view) in the P-V diagram, shown in the third row of Figure~\ref{fig:h36p-v} (PA=51.3\degr\ and centered at $12^{\rm{h}}46^{\rm{m}}56^{\rm{s}}$.608, 51\degr36\arcmin46\arcsec.83).  The two \HI\ peaks are separately visible in velocity space; the one centered at $-$8\arcsec\ and 505 \kms\ creates the left ear of the mouse while the second is well spread out in velocity space and creates the right ear of the mouse and the mouse's main body; the right ear is centered at $\thicksim$485 \kms\ and 8\arcsec.  The spatially larger peak to the southwest has a large range of velocities in the line of sight ($\thicksim$465-540 \kms), while the smaller peak to the northeast is more compact in velocity space in the line of sight ($\thicksim$470-520 \kms).

\section{Discussion: Haro 29}\label{sec:disc29}

	In Figure~\ref{fig:h29x012} the morphology of Haro 29 is visibly disturbed with the curving feature and extension of diffuse gas to the south.  The gas kinematics of the inner region, seen in Figure~\ref{fig:h29x012}, also show that the kinematic major axis of this dense region is not aligned with the stellar major axis of PA=86\degr.  Are these features, both inner and outer, the result of inflow, outflow, a merger, or an encounter with another galaxy? Haro 29 has no obvious companions currently, but a possible explanation for its morphology is some kind of external perturbation, such as a past interaction, a merger, or a hidden companion similar to the one recently discovered near NGC 4449 \citep{delgado12}.  

\subsection{Inner Region}
		
	One viable reason for two \HI\ peaks is star formation occurring between the two peaks consuming, ionizing, and dispersing the gas; it is possible that the gas between the two \HI\ peaks originally had the same column density as the two \HI\ peaks that are seen in the current morphology.  If the stars in Haro 29 were created via a particularly strong starburst, we expect gas consumption, ionization, and the stellar winds of that starburst to have a significant effect on the surrounding gas \citep{d'ercole02}.  In the case that the stars formed in a slower, less violent manner, fewer stars would have created strong stellar winds at any one time. This slower manner of star formation will therefore take much more time to significantly impact the \HI\ morphology than a strong starburst.
	
	Stars in Haro 29 are forming at a high rate and have in the past gigayear been forming stars at a moderate rate.  The stars should therefore have strong stellar winds participating in dispersal.  The FUV and V-band emission is highly concentrated off to the west (right) of the two \HI\ peaks, however, they also have a faint extension between the two \HI\ peaks. Since the dip in \HI\ between the two peaks is small, we cannot rule out recent star formation as a possible reason for the two peak morphology.  
	
	More evidence of the recent star formation affecting the surrounding gas, and possibly having created the two central peaks comes from \citet{schwartz06}; they find that the gas near the brightest stellar regions of Haro 29 has large outflow speeds of $-$191$\pm$92 \kms.  \citet{schwartz06} do this by using stellar clusters as a background source to be able to measure the doppler shift of absorption features in the ultraviolet, giving them the velocity of cold ($\thicksim10^{2}$\ K) and warm ($\la10^{4}$\ K) gases.  Large outflow speeds indicate that the young stellar population is having a large effect on the gas and is a likely reason for the two peak morphology.  
	
	These large outflow speeds could also be affecting the kinematic major axis of the \HI.  The kinematic major axis of the \HI\ in the inner region (seen in Figure~\ref{fig:h29x012}) is noticeably offset from the morphological major axis of the optical galaxy at a PA=86\degr; their position angles are roughly 45-55\degr\ different when the kinematic major axis is drawn by eye.  Cases with such different gas kinematic and photometric major axes are often attributed to mergers, interactions, inflow (from bar instabilities or extragalactic gas), or outflow (NGC 2777: \citealt{hogg98}; NGC 5253: \citealt{kobulnicky08}; NGC 1140 and NGC 1800: \citealt{hunter94}).   Inflow was not found in the study by \citet{schwartz06} and is therefore not a likely cause of the axis offset.  The two \HI\ peaks may also be a result of the gas being lumpy.  This lumpiness may be a result of turbulence causing fragmentation in gas clouds \citep{zhang12}.  Another possible explanation for the two \HI\ peaks is an accretion or merger; however, as seen in the third row of Figure~\ref{fig:h29p-v}, both peaks are participating in solid body motion.  Nonetheless, it is possible that the merger occurred with both components corotating. The offset in axes cannot be ruled out as a result of an advanced merger or an interaction. However, the large outflow speed measured by \citet{schwartz06} does indicate outflow as a possible cause of this discrepancy in the \HI\ kinematic and optical major axes.

\subsection{Outer Region}

	The disturbed kinematics, such as the spreading out of the isovelocity contours seen in  Figure~\ref{fig:h29x012}, and curved morphology of Haro 29 are consistent with the galaxy experiencing a violent event. Such an event could be ram pressure stripping; the distribution of the \HI\ in Figure~\ref{fig:h29x012} shows a high column density gradient along with large velocity dispersions in the southeast of the inner region, while the diffuse gas of the outer region is generally offset to the west of the inner region.  However, Haro 29 is very isolated in the Canes Veneciti Group, so the source of the intergalactic medium that Haro 29 would be traveling through is uncertain.  Another possibility is an interaction or close encounter with another galaxy where the relatively dense \HI\ region near the center has remained in solid body rotation and the diffuse \HI\ in the outer region has been disturbed to the point of being pulled into a slightly different orbit.  \citet{kronberger07} and \citet{bellocchi12} show that the outer disks of ionized gas in massive galaxies have distorted and asymmetric velocity fields when they experience an interaction.  In these massive galaxies, the ionized gas tends to be more thoroughly distributed throughout the entire body of the galaxy.  If interactions are able to distort the velocity fields of the ionized gas in massive galaxies it should have a similar effect on the neutral gas kinematics as well.  \citet{kronberger07} also point out that the less massive galaxies in their models experienced more dramatic asymmetries than massive galaxies.  Therefore, a dwarf sized galaxy will likely have large asymmetries depending on the type of interaction.  Haro 29 has a distorted outer velocity field when compared to the inner velocity field, and also a highly asymmetric morphology, consistent with these interaction models.  \citet{youngblood98} also pointed to the core-halo morphology of Haro 29 as being reminiscent of the models by \citet{noguchi88a, noguchi88b}.  In those models, a central build up of gas results from an interaction that occurred some distant time in the past (about several gigayears).  Haro 29 having experienced a past interaction is therefore consistent with these models.  
	
	The isolation of Haro 29, like many other BCDs, does present a challenge to this theory, but it is possible that Haro 29 has an unknown companion, or, as pointed out by \citet{youngblood98}, that the interaction took place a few gigayears ago, and the other galaxy has traveled a significant distance.  Haro 29's closest companion, NGC 4736 at a deprojected distance of $\thicksim$1.4 Mpc relative to Haro 29 and moving at a velocity of 27 \kms\ relative to Haro 29 would have had to pass nearby to Haro 29 51 Gys ago.  This number is clearly too large to be realistic.
	
	If Haro~29 has had a close encounter with another galaxy, the curving feature or the southern extension in Figure~~\ref{fig:h29x012} may be the beginning of a tidal tail or a bridge.  The gas creating this tidal feature would then come from the outer disk and would not be expected to have a large fraction of the galaxy's total stellar light contained in it.  Using the task \textsc{tvstat} in AIPS, the approximate percent of V-band stellar flux in the inner region (the inner region having been defined in Section~\ref{sec:h29mass}), compared to the entire galaxy disc (within sensitivity limits) is measured to be 73\%.  The majority of the V-band stellar flux is therefore contained in the inner region, while only 24\% of the \HI\ is contained in the inner region.  This is consistent with the outer features being more gas rich than the inner region.  

\begin{figure}[h]
\epsscale{1.11}
\begin{center}
\plottwo{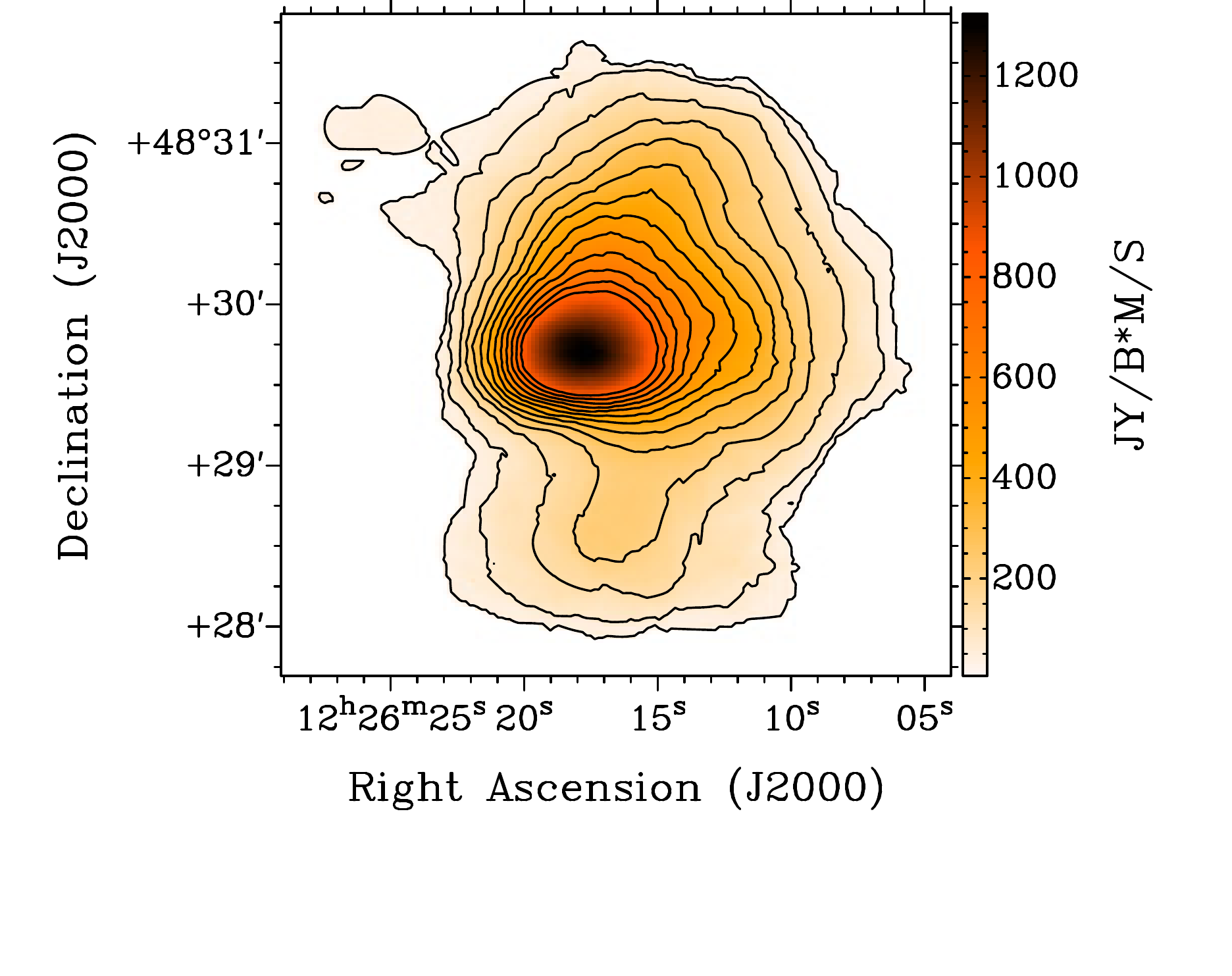}{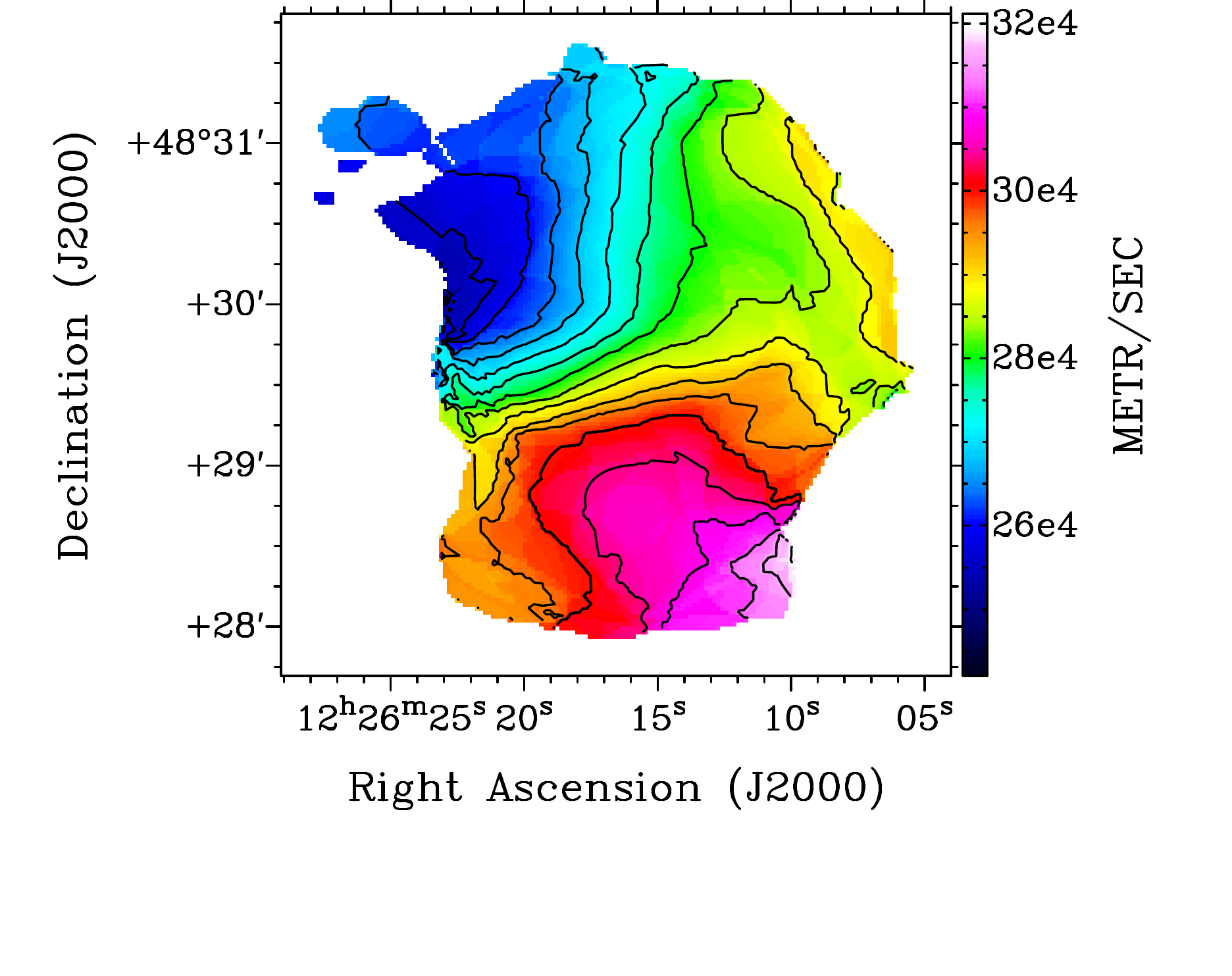}
\end{center}
\figcaption{Haro 29's integrated maps convolved to a $27\arcsec.9\times37\arcsec.2$ beam.  \textit{Left}: Integrated \HI\ intensity map; contour levels are 1$\sigma\times$(2.5, 5, 10, 15, 20, 25, 30, 35, 40, 45, 50, 55, 60) where 1$\sigma=1.43\times10^{19}\ \rm{atoms}\ \rm{cm}^{-2}$.  \textit{Right}: Intensity weighted velocity field; contour levels are 240, 244, 248, 252, 256, 260, 264, 268, 272, 276, 280, 284, 288, 292, 296, 300, 304, 308, 312, 316, and 320 \kms.
\label{fig:h29cvlvial83}}
\end{figure}

To increase the signal-to-noise of some of the more extended features of our \HI\ data, and to search for features seen in \citet{viall83} that were not seen in our natural weighted data, a $27\arcsec.9\times37\arcsec.2$ beam was used to convolve the natural weighted cube.  This beam size was chosen because it is the size of the synthesized beam used to map Haro 29 in \citet{viall83}.  In the integrated \HI\ map (left side of Figure~\ref{fig:h29cvlvial83}), the southern extension is particularly prominent; it has an \HI\ mass of approximately 12\% of the total galaxy \HI\ mass, as measured from the natural weighted cube.  It also does not extend further south in the convolved image at the sensitivity of the data.  Figure~\ref{fig:h29cvlvial83} also reveals another feature; at $2.5\sigma$ or $3.58\times10^{19}\ \rm{atoms}\ \rm{cm}^{-2}$, there is a large piece of separate emission centered at $12^{\rm{h}}26^{\rm{m}}26^{\rm{s}}$, 48\degr31\arcmin7\arcsec.5 and a couple of small pieces of separate emission below that.  The large piece of separate emission is nearly the size of the beam at $\thicksim20\arcsec\times40\arcsec$ and is in the same location as a structure seen in \citet{viall83}  (their Figure 3, Component 7). In Figure 3 of \citet{viall83}, the structure associated with the large piece of tenuous emission seen in our map is unresolved by the beam, but the structure is in the same location as the one seen in the our convolved map, we therefore believe it is possible that this piece of emission seen in Figure~\ref{fig:h29cvlvial83} is part of a real structure.  The origin of this separate structure was not discussed in \citet{viall83}. They did mention the possibility of tenuous gas connecting the piece of emission to the rest of the galaxy, and at our sensitivity the emission between the main body and this separate structure is more extensive making a bridge of tenuous gas seem very likely.  The velocities of this piece of tenuous gas can be seen in the map on the right side of Figure~\ref{fig:h29cvlvial83}.  The velocities in the piece of emission appear consistent with the velocities of the main body closest to the piece of emission.  If it is tidal in origin, then it may have been ripped from the main body of Haro 29 or a galaxy that Haro 29 encountered.  It could also be a foreground/background cloud, or a cloud slowly accreting onto Haro 29.

	More data with high sensitivity are needed on Haro 29 to see if these outer features are more extensive than what is detectable in our current data.  If an interaction has occured, then it is likely that most of these outer features were cause by that interaction and will be more spatially extended by tenuous emission.  High sensitivity, single dish observations to look for extended gas in LITTLE THINGS galaxies are ongoing.  

	The extended emission in the outer region could also be from accretion of intergalactic medium or be part of an outer \HI\ pool that is slowly accreting onto Haro 29 \citep{taylor93, thuan97, vanzee98, wilcots98}.  However, the isolation of Haro 29 makes the source of the intergalactic medium difficult to locate.   	
		
\section{Discussion: Haro 36}\label{sec:disc36}

 The \HI\ kinematic major axis in Haro 36's main body changes position angle dramatically from the southwest to the northeast (see Figure~\ref{fig:h36x012}); the most likely explanation is a violent event. The iso-velocity contours of the main body appear more disturbed in the northeast of the velocity field in Figure~\ref{fig:h36x012} or close to where the main body and northern extension meet.  This could be due to the northern extension overlapping with or impacting the northeast section of the main body.  We attempt to give possible explanations for all of these features here.	

\subsection{Main Body}
  
	Each of the explanations mentioned for Haro 29's two \HI\ peaks are also reasonable explanations for Haro 36's two \HI\ peaks, but the stars in Haro 36 are concentrated directly between the two peaks.  This makes gas consumption, ionization, and dispersal an appealing explanation for two peaks in the gas morphology in Haro 36. However, the two peaks in Haro 36 are also kinematically distinct, as shown earlier in the mouse diagram in Figure~\ref{fig:h36p-v}, pointing to the possibility of accretion or a merger as an explanation for the two peaks.  It is also possible that a hole between the two peaks may be contributing to their appearance in the P-V diagram, giving the illusion of two separate kinematic components.  If there was a hole creating this appearance, then it would appear as an ellipse, a partial ellipse, or a well-defined gap in the P-V diagram \citep{brinks86}.  
	
	To look for a hole in Haro 36 between the two components, the P-V diagram slice was moved perpendicular to the axis created by the two peaks, but no ellipses, partial ellipses, or gaps were seen in the P-V diagrams between the two peaks, which are therefore not separated by a resolved, expanding hole.  If there is a hole instead that is expanding perpendicular to the line of sight or one that has stopped expanding, then it is still possible that a hole may be to blame for the two peak morphology and the appearance of the mouse diagram.  Intense star formation may also have contributed to the kinematics of the \HI\ peaks.  Yet, the right ear of the mouse appears so kinematically distinct from the rest of the P-V diagram, that if removed, the P-V diagram would look solid body with few peculiarities. It therefore seems more likely that the southwest peak is at least partially composed of a kinematically distinct cloud.   

	In the case that one \HI\ peak is a gas cloud that has been accreted onto Haro 36, it may be impacting the central region of the galaxy, encouraging star formation to occur in the gas between the two peaks.  It is also possible that it is a foreground \HI\ cloud that is only seen projected onto the galaxy and is not physically impacting the galaxy.  NGC 1569, a galaxy that is occasionally considered a BCD (e.g., \citealt{devost97, vaduvescu06}), has a similar situation with a kinematically disturbed \HI\ cloud.  In NGC 1569 the disturbed \HI\ cloud has strong non-circular motions and is thought to be colliding with the galaxy resulting in star formation in the gas that it is plowing into \citep{stil02b, johnson12}.  Another galaxy with a possible interacting gas cloud is NGC 5253; \citet{lopez12} suspected the possibility of an independent, interacting \HI\ cloud in NGC 5253 when the P-V diagram going through its southeastern plume revealed a double peaked, sinusoidal shape.  The sinusoidal shape indicated a possible infall and merging scenario between a separate \HI\ body (the southeastern plume) creating one of the peaks in the P-V diagram, and the main body of the galaxy creating the other peak in the P-V diagram.  There is no obvious sinusoidal pattern occurring with the main body and either of the \HI\ peaks in Haro 36, so if the southwest peak in Haro 36 is a distinct cloud interacting with the galaxy, then it does not appear to be infalling in the same manner as the possible distinct \HI\ cloud in NGC 5253.  
	
	The V-band flux is measured using \textsc{tvstat} in AIPS for each peak in Haro 36 to see how the stellar light and approximate stellar mass (assuming a light to mass ratio of $\thicksim$1) compares to the \HI\ mass.  The northeast peak has $\thicksim$16\% of the V-band flux of the southwest peak, and the southwest peak has $\thicksim$19\% of the total V-band flux in the galaxy.  The northeast to southwest peaks' approximate stellar masses then have about the same ratio as their \HI\ masses (the northeast peak has an \HI\ mass $\thicksim$18\% of the southwest peak's mass).  The ratio of each individual peaks' stellar mass to the total stellar mass of the galaxy is nearly double in comparison to the ratio of the individual \HI\ masses of the peaks to the total \HI\ mass of the galaxy (the southwest peak has a mass $\thicksim$9\% of Haro 36's total \HI\ mass). So the ratios between the peaks have stayed nearly the same when comparing stellar mass to \HI\ mass, but the stellar mass appears more concentrated toward the peaks of the galaxy than the \HI.
	
\subsection{The Northern Extension}

	The northern extension of \HI\ in Haro 36 is another interesting feature.  This extension could be a multitude of features: the beginning of a bridge, a tidal tail, or an impacting HI cloud.  In this section we attempt to determine its origin through the use of P-V diagrams.  
	
\begin{figure}[!ht]
\epsscale{0.36}
\begin{center}
\plotone{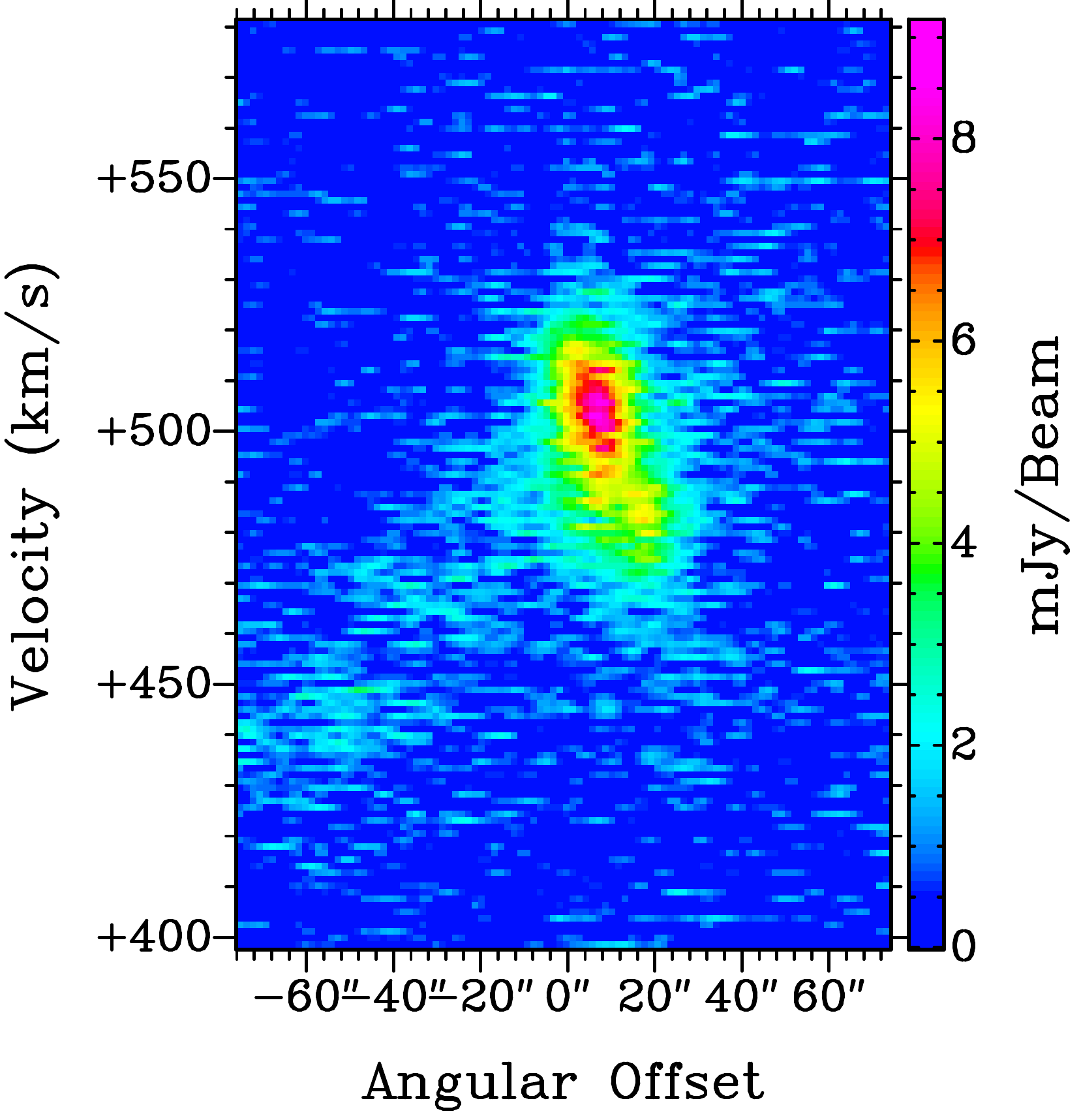}
\epsscale{0.46}
\plotone{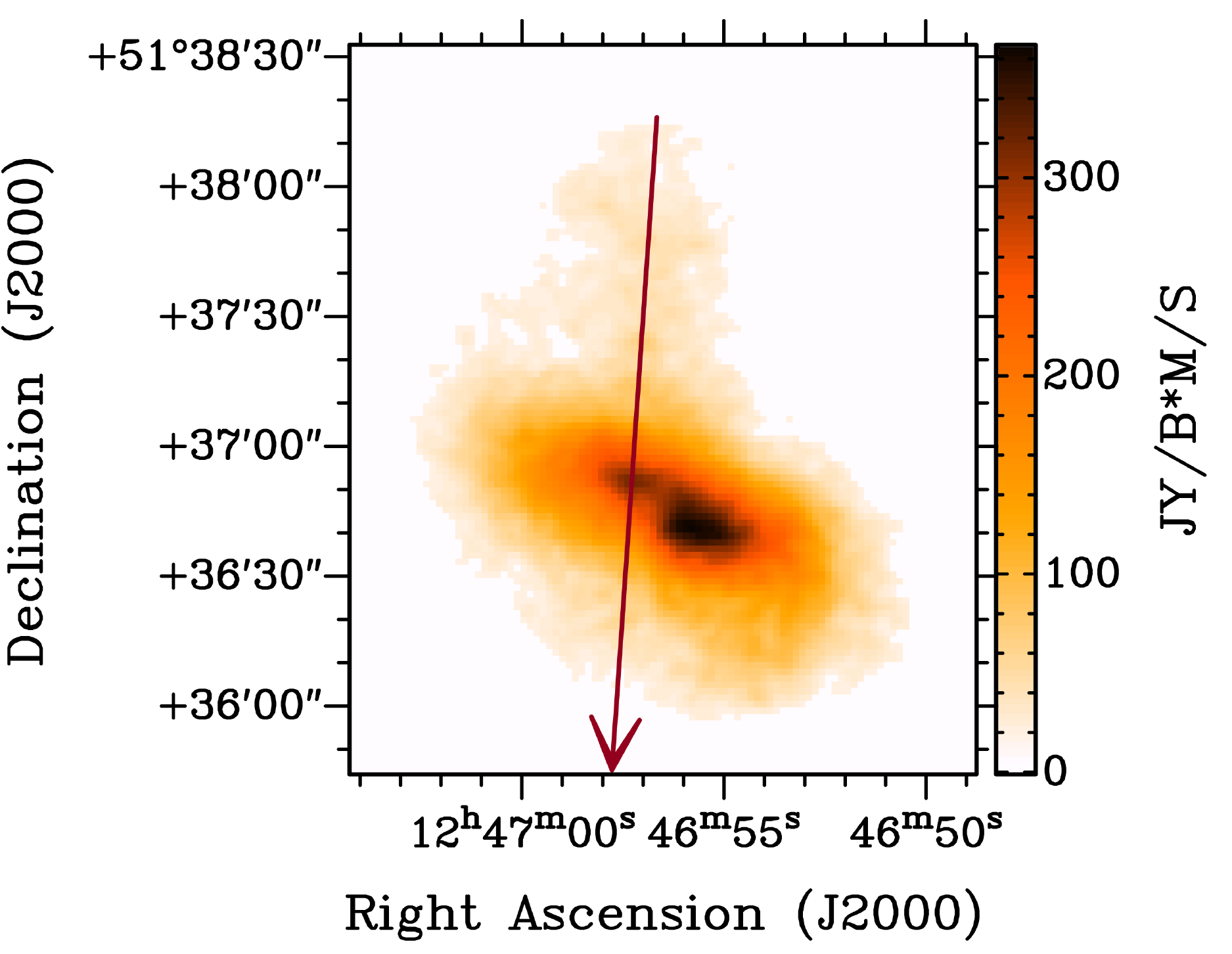}\\
\epsscale{0.36}
\plotone{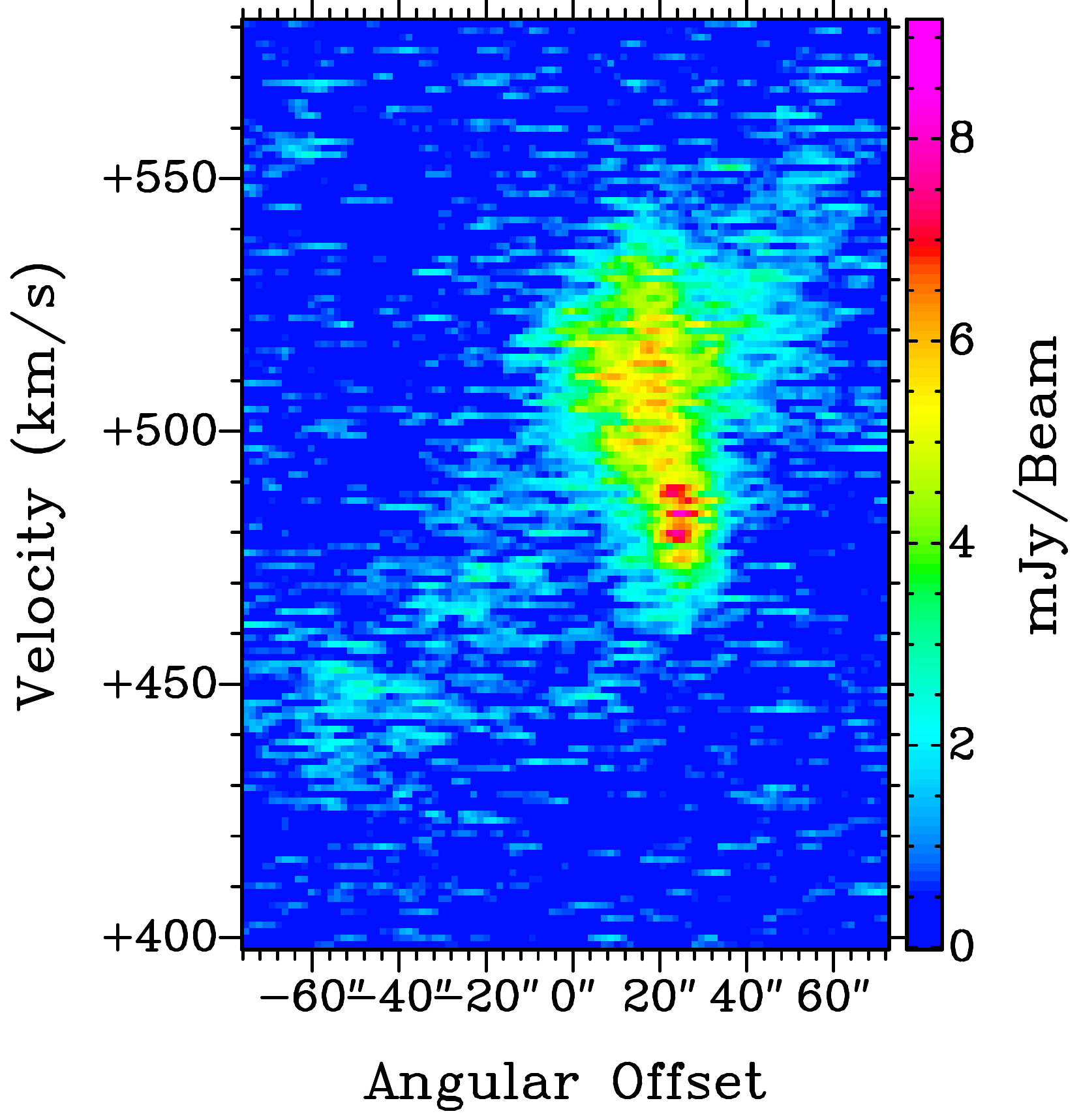}
\epsscale{0.46}
\plotone{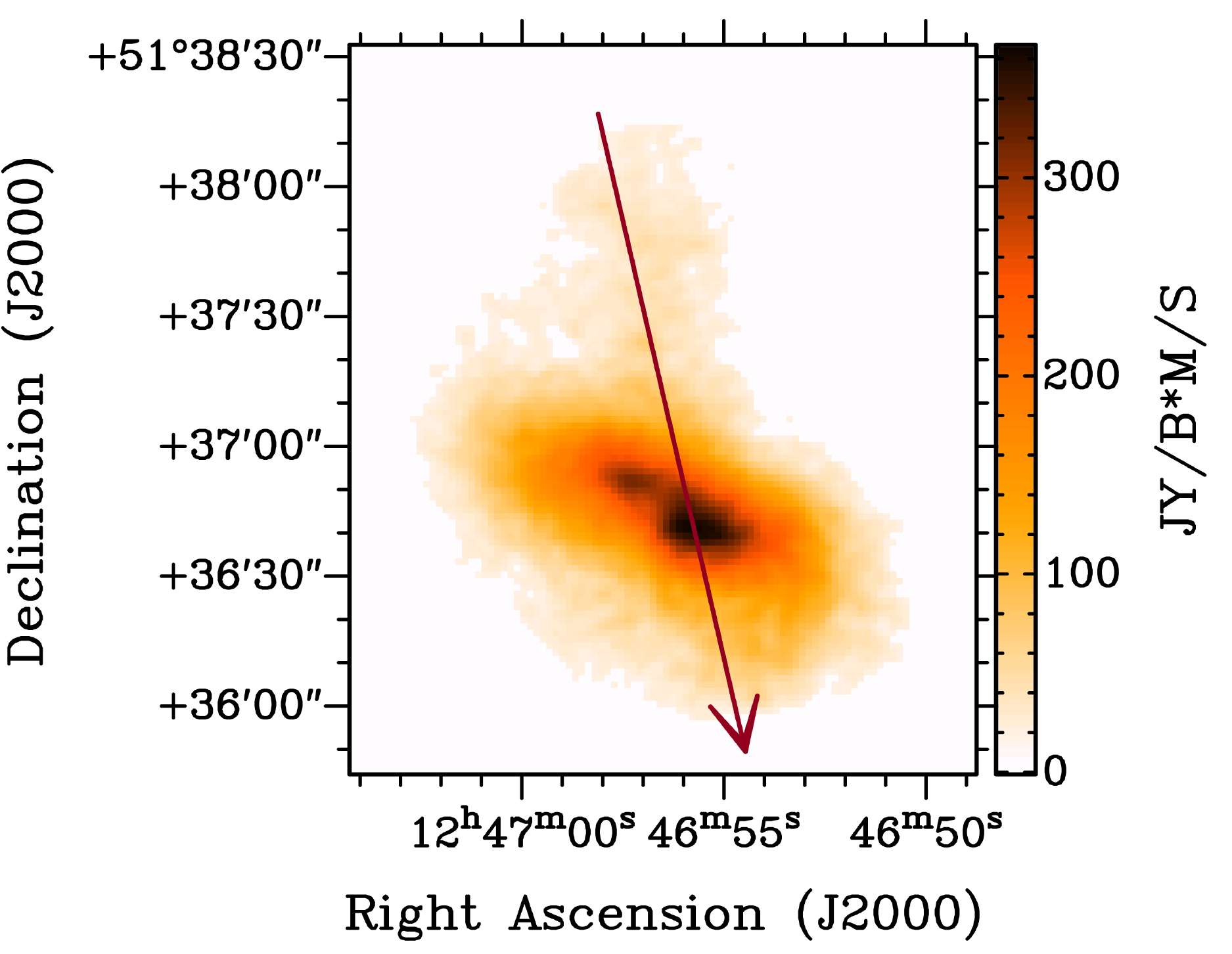}
\end{center}
\figcaption{Haro 36: The left column contains the P-V diagrams and the right column contains the natural-weighted integrated \HI\ map with a red arrow indicating the location of the corresponding slice through the galaxy, and pointing in the direction of positive offset. The first row is a slice through the northeast peak and the northern extension.  The second row is a slice through the southwest peak and the northern extension. \label{fig:h36p-v2}}
\end{figure}

	If an accretion event created the two central \HI\ peaks, then it is possible that the northern extension is also part of that accretion, and is physically connected to one of the central peaks.  If this is the case, then the northern extension would likely display some continuity with one of the peaks in velocity space.  In Figure~\ref{fig:h36p-v2}, slices are made through the northern extension and going through each of the two peaks (for the peak to the northeast a PA=175.9\degr\ and center of $12^{\rm{h}}46^{\rm{m}}57^{\rm{s}}.28$, 51\degr37\arcmin0\arcsec.87 are used; for the peak to the southwest a PA=13\degr\ and center of $12^{\rm{h}}46^{\rm{m}}56^{\rm{s}}$.28, 51\degr37\arcmin3\arcsec.77 are used).  Both P-V slices begin in the north and end in the south to make comparison between the two P-V diagrams simpler.  
	
\begin{figure}
\begin{center}
\epsscale{0.98}
\plottwo{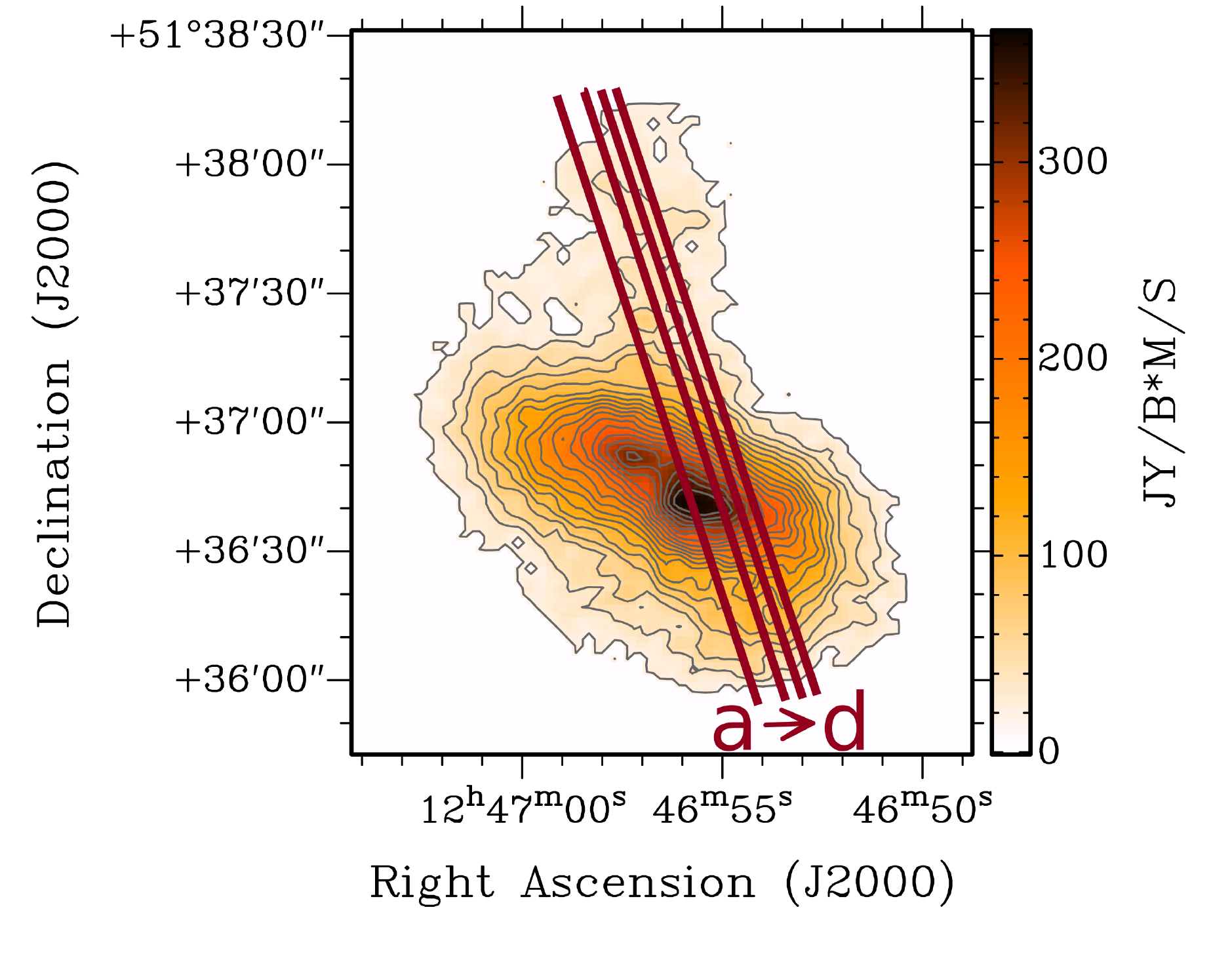}{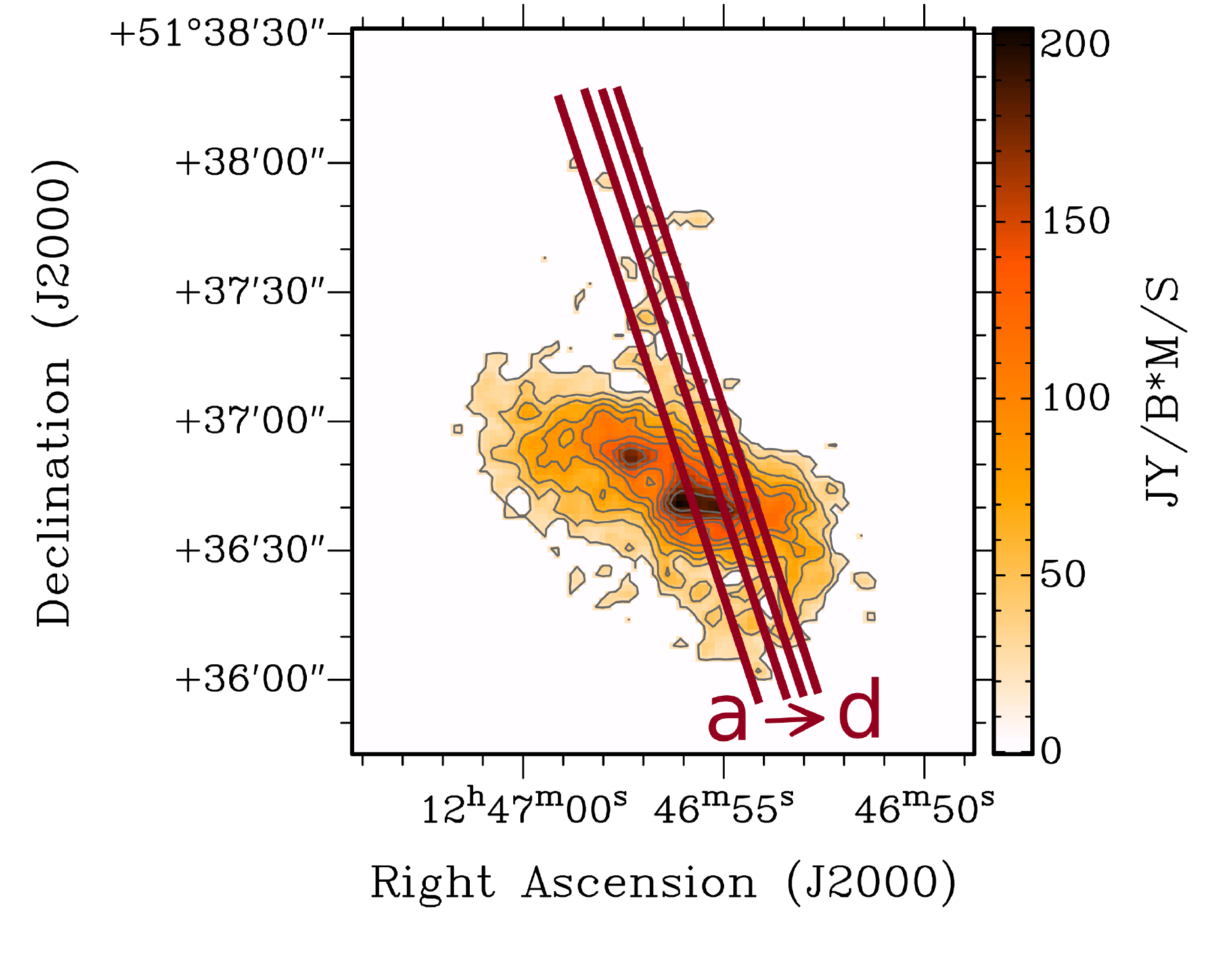}
\plottwo{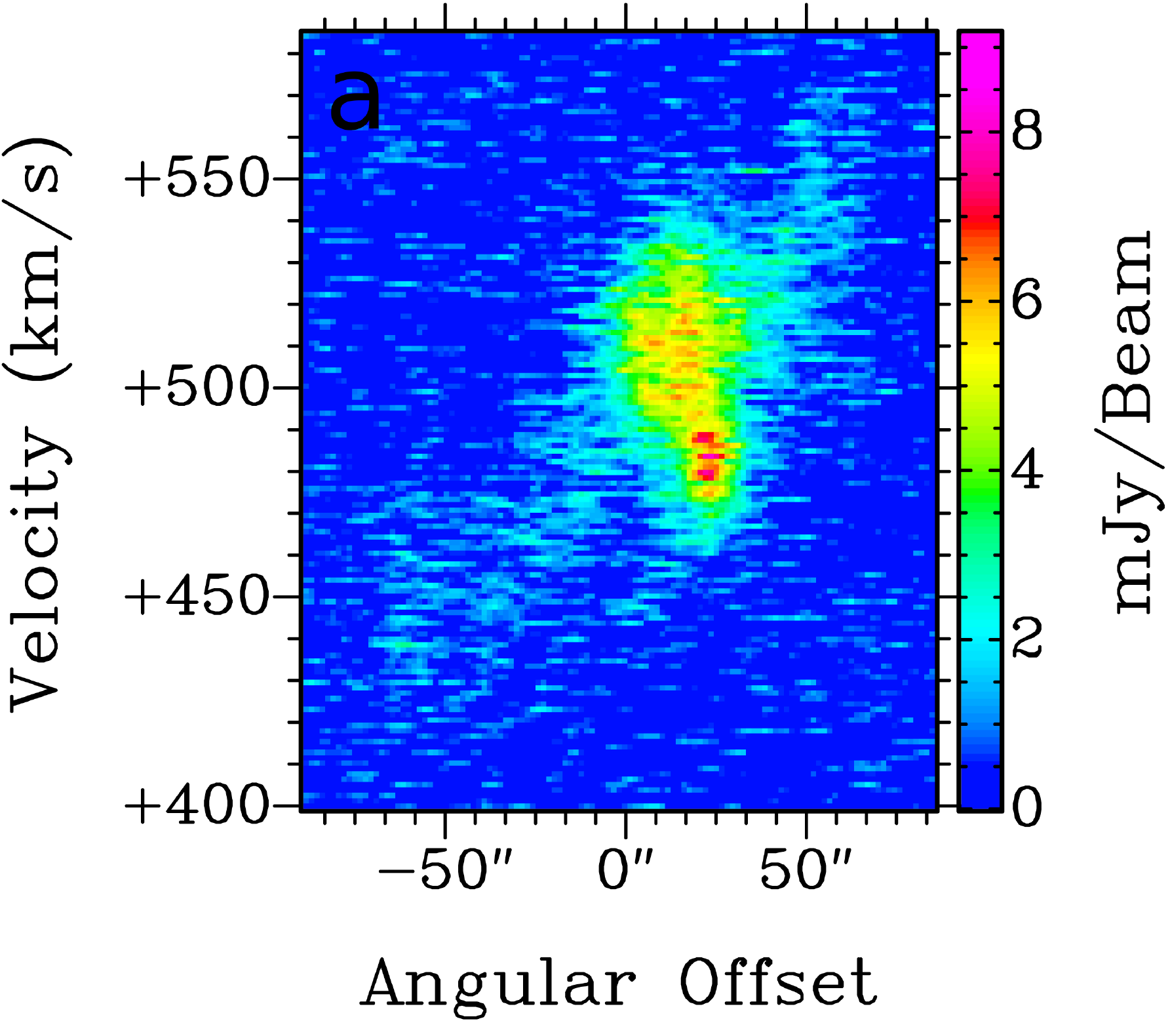}{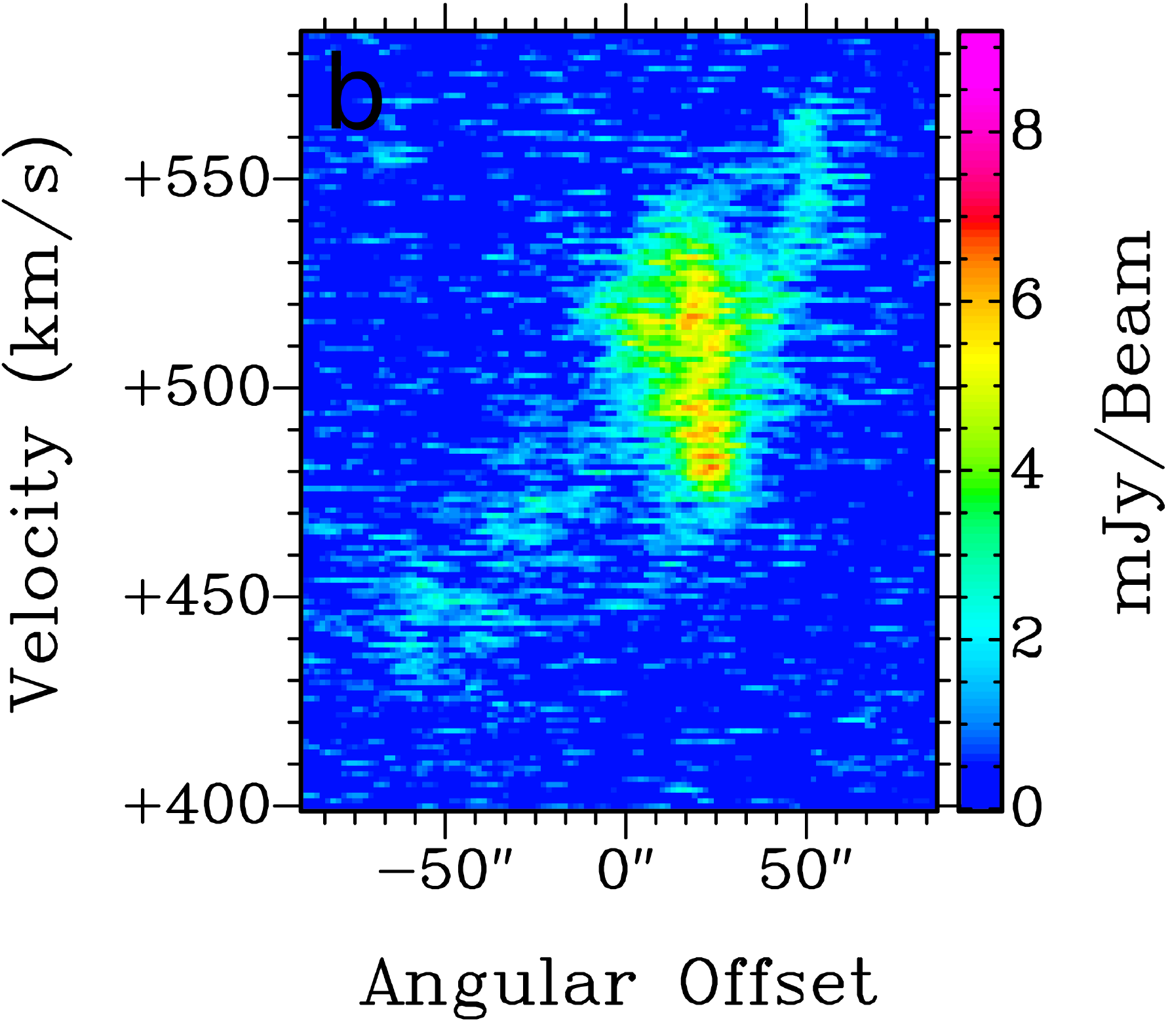}
\plottwo{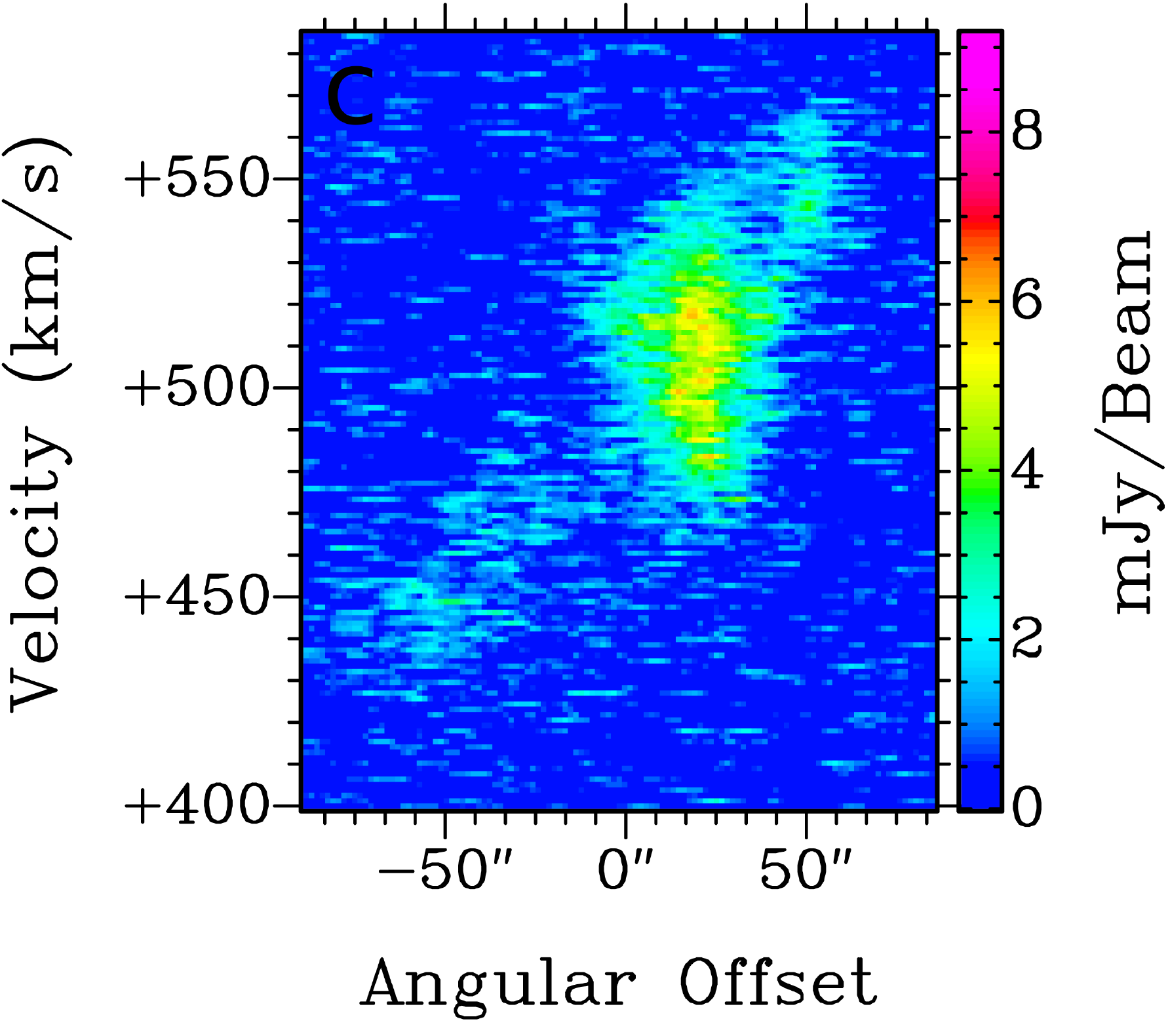}{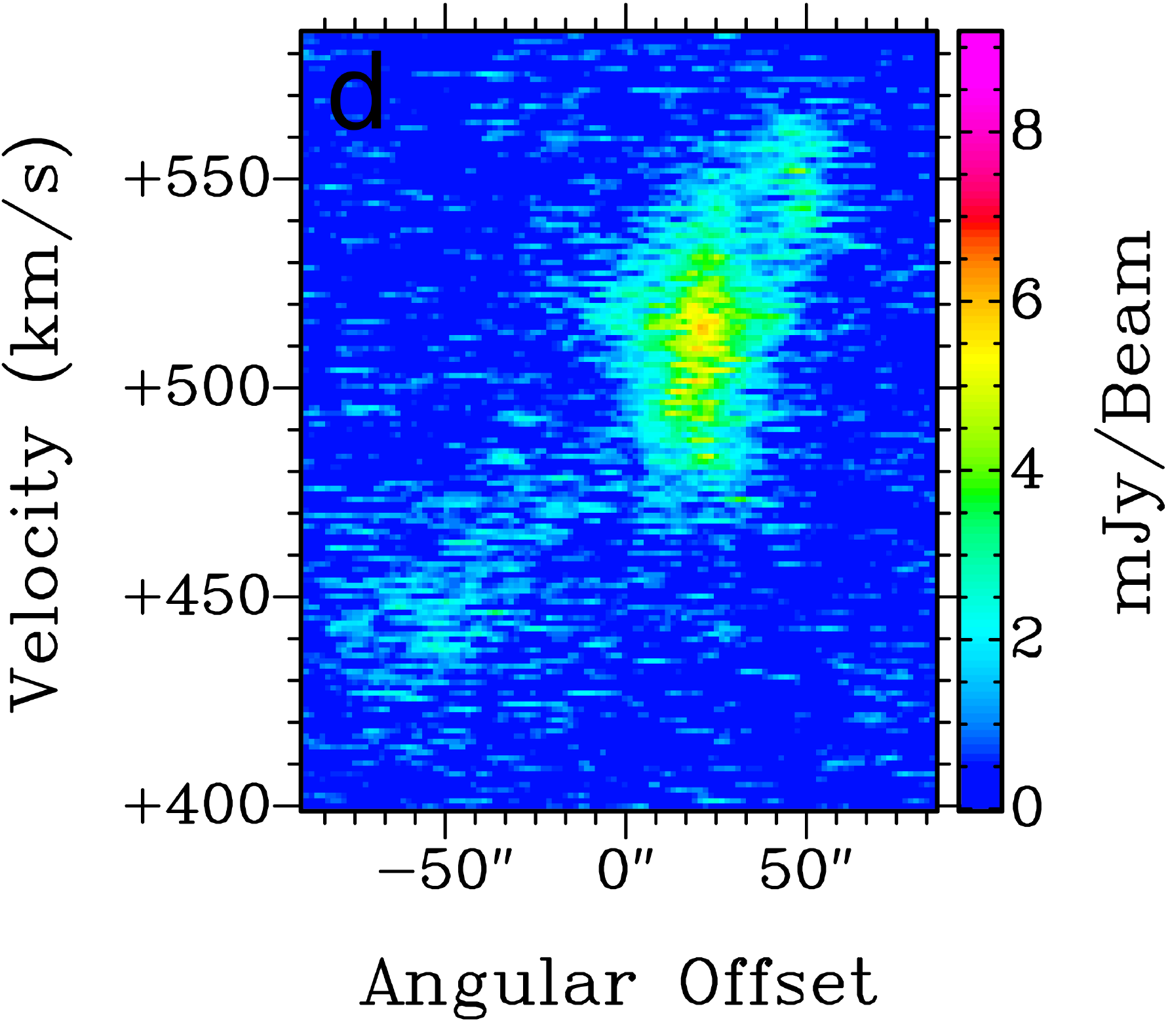}
\end{center}
\figcaption{Haro 36\ \ \textit{Top:} The natural-weighted (left) and robust-weighted (right) integrated \HI\ maps with red lines indicating the approximate locations of the slices through the galaxy used in the P-V diagrams (a-d) located below.   \label{fig:h36p-v3}}
\end{figure}

	In the diagram through the northeast peak at the top of Figure~\ref{fig:h36p-v2}, the northern extension increases in velocity with increasing offset at offsets less than 0 while the main body and \HI\ peak increase in velocity with increasing offset for offsets greater than 0.  There is no obvious connection between the northern extension and the northeast peak in the P-V diagram. The P-V diagram for the peak to the southwest at the bottom of Figure~\ref{fig:h36p-v2} shows relatively the same behavior as the northeast peak but there is a very interesting addition starting at a velocity of $\thicksim$530 \kms\ and an offset of 42\arcsec: a faint continuation of the positive slope seen at lower offsets that represents the northern extension in the same figure.  The bottom panels of Figure~\ref{fig:h36p-v2} suggests that the northern extension has some continuing emission in the southern region of Haro 36's main body as well.   This feature also appears in the mouse diagram as the tail of the mouse at $\thicksim$42\arcsec\, and is further investigated in the next four P-V diagrams in Figure~\ref{fig:h36p-v3}.  These four P-V diagrams represent the P-V slices from north to south that showed a continuation of the slope of the northern extension most prominently.  Going from P-V diagram \textit{a-d}, the slices successively decrease in RA (move to the right) while the center declination is kept constant.  As the slice moves from diagram \textit{a} to diagram \textit{d}, the extended feature at the top-right of the P-V diagram wraps around in a loop on the P-V diagram, leaving a dip in emission in the middle, then the bottom of the loop becomes less prominent and the top of the loop more prominent.  

\begin{figure}[!ht]
\epsscale{0.5}
\plotone{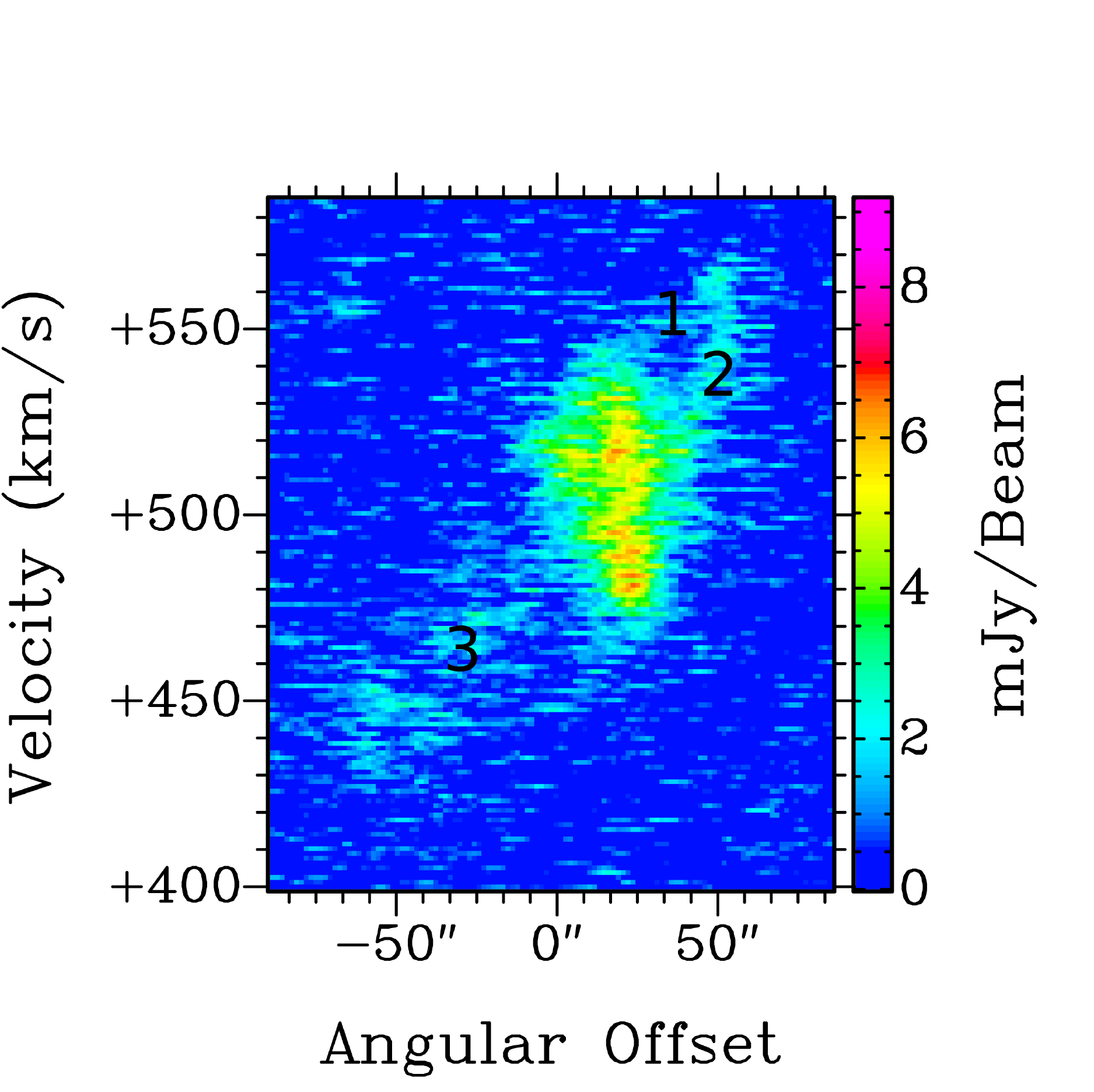}
\figcaption{A labeled version of P-V diagram b from Figure~\ref{fig:h36p-v3}.  \label{fig:h36p-vlabel}}
\end{figure}  

\begin{figure}[!ht]
\epsscale{0.5}
\plotone{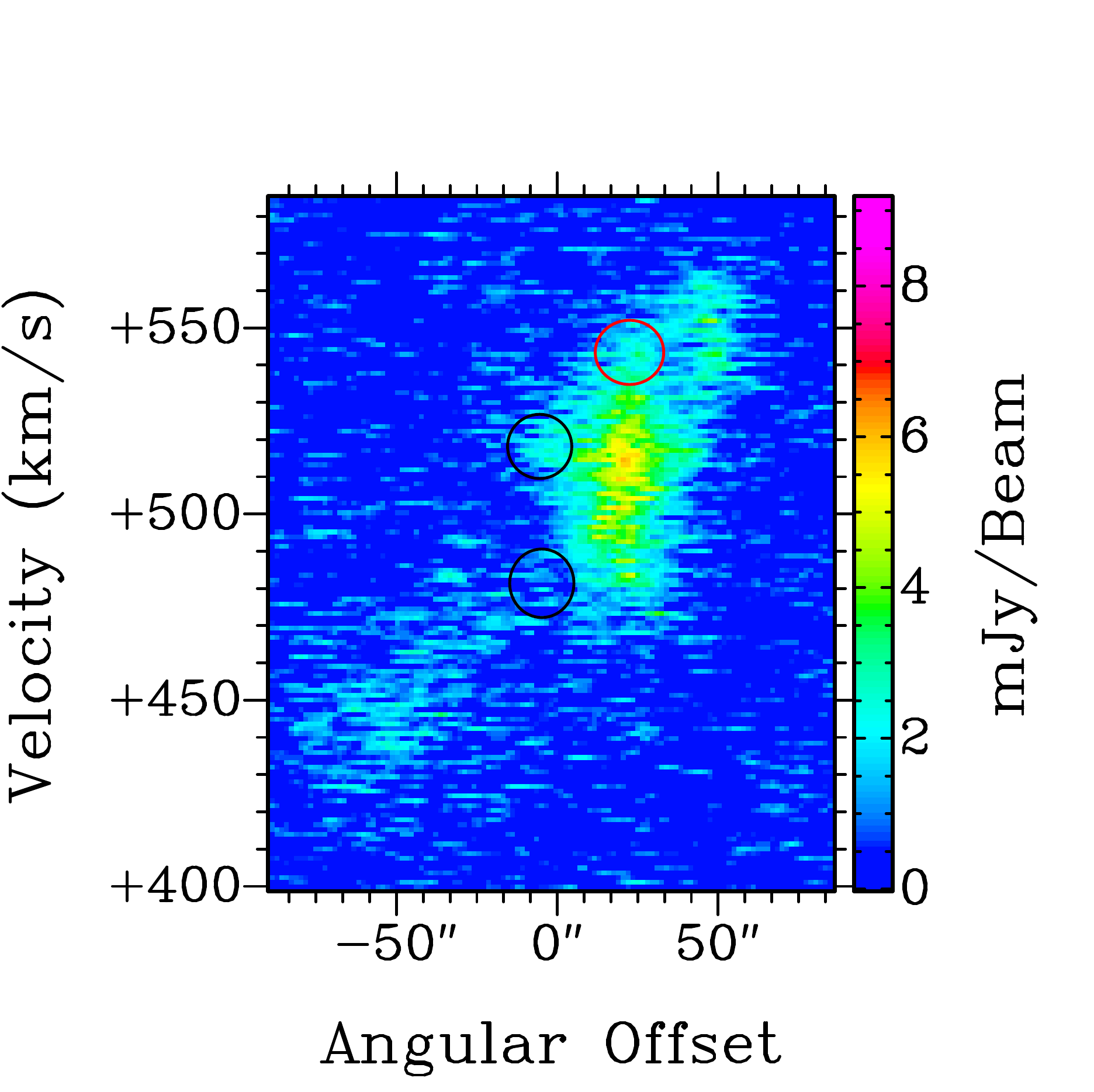}
\figcaption{Figure~\ref{fig:h36p-v3}d; black circles indicate where region 3 ends in velocity space and the location of the disk in velocity space at the same angular offset and the red circle indicates where region 1 ends and the disk meets it in velocity space.  \label{fig:h36p_v_d_labels}}
\end{figure} 

	This feature, if the northern extension and the southern (new) feature are connected, appears to originate from Haro 36 having a tidal tail.  To explore this idea further, the second P-V diagram is labeled in Figure~\ref{fig:h36p-vlabel} to make the discussion of the features simpler.  Region 1 is increasing in velocity with increasing offset, and is redshifted relative to the systemic velocity of the galaxy.  There is then a turn around at $\thicksim$58\arcsec\ in the P-V diagram, marking the beginning of region 2.  The velocity range in region 2 (530-555 \kms) also appears to be increasing with increasing offset but is lower than the velocity range of region 1 (545-555 \kms), and is still redshifted.  In region 3, which is the northern extension, we see this part is decreasing in velocity with increasing negative offsets.

The northern extension (region 3) has lower velocities than much of the main body, as seen throughout Figure~\ref{fig:h36p-v3}.  Also notable is the velocity of the disk where the northern extension meets the disk: it is higher than the velocity of the northern extension in the same location.  Figure~\ref{fig:h36p_v_d_labels} is an annotated version of Figure~\ref{fig:h36p-v3}d to help illustrate this point.  The black circles on the P-V diagram locate both the northern extension (bottom circle) and the disk (top circle) at approximately the same angular offset, where the main body's emission is centered at a velocity of 520 \kms\ while the northern extension's emission is centered on 480 \kms. The systemic velocity of Haro 36 is 502 \kms, meaning at the same location, much of the disk is redshifted with respect to the systemic velocity, while the northern extension is blueshifted. There is also a significant gap in velocity between the two regions at the same angular offset, from $\thicksim$490 \kms\ to $\thicksim$505 \kms.   This behavior suggests that the main body and the northern extension are likely separate structures.  In contrast, the red circle indicates the location where the disk meets region 1 in the P-V diagram.  At this angular offset, where region 1 and the disk meet, there is no gap between them in velocity: the emission is continuous there. Therefore, we cannot assume that region 1 is not connected to the disk.  Region 2 is more difficult; as seen in Figure~\ref{fig:h36p_v_d_labels}, there is a small gap where the disk and region 2 meet, but that could be due to the emission from region 2 not being prominent in the last P-V diagram.  In Figures~\ref{fig:h36p-v3}a and b, region 2 and the disk seem to be more continuous in velocity space, therefore, region 1 or region 2 may be spatially closest to the disk.  

If region 1, 2, and 3 are all physically connect to one another, then can we tell if the the gas connecting regions 1/2 to region 3 is in the foreground or the background of the main body of Haro 36? An interesting piece of evidence for this position of the tail is in the P-V diagrams b and c in Figure~\ref{fig:h36p-v3}; at an angular offset of $\thicksim$16\arcsec.7 and velocities of 495 \kms\ to 510 \kms, there is a slope in the yellow section of the P-V diagram that continues the region 2 to region 3 slope.  This may be a contribution from the gas connecting these features going in front of Haro 36 and slightly distorting the velocities seen in the main body of the galaxy. The distortion of slope in the P-V diagram also suggests that if regions 1, 2, and 3 are all physically connected, then they form a tidal tail that starts in the southwest with region 1 receding from us on the backside of the galaxy, curves around to region 2, and ends in the northern extension approaching us on the near side of the galaxy.  In that case, the tidal tail on the line of sight through the galaxy would be on the near side of the galaxy as well. The looping structure in the southern part of the robust image (top right of Figure~\ref{fig:h36p-v3}) could be the southern end of the tail being partially resolved, with the west side of the loop associated with region 1 and the east side of the loop associated mostly with region 2.

Assuming that this is a tidal tail, we can also say that the motion of the tail relative to the disk points to the majority of the tidal tail being in front of the galaxy, as opposed to behind it.  With the main body being blueshifted to the east and redshifted to the west (see the top right panel of Figure~\ref{fig:h36x012}), the northern extension would be most likely expanding away from the main body of the galaxy \citep{toomre72}, towards us, placing the portion of the tail crossing the main body again on the near side of the galaxy.  The tail being in the line of sight of the main body could also be giving rise to some of the higher velocity dispersions in the main body of the galaxy. The highest velocity dispersions centered near $12^{\rm{h}}46^{\rm{m}}58^{\rm{s}}$, 51\degr36\arcmin25\arcsec\ in the southeast of the main body (see the red and orange regions in the bottom of Figure~\ref{fig:h36x012}) do not appear to be associated with the tidal tail seen in Figure~\ref{fig:h36p-v3}.  However, the velocity dispersions above 19.74 \kms\ throughout the western side of Haro 36's main body (see the yellow-green regions in the bottom of Figure~\ref{fig:h36x012}) appear to be showing the location of the tail as it moves into the line of sight of the main body between the northern extension and the southern region that makes up regions 1 and 2. The cause of the high velocity dispersions in red and orange, centered near $12^{\rm{h}}46^{\rm{m}}58^{\rm{s}}$, 51\degr36\arcmin25\arcsec\ remains unknown.

\begin{figure}[h]
\epsscale{0.5}
\plotone{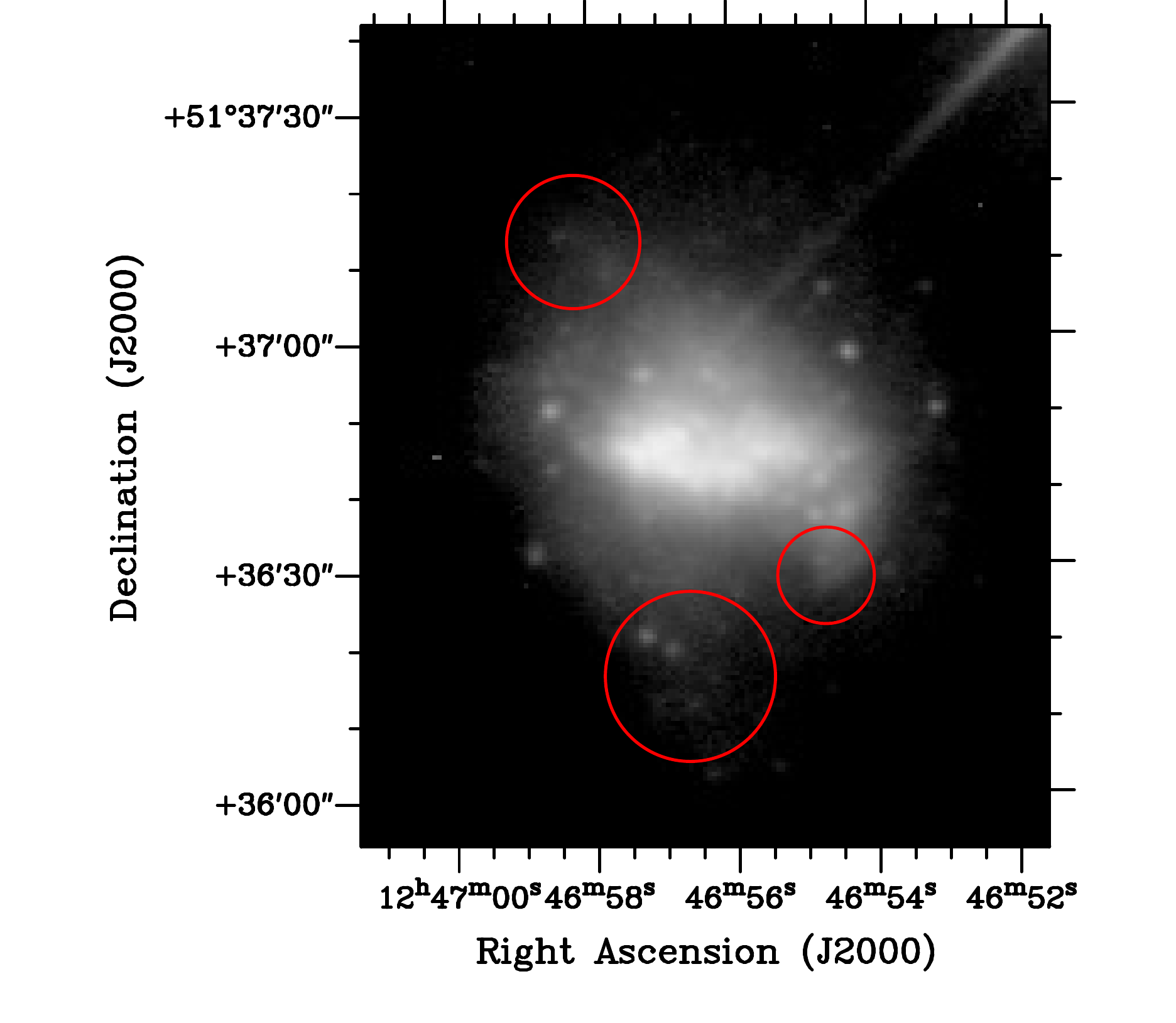}
\figcaption{The V-band emission seen in Figure~\ref{fig:h36v_fuv}; the red circles indicate the location of features that may be a result of the tidal tail in Haro 36. \label{fig:h36v_labels}}
\end{figure}

The optical component of the galaxy may also be showing signs of a tidal tail.  Figure~\ref{fig:h36v_labels} is the V-band emission of Haro 36 with the red circles indicating features that may be associated with the tidal tail. The circle furthest south in the image most prominently elongates the outer isophotes' major axis in the north-south direction, while the other two circled features are more subtle.  The northernmost and southernmost circled features do not appear to be well aligned with the tidal tail \HI\ features (see Figure~\ref{fig:h36x0r}).  The third, most western circled feature does, however, appear to possibly be associated with the looping feature of the robust-weighted map on the top right of Figure~\ref{fig:h36p-v3}.  This feature could be associated with the tidal tail seen in the \HI.

The tidal tail in Haro 36 shows that this galaxy is not truly isolated; it has likely interacted with something in the past.  The northern extension could be the beginning of a bridge, while one of the central peaks could be a gas cloud coming from the same interaction that created the northern extension, and is now falling into the center of the galaxy.  The two central peaks could also be from a merger and the northern extension from gas that was ejected from the main body during the beginning of the merger.  If the northern extension is the beginning of a larger structure, such as a bridge, then seeing the full extent of it with single dish observations will certainly help to classify it and possibly help determine the origin of the two kinematically separate peaks, if they are related.

\section{Conclusions}\label{sec:concl}

	We have presented the LITTLE THINGS \HI\ data for two BCDs, Haro 29 and Haro 36.  These data are combined from the B, C, and D arrays to make maps of high angular resolution ($\thicksim$7\arcsec) and high velocity resolution ($\thicksim$1.3 \kms).   Both of these galaxies have morphologically and kinematically disturbed \HI.  
	
	Haro 29 has two \HI\ peaks in the densest \HI\ region, where the kinematics appear the most organized at the resolution of our data.  Both peaks appear to be participating in solid body rotation together, as evident by a P-V diagram running through both of them.  The stars are highly concentrated to the west of both peaks, extending to just between the two peaks.  The large outflow velocities near the brightest stellar regions make the recent star formation the most likely cause of the two peaks through consumption, ionization, and dispersal of the gas.   The morphology and kinematics appear to be more disturbed outside of the densest \HI.  It is unclear what has caused these disturbances in the outer regions of Haro 29, but the asymmetric morphology, distorted outer kinematics, and misalignment of the kinematic major axis and optical major axis together make the scenario of an interaction or merger with another object possible.

	Haro 36 has three components in \HI: a main body, a northern extension, and a small looping structure in the south.  Within the main body, there are two central \HI\ peaks with the star forming region falling directly between the two.  The main body, however, appears highly disturbed with the two peaks being kinematically distinct.  These two peaks are so kinematically different from each other that a merger, accretion, or foreground gas cloud scenario is possible, although formation of the two \HI\ peaks by intense star formation cannot be ruled out.  The northern extension appears to be composed of gas with low velocity dispersion, as seen in the velocity dispersion field, and it has very different kinematics than the main body which are most evident in the P-V diagrams of Haro 36.  Haro 36 has a likely tidal tail starting at the southern looping structure on the far side and ending as the tenuous extension on the near side to the north indicating that this galaxy is not likely truly isolated. To search for and map large-scale \HI\ structures in the outer \HI\ pools of BCDs, deep single-dish observations are currently scheduled with the Robert C. Byrd Green Bank Telescope for both galaxies.

\acknowledgments
We would like to thank Elias Brinks for his invaluable advice.  We would also like to thank the anonymous referee for their helpful advice. This project was funded in part by the National Science Foundation under grant numbers AST-0707563 AST-0707426, AST-0707468, and AST 0707835 to Deidre A. Hunter, Bruce G. Elmegreen, Caroline E. Simpson, and Lisa M. Young.  The GALEX work was funded by NASA through grant NNX07AJ36G and by cost-sharing from Lowell Observatory. This research has made use of the NASA/IPAC Extragalactic Database (NED) which is operated by the Jet Propulsion Laboratory, California Institute of Technology, under contract with the National Aeronautics and Space Administration (NASA).

\clearpage
\begin{deluxetable}{lccccccc}
\tablecaption{Basic Galaxy Information\label{tab:galinfo}}
\tabletypesize{\scriptsize}
\tablewidth{0pt}
\tablehead{
\colhead{Galaxy Name} &  \colhead{R.A.} &  \colhead{Dec} &  \colhead{Distance\tablenotemark{a}} & \colhead{Systemic} & \colhead{$\rm{R}_{\rm{D}}$\tablenotemark{b}} & \colhead{$\rm{log\ SFR}_{\rm{D}}$\tablenotemark{c}} & \colhead{$\rm{M}_{\rm{V}}$\tablenotemark{d}} \\ \colhead{} & \colhead{(J2000)} & \colhead{(J2000)} & \colhead{(Mpc)} & \colhead{Velocity (\kms)} & \colhead{(kpc)} & \colhead{($\rm{M}_{\sun}\ \rm{yr}^{-1}\ \rm{kpc}^{-2}$)} &  \colhead{(mag)}}

\startdata

Haro 29 & $12^{\rm{h}}26^{\rm{m}}15^{\rm{s}}.9$ & 48\degr29\arcmin37\arcsec & 5.8 & 281 & $0.3\pm0.01$ & $-1.07\pm0.01$ & $-14.7$ \\  
Haro 36 & $12^{\rm{h}}46^{\rm{m}}56^{\rm{s}}$.4 & 51\degr36\arcmin47\arcsec & 9.3 & 502 & $0.69\pm0.01$ & $-1.55\pm0.01$ &  $-15.9$ \\

\enddata

\tablenotetext{a}{Haro 29's distance is from \citet{schulte01} and Haro 36's distance is from \citet{hunter12}}
\tablenotetext{b}{$\rm{R}_{\rm{D}}$ is the V-band disk scale length \citep{hunter06}.}
\tablenotetext{c}{$\rm{SFR}_{\rm{D}}$ is the star formation rate, measured from FUV data, normalized to an area of $\pi \rm{R}_{\rm{D}}^{2}$ \citep{hunter12}}
\tablenotetext{d}{\citet{hunter12}}

\end{deluxetable}
\clearpage

\clearpage
\begin{deluxetable}{lcccc}
\tablecaption{Observing Information \label{tab:obsinfo}}
\tabletypesize{\scriptsize}
\tablewidth{0pt}
\tablehead{
\colhead{Galaxy Name} & \colhead{Array} &  \colhead{Date Observed} &  \colhead{Project ID} & \colhead{Time on Source (hours)}}
\startdata

\multirow{4}{*}{Haro 29} & B & 08 Jan 28, 08 Jan 30 & AH927 & 9.2\\
& CnB & 08 Feb 7 & AH927 & 1.7\\
& C & 08 Mar 23, 08 Apr 19 & AH927 & 5.9\\
& D & 08 Jul 6, 08 Jul 29 & AH927 &  1.7\\ \hline
\multirow{3}{*}{Haro 36} & B & 08 Jan 15, 08 Jan 21, 08 Jan 27 & AH927 & 10.3\\
& C & 08 Mar 23, 08 Apr 15 & AH927 & 5.7\\
& D & 08 Jul 8, 08 Jul 24, 08 Jul 25 & AH927 &  2.2\\ \hline

\enddata
\end{deluxetable}

\clearpage
\begin{deluxetable}{llccccc}
\tablecaption{Map Information\label{tab:mapinfo}}
\tabletypesize{\tiny}
\tablewidth{0pt}
\tablehead{
\colhead{Galaxy} & \colhead{Weighting} &  \colhead{Synthesized} &  \colhead{Linear} & \colhead{Velocity Resolution} & \colhead{RMS per Channel} & \colhead{Integrated $\sigma$\tablenotemark{a}} \\  \colhead{Name} & \colhead{Scheme} & \colhead{Beam Size (\arcsec)} & \colhead{Resolution (pc)} & \colhead{(\kms)} & \colhead{(mJy $\rm{beam}^{-1}$)} & \colhead{(mJy\ $\rm{beam}^{-1}$ \kms)}} 

\startdata

\multirow{3}{*}{Haro 29} & Robust (r=0.5) & 6.83 x 5.55 & 200 & 2.58 & 0.52 & 7.96\\  
& Natural & 12.42 x 7.92 & 360 & 2.58 & 0.48 & 7.39\\  
& Convolved & 25 x 25 & 720 & 2.58 & 1.16 & 18.25\\ 
& Natural (no Hanning) & 12.42 x 7.92 & 360 & 1.81 & 0.78 & 7.97 \\ \hline
\multirow{3}{*}{Haro 36} & Robust (r=0.5) & 6.96 x 5.77 & 310 & 2.58 & 0.52 & 10.14\\  
& Natural & 12.43 x 8.91 & 560 & 2.58 & 0.47 & 9.17\\
& Natural (no Hanning) & 12.43 x 8.91 & 560 & 1.81 & 0.78 &  10.38\\

\enddata

\tablenotetext{a}{Calculated using Integrated\ $\sigma=\sqrt{\rm{N}}\ \rm{\Delta V}\ \sigma_{\rm{chan}}$ where N is the number of channels with signal, $\Delta$V is the channel width, and $\sigma_{\rm{chan}}$ is the average noise per channel \citep{deblok08}.}

\end{deluxetable}

\end{document}